\documentclass[12pt]{revtex4}

\usepackage[dvips]{graphicx}

\newcommand{\lsim}{\lower0.6ex\vbox{\hbox{$ \buildrel{\textstyle 
<}\over{\sim}\ $}}}
\newcommand{\gsim}{\lower0.6ex\vbox{\hbox{$ \buildrel{\textstyle 
>}\over{\sim}\ $}}}

\newcommand{\eg}{{\it e.g.,~}}
\newcommand{\pinocchio}{{\tt PINOCCHIO~}}

\newcommand{\beq}{\begin{equation}}
\newcommand{\eeq}{\end{equation}}

\newcommand{\hMsun}{\ h^{-1}\mathrm{M}_{\odot}}
\newcommand{\hMpc}{\ h^{-1}\mathrm{Mpc}}

\newcommand{\rhomean}{\rho_{\mathrm{M}}}

\newcommand{\Omegam}{\Omega_{\mathrm{M}}}
\newcommand{\Omegal}{\Omega_{\mathrm{\Lambda}}}

\newcommand{\Obhh}{\Omega_{\mathrm{B}}h^2}

\newcommand{\dd}{\mathrm{d}}

\newcommand{\avg}[1]{\langle #1 \rangle}
\newcommand{\abs}[1]{\vert #1 \vert}
\newcommand{\Vwindow}{V_{\mathrm{W}}}
\newcommand{\Rw}{R_{\mathrm{W}}}

\newcommand{\dc}{\delta_{\mathrm{c}}}
\newcommand{\dell}{\delta_{\mathrm{c}}^{\mathrm{ell}}}
\newcommand{\Mstar}{M_{\star}}

\newcommand{\dv}{\delta_{\mathrm{v}}}
\newcommand{\dcv}{\delta_{\mathrm{c}}^{\mathrm{v}}}
\newcommand{\nuv}{\nu_{\mathrm{v}}}
\newcommand{\gcv}{\gamma_{\mathrm{c,v}}}
\newcommand{\fv}{f_{\mathrm{v}}}
\newcommand{\DD}{\mathcal{D}}

\newcommand{\bh}{b_{\mathrm{h}}}
\newcommand{\deltahl}{\delta_{\mathrm{halo}}^{\mathrm{L}}}
\newcommand{\deltah}{\delta_{\mathrm{halo}}}

\newcommand{\DS}{\Delta S}
\newcommand{\Ddelta}{\Delta \delta}

\newcommand{\pDdelta}{(\pDdelta)}
\newcommand{\w}{\omega}

\newcommand{\wf}{\omega_{\mathrm{F}}}
\newcommand{\af}{a_{\mathrm{F}}}

\newcommand{\Mmin}{M_{\mathrm{min}}}
\newcommand{\Mrem}{M_{\mathrm{rem}}}
\newcommand{\fstep}{f_{\mathrm{step}}}

\newcommand{\Nscript}{\mathcal{N}}

\newcommand{\erf}{\mathrm{erf}}
\newcommand{\erfc}{\mathrm{erfc}}
\newcommand{\dirac}{\delta_{\mathrm{D}}}

\begin{document}

%
%
%

\title{
The Excursion Set Theory of Halo Mass Functions, 
Halo Clustering, and Halo Growth
}

\author{Andrew R. Zentner
\footnote{
National Science Foundation Fellow}
}
\affiliation{Kavli Institute for Cosmological Physics,
Department of Astronomy and Astrophysics, \& 
The Enrico Fermi Institute, 
The University of Chicago, 
Chicago, IL, 60605 USA; 
\\
zentner@kicp.uchicago.edu}


\begin{abstract}
I review the excursion set theory with particular attention 
toward applications to cold dark matter halo formation and growth, 
halo abundance, and halo clustering.  After a brief introduction to 
notation and conventions, I begin by recounting the heuristic argument 
leading to the mass function of bound objects given by Press \& Schechter.  
I then review the more formal derivation of the Press-Schechter 
halo mass function that makes use of excursion sets of the density field.  
The excursion set formalism is powerful and can be applied to 
numerous other problems.  
I review the excursion set formalism for describing 
both halo clustering and bias and the properties of void regions.  
As one of the most enduring legacies of the excursion set approach 
and one of its most common applications, I spend considerable time 
reviewing the excursion set theory of halo growth.  This section of 
the review culminates with the description of two Monte Carlo methods 
for generating ensembles of halo mass accretion histories.  In the 
last section, I emphasize that the standard excursion set approach 
is the result of several simplifying assumptions.  Dropping these 
assumptions can lead to more faithful predictions and open excursion 
set theory to new applications.  One such assumption is that the 
height of the barriers that define collapsed objects is a constant 
function of scale.  I illustrate the implementation of the excursion 
set approach for barriers of arbitrary shape.  One such application 
is the now well-known improvement of the excursion set mass function 
derived from the ``moving'' barrier for ellipsoidal collapse.  
I also emphasize that the statement that halo accretion histories 
are independent of halo environment in the excursion set approach 
is not a general prediction of the theory.  
It is a simplifying assumption.  
I review the method for constructing correlated 
random walks of the density field in the more general case.  
I construct a simple toy model to illustrate that 
excursion set theory (with a constant barrier height) 
makes a simple and general prediction for the relation between 
halo accretion histories and the large-scale environments of halos:  
regions of high density preferentially contain late-forming halos 
and conversely for regions of low density.  I conclude with a 
brief discussion of the importance of this prediction relative to 
recent numerical studies of the environmental dependence of halo properties.
\end{abstract}

\maketitle


\section{Introduction}
\label{section:intro}

In the standard, hierarchical, cold dark matter (CDM) paradigm of 
cosmological structure formation, galaxy formation begins with 
the gravitational collapse of overdense regions 
into bound, virialized halos of dark matter (DM).  
The average density of dark matter outweighs that 
of baryonic matter by roughly six to one.  
Bound in the potential wells of dark matter halos, 
baryons proceed to cool, condense, and form galaxies.  
Understanding the fundamental properties and abundances of 
these dark matter halos is the first, necessary step in 
understanding the properties of galaxies.

In this manuscript, I review the excursion set 
(also referred to as ``extended Press-Schechter'') approach to 
dark matter halo formation and halo clustering.  
I begin with some preliminaries that 
serve to define my notation and several common 
conventions in \S~\ref{section:notation}.  I review 
the heuristic argument that Press \& Schechter~\cite{press_schechter74}
used to derive their analytic mass function in \S~\ref{section:PS}.
In this section I also draw attention to the most 
obvious weakness of the Press-Schechter argument, 
often referred to as the cloud-in-cloud problem.  
In \S~\ref{section:excursion} I review the theory 
of excursion sets of the density contrast field and 
the manner in which the excursion set theory solves 
the cloud-in-cloud problem.  The standard excursion 
set theory then yields a halo mass function 
identical to the Press-Schechter proposal.

The excursion set approach is very powerful and yields 
a great deal of insight into numerous aspects of 
halo formation and halo clustering.  
Following the mass function, one of 
the most immediate applications of the excursion 
set approach is to predict the clustering properties 
of dark matter halos relative to the dark matter.  
I review the issue of halo bias in the context of the 
excursion set theory in \S~\ref{section:bias}.  The 
bias relation derived in \S~\ref{section:bias} was 
derived prior to widespread use of excursion set theory, 
but this section also partially serves as a warm-up for the more 
complex arguments that follow.  In \S~\ref{section:voids}, I 
summarize the excursion set theory of the void population.  
The content of \S~\ref{section:bias} and \S~\ref{section:voids} 
contain some redundant information, but my intention is that 
repetition of the basic arguments in these sections will 
reinforce the logic and methods of the excursion set approach.

In \S~\ref{section:trees}, I review excursion 
set predictions for halo conditional mass functions, 
halo accretion rates, and halo formation 
times.  The culmination of this review is the predominant 
algorithm for generating Monte Carlo merger histories 
on a halo-by-halo basis using the excursion set 
theory.  This algorithm, refined most recently by 
Somerville \& Kolatt~\cite{somerville_kolatt99} and 
Cole et al.~\cite{cole_etal00} to similar effect, 
is the subject of \S~\ref{section:mt}.

Recent advancement in both theoretical methods and 
observational data have emphasized the weaknesses of the 
simplest implementation of the excursion set model.  
Nevertheless, the basic idea of the excursion set model 
is extremely valuable because it provides a toolbox for 
making quick estimates for numerous halo and galaxy 
properties.  In addition, and perhaps more importantly, 
the excursion set approach remains one of the principle 
frameworks for qualitative reasoning and understanding of 
the complex results of direct, cosmological, numerical 
simulations.  The algorithm for generating merger trees 
is one of the most widely used aspects of excursion 
set theory and is likely to remain one of the primary 
legacies of excursion set theory.  These excursion set 
halo merger trees form the basis of countless semi-analytic 
explorations of the physical processes involved in galaxy 
formation (examples are 
Refs.~\cite{somerville_primack99,cole_etal00,kauffmann_haehnelt00}).  
The excursion set halo merger trees are considerably 
less costly to produce than direct $N$-body simulation, 
so they continue to be used to make general points 
(there are many examples, but an 
elegant recent example is Ref.~\cite{neistein_etal06}) 
and to study the effects of varying cosmological parameters 
(\eg Ref.~\cite{zentner_bullock03}).  Moreover, excursion set 
merger histories are also the basis of many models that 
aim to expand upon simulation results, either by 
extending results beyond the resolution of current simulations 
or building statistically-large halo samples in regimes 
where simulations have not, by modeling nonlinear dynamics 
in simplified ways 
(\eg Refs.~\cite{zentner_bullock03,benson_etal04,koushiappas_etal04,zentner_etal05,penarrubia_benson05,taylor_babul05a,taylor_babul05b}).

The simplest implementation of the 
excursion set theory, which is the primary focus of these 
lectures, is certainly not the only possible implementation 
of the excursion set approach.  Excursion set theory can 
be modified to approximate numerous physical effects that 
are ignored in the derivations of the most basic set of 
equations.  
I address extensions of the excursion set models 
and proposed improvements in the 
final section of this review, \S~\ref{section:beyond}.  
This section highlights areas where excursion set theory 
can be expanded and points toward the principle references 
where these extensions have been developed.  I focus primarily 
on generalized forms for the barrier shape and excursion 
sets of correlated random walks.  An interesting aspect 
of this section is the calculation of a general excursion 
set prediction for the dependence of halo formation 
times on the large-scale environments of halos.  
I close with a mention of the 
\pinocchio algorithm, developed by 
P. Monaco and collaborators primarily in 
Refs.~\cite{monaco_etal02,monaco_etal02b,taffoni_etal02}, 
which makes use of Lagrangian perturbation theory 
together with some of the ideas of the excursion set 
approach to produce, of many things, halo merger histories.

My primary aims are to present a relatively 
concise, pedagogical review of the basic logic and techniques of 
excursion set theory that can be used as a springboard 
to new studies and to collect the most important results 
of excursion set theory into a single manuscript, with a 
single notation.  As such, 
the reference list contained in this manuscript is by 
no means an exhaustive review of the literature on 
analytic halo formation or even the excursion set 
approach.  Many important contributions have been 
omitted in the interest of brevity and to limit the 
scope of the lectures.  However, the references 
should be sufficient to develop the fundamental logic of 
the excursion set approach and the basic tools needed 
to extend this approach to new problems.  
As a minimal approach, the basic development of 
the excursion set approach can be followed through 
the series of papers by 
Bardeen et al.~\cite{bardeen_etal86}, 
Bond et al.~\cite{bond_etal91}, 
Lacey \& Cole~\cite{lacey_cole93,lacey_cole94}, 
Sheth \& Lemson~\cite{sheth_lemson99}, 
and Somerville \& Kolatt~\cite{somerville_kolatt99}.  
The various manuscripts contained in the collection edited by 
Wax~\cite{wax_54} are an extremely valuable resource for 
understanding stochastic processes and deriving many 
of the basic results of excursion set theory (including 
results not contained in this review).  Of the 
articles reprinted in the Wax collection, 
those by Chandrasekhar~\cite{chandrasekhar43}, and 
Rice~\cite{rice_44,rice_45} are probably of the 
most immediate interest.

\section{Notation and Conventions}
\label{section:notation}

In the following, I consider fluctuations in the density field 
$\rho(\vec{x})$ described by the density contrast 
$\delta(\vec{x}) \equiv [\rho(\vec{x})-\rhomean]/\rhomean$, 
where $\rhomean$ is the mean mass density in the universe and 
$\vec{x}$ is a comoving spatial coordinate.  
In the standard paradigm, the universe is endowed with primordial 
density fluctuations during an epoch of cosmological inflation and 
the primordial density contrast is a statistically homogeneous and 
isotropic Gaussian random field.  This means that the joint probability 
distribution of the density contrast at a set of points in space is 
given by a multivariate Gaussian distribution.  
Homogeneity requires that the mean $\avg{\delta(\vec{x})}$, 
of the distribution and the two-point function 
$\avg{\delta(\vec{x}_1) \delta(\vec{x}_2)} \equiv \xi(\vec{x}_1,\vec{x}_2)$ 
be invariant under translations.  The two-point function is then 
a function only of the separation vector between two points,  
$\xi(\vec{x}_1,\vec{x}_2) = \xi(\vec{x}_1-\vec{x}_2)$.  
Isotropy requires that $\xi(\vec{x})$ is invariant under 
rotations as well, so the two-point correlation function is 
only a function of the distance between two points, 
$\xi(\vec{x}_1,\vec{x}_2) = \xi(\abs{\vec{x}_1-\vec{x}_2})$.

The Fourier transform of the density contrast is given by the 
convention
\beq
\delta(\vec{k}) =  \int \dd^3 x \ \delta(\vec{x}) e^{i\vec{k}\cdot \vec{x}}
\eeq
with the inverse transform
\beq
\delta(\vec{x}) = 
\frac{1}{(2\pi)^3} \int \dd^3 k \ \delta(\vec{k}) e^{-i\vec{k}\cdot\vec{x}}.
\eeq
Notice that the $\delta(\vec{k})$ have dimensions of volume and 
that for a real-valued field $\delta(\vec{x})$, the Fourier 
coefficients obey the relation $\delta(-\vec{k}) = \delta^{*}(\vec{k})$.  
We have implicitly assumed that there is some very large cut-off 
scale $L \equiv V^{1/3}$ that renders the integral 
$\int \abs{\delta(\vec{x})} \dd^3 x$ finite and that this 
scale is much larger than any other scale of interest so that 
it plays no meaningful role.  
Using these conventions, one can compute the two-point function 
$\xi(\vec{r}) \equiv \avg{\delta(\vec{x})\delta(\vec{x}+\vec{r})}$ 
in terms of the Fourier coefficients, where the average is taken 
over all space.  The two-point function is a function only of the 
amplitude of $\vec{r}$ due to isotropy, and the result is 
\beq
\label{eq:xiofr}
\xi(r) = 
\frac{1}{2\pi^2}\int k^3 V^{-1}\abs{\delta(k)}^2 \ \frac{\sin(kr)}{kr} \ \dd \ln k.
\eeq
The correlation function is the Fourier transform of the power spectrum
\beq
P(k) \equiv V^{-1} \avg{\abs{\delta(k)}^2},
\eeq
where the average is over an ensemble of universes with the same statistical 
properties.  The power spectrum has dimensions 
of volume and so a quantity that lends itself more easily to 
direct interpretation is the dimensionless combination 
\beq
\Delta^2(k) \equiv k^3P(k)/2\pi^2.
\eeq
The correlation function $\avg{\delta^2({\vec{x}})}$ is simply 
the mass variance.  From Eq.~(\ref{eq:xiofr}), $\Delta^2(k)$ 
is the contribution to the mass variance from modes 
in a logarithmic interval in wavenumber, 
so that $\Delta^2(k) \sim 1$ indicates order unity fluctuations 
in density on scales of order $\sim k$.

In the standard, cold dark matter (CDM) model, $\Delta^2(k)$ increases 
with wavenumber (at least until some exceedingly small scale 
determined by the physics of the production of 
the CDM in the early universe), but we observe 
the density field smoothed with some resolution.  Therefore, 
a quantity of physical interest is the density field smoothed on a 
particular scale $\Rw$,  
\beq
\label{eq:smoothdelta}
\delta(\vec{x};\Rw) \equiv 
\int \dd^3x' \ W(\abs{\vec{x}'-\vec{x}};\Rw) \delta(\vec{x}')
\eeq
The function $W(x;\Rw)$ is the window function that weights the 
density field in a manner that is relevant for the particular 
application.  According to the convention used in 
Eq.~(\ref{eq:smoothdelta}), the window function (sometimes 
called {\em filter} function) has units of inverse volume by 
dimensional arguments.  It is also useful to think of a window as 
having a particular window volume $\Vwindow$.  The window volume 
can be obtained operationally by normalizing $W(x)$ such 
that it has a maximum value of unity and is dimensionless.  
Call this new dimensionless window function $W'(x)$.
The volume is given by integrating to give 
$\Vwindow = \int \dd^3x W'(x)$.  In this way, one thinks of 
the window weighting points in the space by different amounts.  
It should be clear that $W(x) = W'(x)/\Vwindow$.  Roughly speaking, 
the smoothed field is the average of the density fluctuation in a region 
of volume $\Vwindow \sim \Rw^3$.  
The Fourier transform of the smoothed field is 
\beq
\label{eq:smoothk}
\delta(\vec{k};\Rw) \equiv W(\vec{k};\Rw) \delta(\vec{k}), 
\eeq
where $W(\vec{k};\Rw)$ is the Fourier transform of the window function.

The most natural choice of window function is probably a simple 
sphere in real space.  The window function is then  
\beq
\label{eq:thwindow}
W(x;\Rw) =  \begin{array}{lr}
			\frac{3}{4\pi\Rw^3} & \mbox{($x \le \Rw$)}\\
			0 & \mbox{($x > \Rw$)}
		    \end{array} .
\eeq
In this case, the smoothed field is the 
average density in spheres of radius $\Rw$ about point 
$\vec{x}$.  The window volume is simply $\Vwindow=4\pi\Rw^3/3$.  
However, this choice of window has the undesirable 
property that the sharp transition in configuration space leads 
to power on all scales in Fourier space.  Therefore, it is often 
convenient to smooth the boundary in real space.  As I discuss 
below, it is often convenient to introduce particular window functions 
to ensure that the smoothed field has particular properties.  The 
three most commonly used window functions are the real-space 
tophat window of Eq.~(\ref{eq:thwindow}), with Fourier transform 
\beq
\label{eq:thwindowk}
W(k;\Rw) = \frac{3[\sin(kR\Rw)-k\Rw \cos(k\Rw)]}{(k\Rw)^{3}},
\eeq
the Fourier-space tophat window, and the Gaussian window.  
The Fourier-space tophat is defined in Fourier space as 
\beq
\label{eq:sharpkk}
W(k;\Rw) = \begin{array}{lr}
		1 & \mbox{($k \le \Rw^{-1}$)}\\
		0 & \mbox{($k > \Rw^{-1}$)}
	\end{array}, 
\eeq
and is 
\beq
\label{eq:sharpkr}
W(x;\Rw) = \frac{1}{2\pi^2\Rw^3}
\frac{(\sin(x\Rw^{-1})-x\Rw^{-1}\cos(x\Rw^{-1}))}{(x\Rw^{-1})^3}
\eeq
in real space.  A disadvantage of this window is that it does not 
have a well-defined volume.  
This concern creeps up repeatedly in what follows.
The Gaussian window is 
\beq
\label{eq:gaussianr}
W(x;\Rw) = \frac{\exp(-x^2/2\Rw^2)}{(2\pi)^{3/2}\Rw^3}
\eeq
with a Fourier transform that also has the form of a Gaussian 
\beq
\label{eq:gaussiank}
W(k;\Rw) = \exp(-k^2/2\Rw^{-2}),
\eeq
with a width that is the reciprocal of its width in real space.
The volume of the Gaussian window is $\Vwindow=(2\pi)^{3/2}\Rw^3$.

\begin{figure}[t]
\begin{center}
\includegraphics[height=7.0cm]{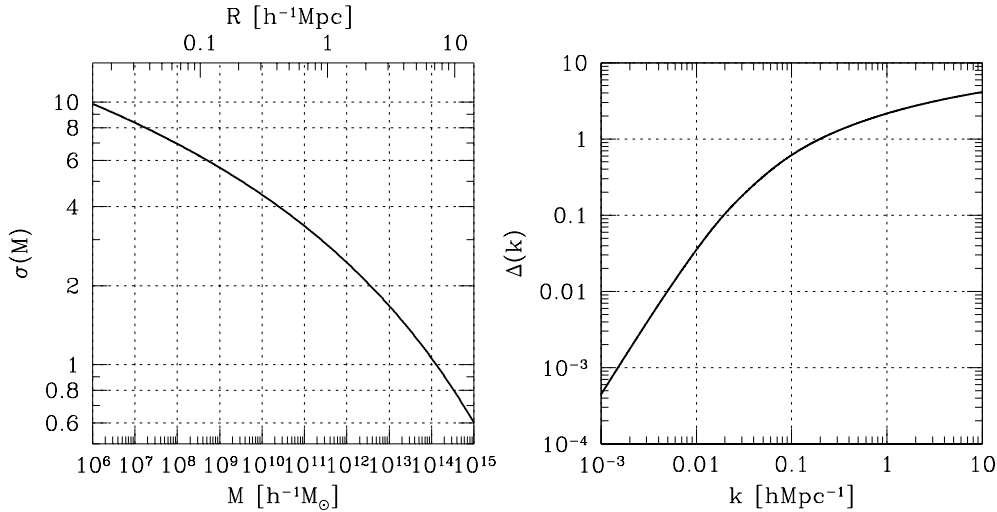}
\caption{
Power spectra in the standard $\Lambda$CDM cosmology with 
$\Omegam=1-\Omegal=0.3$, $h=0.7$, $\sigma_8=0.93$, and 
$\Obhh=0.022$.  The left panel shows the mass variance 
smoothed with a real space tophat window as a function of 
the smoothing mass or smoothing radius 
[Eq.~(\ref{eq:smoothvariance})].  The right panel 
shows the {\rm rms} density fluctuation per logarithmic 
interval of wavenumber as a function of wavenumber 
[Eq.~(\ref{eq:xiofr})].
\label{fig:ps}
}
\end{center}
\end{figure}

The density fluctuation field is assumed to be a Gaussian random 
variable so the smoothed density fluctuation field $\delta(\vec{x};R)$ is 
then a Gaussian random variable as well because it represents a 
sum of Gaussian random variables.  The variance of 
$\delta(\vec{x};R)$ can be computed in the same way as before 
[see Eq.~(\ref{eq:xiofr})] and is 
\beq
\label{eq:smoothvariance}
\sigma^2(R) = \avg{\delta^2(\vec{x};R)} 
= \int \dd \ln k \ \Delta^2(k) \abs{W(k;R)}^2.
\eeq
Thus, the probability of attaining a value of $\delta(\vec{x};R)$ 
between $\delta$ and $\delta + \dd \delta$ is 
\beq
\label{eq:Pdelta}
P(\delta;R)\dd \delta 
= \frac{1}{\sqrt{2\pi\sigma^2(R)}}\exp[ -\delta^2/2\sigma^2(R) ]\dd \delta.
\eeq
It is common to refer to the smoothing scale as either a length, 
as above, or a mass given by the mean density multiplied by the 
volume of the window $M = \rhomean\Vwindow$.  The left panel of 
Fig.~\ref{fig:ps} depicts a standard model power spectrum 
expressed as the variance of the smoothed density field, 
smoothed with a real-space tophat, and shows the smoothing scale 
in terms of both mass and length.  The right panel of 
Fig.~\ref{fig:ps} shows the {\em rms} density fluctuation per 
logarithmic interval in wavenumber $\Delta(k)$.  The power spectra 
were computed assuming $\Omegam=1-\Omegal=0.3$, $h=0.7$, 
$\sigma_8=0.93$, $\Obhh=0.022$, and a spectrum of primordial 
fluctuations with $P_{\mathrm{prim}}(k) \propto k$.  
I have used the transfer function of 
Eisenstein \& Hu~\cite{eisenstein_hu99} to process the scale-invariant 
primordial spectrum into the post-recombination spectrum of 
density fluctuations out of which nonlinear structures form.  

\section{The Press-Schechter Mass Function}
\label{section:PS}

Press \& Schechter~\cite{press_schechter74} 
derived a relation for the mass spectrum 
of virialized objects from the hierarchical density field.  
In hierarchical models, there is structure on all scales and 
the variance $\sigma^2(R) \rightarrow \infty$ as the 
smoothing scale $R \rightarrow 0$.  Press \& Schechter 
essentially assumed that objects will collapse on some small 
scale once the smoothed density contrast on this scale exceeds 
some threshold value, but that the nonlinearities introduced by these 
virialized objects do not affect the collapse 
of overdense regions on much larger scales.  The collapsed 
objects act only as resolution elements that trace 
the larger-scale fluctuations which 
may collapse at some later time.  Strictly speaking, 
this is not correct; however, it is approximately true  
when primordial power spectra are sufficiently shallow 
that the additional large-scale 
power generated by nonlinearities is small 
compared to the primordial fluctuations on these scales 
(a study in one-dimension is Ref.~\cite{williams_etal91}).  
Moreover, this assumption leads to a simple parsing of 
the ingredients in the formation of nonlinear structure.  
The first ingredient is a characterization of the 
statistical properties of the primordial density fluctuations.  
In the standard picture this is set during the inflationary epoch.  
The second ingredient is the evolution of overdensities according 
to linear perturbation theory.  This is encapsulated in the 
growth function $D(a) = \delta(k,a)/\delta(k,a=1)$ which is specified 
by the evolution of the background cosmology 
(\eg Refs.~\cite{carroll_etal92,bildhauer_etal92,lacey_cole93}).  
The last ingredient is the threshold for collapse into a 
virialized object and is determined by examining the 
nonlinear collapse of spherical overdensities.

Press \& Schechter stated that the 
likelihood for collapse of objects of a specific size or 
mass ($R \propto M^{1/3}$) could be computed by examining 
the density fluctuations on the desired scale.  They continued by 
using a model for the collapse of a spherical tophat overdensity 
to argue that collapse on scale $R$ should occur roughly when the 
smoothed density on that scale exceeds a critical value 
$\dc$, of order unity, independent of $R$.

The implementation of the Press-Schechter 
prescription is simple.  The mass within a region in which 
the smoothed density fluctuation (dictated by the 
linear theory) is the critical value $\dc$, at some 
redshift $z$, corresponds to an object that has just virialized 
with mass $M(R)$.  The relationship between mass and smoothing 
scale is set by the volume of the window function.  As an 
example, the relationship is $M=4\pi\rhomean R^3/3$ for a 
tophat window and $M=(2\pi)^{3/2} \rhomean R^3$ for a Gaussian 
window.  Further, any region that exceeds the 
critical density fluctuation threshold, will 
meet that threshold when smoothed on some larger scale $R'>R$.  
Consequently, the cumulative probability for a region to have a 
smoothed density above threshold gives the fractional volume 
occupied by virialized objects larger than the smoothing 
scale $F(M)$.  Integrating Eq.~(\ref{eq:Pdelta}), 
this probability is 
\beq
\label{eq:FofM}
F(M) = \int_{\dc}^{\infty} P(\delta;R)\dd \delta 
= \frac{1}{2} \erfc \Bigg(\frac{\nu}{\sqrt{2}}\Bigg), 
\eeq
where $\erfc (x)$ is the complementary error function, and 
$\nu \equiv \dc/\sigma(M)$ is the height of the 
threshold in units of the standard deviation of the smoothed 
density distribution.  In this model, collapse of mass $M$ 
is defined so that it occurs when the smoothed density 
fluctuation is $\dc$ on the appropriate scale.  Thus there 
is a typical scale that is collapsing at the present epoch, $\Mstar$, 
when the variance is $\sigma(\Mstar)=\dc$.

In the hierarchical power spectra that we consider, 
$\sigma(R)$ becomes arbitrarily large as $R$ becomes 
arbitrarily small.  Thus, $F(0)$ in Eq.~(\ref{eq:FofM}) should 
give the fraction of all mass in virialized objects; 
however, $\erfc(0) = 1$ so that Eq.~(\ref{eq:FofM}) states 
that only half of the mass density of the universe is contained 
in virialized objects.  Press \& Schechter noted this 
as a problem associated with not counting underdense regions 
in the integral Eq.~(\ref{eq:FofM}).  These authors 
argued that underdense regions will 
collapse onto overdense regions and multiplied 
$F(M)$ in Eq.~(\ref{eq:FofM}) by a factor of two in order to 
account for all mass.  Though the sense of this effect is 
certainly such that more mass will be contained in bound objects, 
that this should lead to precisely a factor of two 
increase in $F(M)$ is far from convincing.

Proceeding with this extra factor of two, the number of virialized 
objects with masses between $M$ and $M + \dd M$ is 
\beq
\label{eq:mf1}
\frac{\dd n}{\dd M}\dd M = 
\frac{\rhomean}{M} \Bigg| \frac{\dd F(M)}{\dd M}\Bigg| \dd M.
\eeq
In terms of the mass variance, this is 
\begin{eqnarray}
\label{eq:mf2}
\frac{\dd n}{\dd M}\ \dd M & = &
\sqrt{\frac{2}{\pi}} \frac{\rhomean}{M^2} 
\frac{\dc}{\sigma} \Bigg| \frac{\dd \ln \sigma}{\dd \ln M}\Bigg| 
\exp\Bigg( -\frac{\dc^2}{2\sigma^2} \Bigg) \ \dd M \nonumber \\
 & = & 
\sqrt{\frac{2}{\pi}}\frac{\rhomean}{M^2}\ \nu
\frac{\dd \ln \nu}{\dd \ln M}\ \exp\Bigg(-\frac{\nu^2}{2}\Bigg)\ \dd M.
\end{eqnarray}
Without regard to the details of the shape of the power spectrum, 
$\sigma(M)$ or $\nu(M)$, the mass function is close to a power 
law $\dd n/\dd M \propto M^{-2}$ for $M \ll \Mstar$ and is 
exponentially cut-off for $M \gsim \Mstar$.

\section{Excursion Set Theory of the Mass Function}
\label{section:excursion}

A weakness of the Press-Schechter approach is 
that it does not account for the fact that at a 
particular smoothing scale $\delta(\vec{x};R)$ may 
be less than $\dc$, yet it may be larger than 
$\dc$ at some {\em larger} smoothing scale $R'>R$.  
It seems natural that this larger volume should collapse 
to form a virialized object, overwhelming the more diffuse 
patches within it.  Clearly, the sense of including this effect is 
to increase the fraction of mass in collapsed objects 
and mitigate the factor of two discrepancy in the 
Press-Schechter formulas.  In the literature, 
the issue of regions below threshold on a particular 
scale, but above threshold on a larger scale is referred 
to as the ``cloud-in-cloud'' problem.  The cloud-in-cloud
problem is closely tied to the Press-Schechter factor of 
two as I discuss below.

To solve the cloud-in-cloud problem, it is necessary to compute the 
{\em largest} value of the smoothing scale for which the density 
threshold is exceeded.  This was done in a formal way by 
Bond et al.~\cite{bond_etal91} 
(see also Refs.~\cite{epstein83,peacock_heavens90,bower91}) 
and I follow their approach quite closely.  
The development of Bond et al. in turn follows closely the 
elegant review of stochastic processes by 
Chandrasekhar~\cite{chandrasekhar43}, 
and makes use of results from both Ref.~\cite{rice_44} and 
Ref.~\cite{adler81}.

In what follows, I consider the density contrast field evaluated 
at some early time far before any scales of interest have 
approached the nonlinear regime, but extrapolated to the present 
day using linear perturbation theory.  In this way, the 
density contrast field does not need to obey the physical 
constraint $\delta \ge -1$ because this is merely the linear 
extrapolation of a density contrast of much smaller magnitude.  
In addition, all coordinates are Lagrangian coordinates 
defined at the same early time so that the position of each 
mass element is independent of time.  The excursion set theory 
is, at its foundation, a set of rules for assigning mass elements 
to virialized objects of various sizes.  Where needed, I will 
introduce the necessary conversions that map the initial 
Lagrangian sizes of overdense and underdense regions onto 
Eulerian coordinates, thereby accounting for the contraction 
or expansion of such regions respectively.

Consider again evaluating the field $\delta(\vec{x};R)$ at various 
values of the smoothing scale $R$, at a single point $\vec{x}$.  For 
now, I will suppress the argument $\vec{x}$ and consider $\delta(R)$ as 
a function of smoothing scale at a single point in space.  For very large $R$, 
$\sigma(R) \ll \dc$ so the probability that the region lies above the 
boundary $\dc$ is vanishingly small.  With decreasing $R$, 
the standard deviation becomes larger and $\delta(R)$ will eventually 
reach $\dc$ at the first up-crossing of the boundary.  The problem is 
to compute the probability that the first up-crossing of the barrier 
at $\dc$ occurs on a scale $R$.  For simplicity in what follows, let 
$S \equiv \sigma^2(R)$ and let the value of $S$ serve to denote the 
smoothing scale by exploiting the fact that $S$ is a monotonically 
decreasing function of $R$.  I will then refer to the density contrast 
on that scale as $\delta(S)$.  Throughout this discussion, it is 
important to remember that increasing $S$ corresponds to decreasing 
$R$.  The problem is to compute the 
probability that the first up-crossing through the barrier occurs 
between a value $S$ and $S+\dd S$.

Consider starting at a large smoothing scale, or small $S=S_1$, where 
$\delta(S_1) \equiv \delta_1 < \dc$.  For a given change in the filtering scale 
$\DS$, there is some distribution for the probability 
of reaching $\delta_2$ after an increment $\DS = S_2 - S_1 > 0$.  
In general, this probability distribution may depend not only on the 
size of the step $\DS$, but the value of the density field on 
other scales.  
If the probability distribution for $\delta_2$ after an increment 
$\DS$ depends upon other points on the curve $\delta(S)$, 
solving for the probability distribution of $\delta$ at 
a given $S$ is nontrivial.  An important special case is when the 
smoothing window used to define $\delta(S)$ is a k-space tophat as in 
Eqs.~(\ref{eq:sharpkk})-(\ref{eq:sharpkr}).  In that case, increasing 
the filter scale corresponds to adding a set of independent Fourier 
modes to the smoothed density.  These modes have not played a 
role in determining $\delta(S)$ at other smoothing scales.  
In this special case, the transition 
probability for a change in density $\Ddelta$ 
associated with a change in filtering scale $\DS$ 
is Gaussian with zero mean and variance $S_2 - S_1 = \DS$, 
independent of the starting point $\delta_1(S_1)$.

\begin{figure}[t]
\begin{center}
\includegraphics[height=10.0cm]{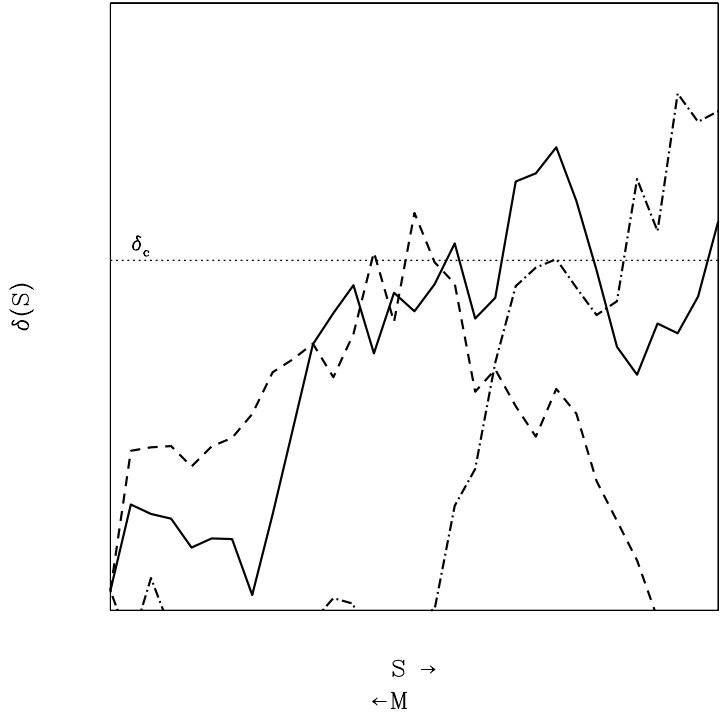}
\caption{
Three examples of random walks of $\delta(S)$ assuming each s
step is independent as in the case of a sharp k-space window 
function.  The axes are arbitrary.  The horizontal dotted line 
represents some threshold value $\dc$.  Notice that trajectories 
may penetrate the ``barrier'' at $\dc$ many times.  
\label{fig:traj}
}
\end{center}
\end{figure}

Following Bond et al.~\cite{bond_etal91}, it is common to refer to a 
sequence of $\delta(S_\mathrm{i})$ given by many subsequent increases 
of the smoothing scale by increments $\DS_{\mathrm{i}}$ as a 
trajectory for $\delta(S)$.  
In the case of k-space tophat filtering of the density field, 
each trajectory of $\delta(S)$ executes a Brownian random 
walk.  Three examples of such uncorrelated random walks are 
shown in Fig.~\ref{fig:traj}.  Notice that trajectories pierce 
the ``barrier'' at $\dc$ many times and drop below $\dc$ 
between subsequent up-crossings.  The aim of the excursion set 
approach is to solve the cloud-in-cloud problem by determining 
the largest smoothing scale $R$ or $M$, or equivalently the 
smallest value of the variance $S$, at which a trajectory 
penetrates the barrier at $\dc$.  

In the case of k-space tophat filtering, the probability of a 
transition from $\delta_1$ to $\delta_2 = \delta_1 + \Ddelta$ is 
\beq
\label{eq:transition}
\Pi(\delta_2,S_2)\ \dd\delta_2 = 
\Psi(\Ddelta;\DS) \ \dd (\Ddelta),
\eeq
where
\beq
\label{eq:Ptransition}
\Psi(\Ddelta;\DS)\ \dd (\Ddelta) 
= \frac{1}{\sqrt{2\pi\DS}} 
\exp \Bigg( - \frac{(\Ddelta)^2}{2(\DS)^2}\Bigg)\ \dd (\Ddelta)
\eeq
is the Gaussian transition probability.  Taking $S_1=0$, 
$S_2$ the smoothing scale of interest, and finding the 
probability of $\delta_2 \ge \dc$ returns the Press-Schechter 
probability for being in a collapsed object.  The fact that 
some regions will exceed $\dc$ for a smaller change in $S$ and 
then fall below $\dc$ by $\DS$ has not yet been accounted for.

Consider now, relating the distribution of $\delta$ at one value of 
the smoothing scale $\Pi(\delta,S)$ to the distribution on a 
subsequent step $\Pi(\delta,S+\DS)$ 
(smaller smoothing scale, larger $S$).  
This is  
\beq
\label{eq:markoff}
\Pi(\delta,S+\DS) = 
\int \dd(\Ddelta)\ \Psi(\Ddelta;\DS) \Pi(\delta-\Ddelta,S).
\eeq
Taylor expanding Eq.~(\ref{eq:markoff}) for small transitions, keeping 
terms up to $(\Ddelta^2)$, and integrating each term yields 
\beq
\label{eq:inter}
\frac{\partial \Pi}{\partial S} 
= \lim_{\DS \to 0} 
\Bigg( \frac{\avg{(\Ddelta)^2}}{2\DS} \frac{\partial^2\Pi}{\partial\delta^2} 
- \frac{\avg{\Ddelta}}{\DS} \frac{\partial \Pi}{\partial \delta} \Bigg).
\eeq
Using the fact that the transition probability is a Gaussian with 
$\avg{\Ddelta}=0$ and $\avg{(\Ddelta)^2}=\DS$ reveals 
\beq
\label{eq:heateq}
\frac{\partial \Pi}{\partial S} 
= \frac{1}{2}\frac{\partial^2 \Pi}{\partial \delta^2}
\eeq
as the relation governing the evolution of the probability 
distribution $\Pi$ with smoothing scale.  The probability 
distribution $\Pi$ for trajectories $\delta(S)$ that {\em never} 
exceed $\dc$ prior to some specific value of $S$ can be 
obtained by solving Eq.~(\ref{eq:heateq}) with the appropriate 
boundary conditions.  The first boundary condition is that 
$\Pi(\delta,S)$ is finite as $\delta \to -\infty$.  Imagining 
each trajectory as absorbed and removed from the sample of 
trajectories at the moment it first crosses the threshold 
at $\dc$, then $\Pi(\dc,S) = 0$ provides the second boundary 
condition.

Eq.~(\ref{eq:heateq}) may have a familiar form as it is 
also an equation describing diffusion with no drift 
or a one-dimensional heat equation describing 
heat flow in a long, thin, semi-infinite bar.  
The boundary condition at $\delta = \dc$ is  
analogous to fixing the temperature to zero at this end of the bar.  
Take $\delta(S_0)=\delta_0$ as an arbitrary starting point for 
the random walk trajectories so that the initial condition is 
$\Pi(\delta_0,S_0) = \dirac(\delta_0)$ where 
$\dirac(x)$ is the Dirac delta function.  Eq.~(\ref{eq:heateq}) 
can now be solved using familiar techniques.  It is 
easiest to work in a shifted variable 
$\gamma \equiv \dc - \delta$ so that the finite 
boundary is at $\gamma = 0$.

The first step is to Fourier transform Eq.~(\ref{eq:heateq}) in 
the variable $\gamma$ (with conjugate $\omega$).  Thus we define 
the Fourier transform of the probability distribution 
\beq
\label{eq:fourierpi}
\tilde{\Pi}(\omega,S) = \int \dd \gamma\ \Pi(\gamma,S) e^{i\omega\gamma}.
\eeq
This gives an equation for the transform 
\beq
\label{eq:heateqF}
\frac{\partial \tilde{\Pi}}{\partial S} = -\frac{\omega^2}{2}\tilde{\Pi}
\eeq
with solution
\beq
\label{eq:heatsF}
\tilde{\Pi}(\omega,S) = c(\omega) \ \exp\Bigg(\frac{-\omega^2}{2}S\Bigg).
\eeq
The boundary condition at $\gamma=0$ ($\delta=\dc$) guarantees 
that $c(\omega)$ is an odd function, so 
\beq
\label{eq:heat1}
\Pi(\delta,S) 
= \int_{0}^{\infty} c(\omega)\sin(\omega\gamma)\ e^{-\omega^2 S/2}\ \dd \omega.
\eeq
Applying the initial condition to Eq.~(\ref{eq:heat1}) gives 
\beq
\label{eq:heatcw}
c(\omega)=\frac{2}{\pi}\sin(\omega\gamma_0)
\exp\Bigg(\frac{\omega^2}{2}S_0\Bigg), 
\eeq
where $\gamma_0 \equiv \delta_0-\dc$.  
The final solution then follows by inverting the transform, so 
\beq
\label{eq:heatinv}
\Pi(\gamma,S) = 
\frac{2}{\pi}\int_{0}^{\infty} 
\sin(\omega\gamma_0) \sin(\omega\gamma)\ 
\exp\Bigg(\frac{-(S-S_0)}{2}\omega^2\Bigg)\ \dd \omega, 
\eeq
from which 
\beq
\label{eq:heatsolved}
\Pi(\delta,S) = 
\frac{1}{\sqrt{2\pi\DS}}
\Bigg[\exp\Bigg(-\frac{(\Ddelta)^2}{2\DS}\Bigg) 
- \exp\Bigg(-\frac{[2(\dc-\delta_0)-\Ddelta]^2}{2\DS}\Bigg)\Bigg],
\eeq
where $\DS = S - S_0$ and $\Ddelta = \delta - \delta_0$.  
The solution to Eq.~(\ref{eq:heateq}) in Eq.~(\ref{eq:heatsolved}) 
depends on the initial condition only with regard 
to the height of the barrier at $\dc$ in 
relation to the starting position at $\delta_0$.  
The first term in Eq.~(\ref{eq:heatsolved}) is the Gaussian 
distribution that represents the points above threshold at 
$S$ while the second term accounts for the trajectories that 
have been removed because they crossed above threshold at a 
filter scale $S' < S$ but would have crossed back below 
the threshold by $S$.  Chandrasekhar~\cite{chandrasekhar43} 
gives an elegant solution to this ``absorbing barrier'' 
problem almost upon inspection, exploiting 
the fact that random walks that hit the barrier 
have equal probability of continuing upward after their 
first encounter with the barrier as they do of continuing 
downward.  Upon making that realization, the solution above can 
be obtained using the method of images and subtracting 
the contribution from another source beginning at 
$\Pi(\delta,S=S_0) = \dirac(\delta-2\dc)$, a shift of 
$\dc$ {\em above} the threshold.  The derivation given here 
is most useful because it contains the logic needed to 
generalize the problem to more complicated processes 
and barriers.

The fraction of trajectories that have crossed above the 
threshold at or prior to some scale $S(M)$ is the complement 
of the $\Pi$ distribution, 
\beq
\label{eq:Fexcursion}
F(S) 
= 1 - \int_{-\infty}^{\dc} \Pi(\delta,S)\ \dd \delta 
= \erfc\Bigg(\frac{\dc-\delta_0}{\sqrt{2}\DS}\Bigg).
\eeq
Taking $S_0 = 0$ and $\delta_0 = 0$ to indicate a starting 
value at very large smoothing scale, 
Eq.~(\ref{eq:Fexcursion}) is precisely the value given by 
Press \& Schechter~\cite{press_schechter74}, 
but without having to introduce an 
arbitrary factor of two to achieve the correct normalization.  
The second term in Eq.~(\ref{eq:heatsolved}) that accounts 
for trajectories that crossed threshold at some large scale 
and then traversed back back below threshold by $S$ accounts for 
the probability missing from the Press-Schechter 
argument.

The differential probability for a first piercing of the 
threshold then follows by differentiation, 
\begin{eqnarray}
\label{eq:fdelta}
f(S|\delta_0,S_0)\dd S \equiv \frac{\dd F}{\dd S}\dd S
 & = & \Bigg(\int_{-\infty}^{\dc} 
\frac{\partial \Pi}{\partial S}\dd \delta \ \Bigg) \dd S \nonumber\\
 & = & 
\frac{1}{2}\Bigg[\frac{\partial \Pi}{\partial \delta}\Bigg]_{-\infty}^{\dc} 
\ \dd S \nonumber \\
 & = & \frac{\dc-\delta_0}{\sqrt{2\pi}\DS^{3/2}}\ 
\exp\Bigg[\frac{(\delta_c-\delta_0)^2}{2\DS}\Bigg]\ \dd S.
\end{eqnarray}
Eq.~(\ref{eq:fdelta}) is the 
fundamental relation as it gives the probability for first  
crossing a threshold given any starting point and any change 
in filtering scale $\DS$.  The function $f(S|\delta_0,S_0)$ is 
often referred to as the ``first-crossing distribution.''  
In cases where the random walk starts from 
$S_0=0$ with $\delta_0=0$ these arguments are often 
suppressed and the first-crossing distribution is written simply $f(S)$.  
The fraction of mass in collapsed objects in a narrow range of 
masses is obtained by setting $S_0=0$ and 
$\delta_0=0$ and rewriting Eq.~(\ref{eq:fdelta}) 
in terms of the mass that corresponds to the variance 
$M(S)$, 
\beq
\frac{\dd F}{\dd M} =  
\frac{1}{\sqrt{2\pi S}}\ \frac{\dc}{S} 
\Bigg|\frac{\dd S}{\dd M}\Bigg| \ \exp\Bigg(-\frac{\dc^2}{2S}\Bigg).
\eeq
The mass function follows by using Eq.~(\ref{eq:mf1}) and 
substituting $\sigma^2 = S$ results in 
\beq
\label{eq:mfexcursion}
\frac{\dd n}{\dd M} = 
\sqrt{\frac{2}{\pi}}\ \frac{\rhomean}{M^2}\ \frac{\dc}{\sigma}
\Bigg|\frac{\dd \ln \sigma}{\dd \ln M}\Bigg|\ 
\exp \Bigg(-\frac{\dc^2}{2\sigma^2}\Bigg).
\eeq
Again, this is precisely the Press-Schechter mass function 
with no {\it ad hoc} factor of two.  

\begin{figure}[t]
\begin{center}
\includegraphics[height=12.0cm]{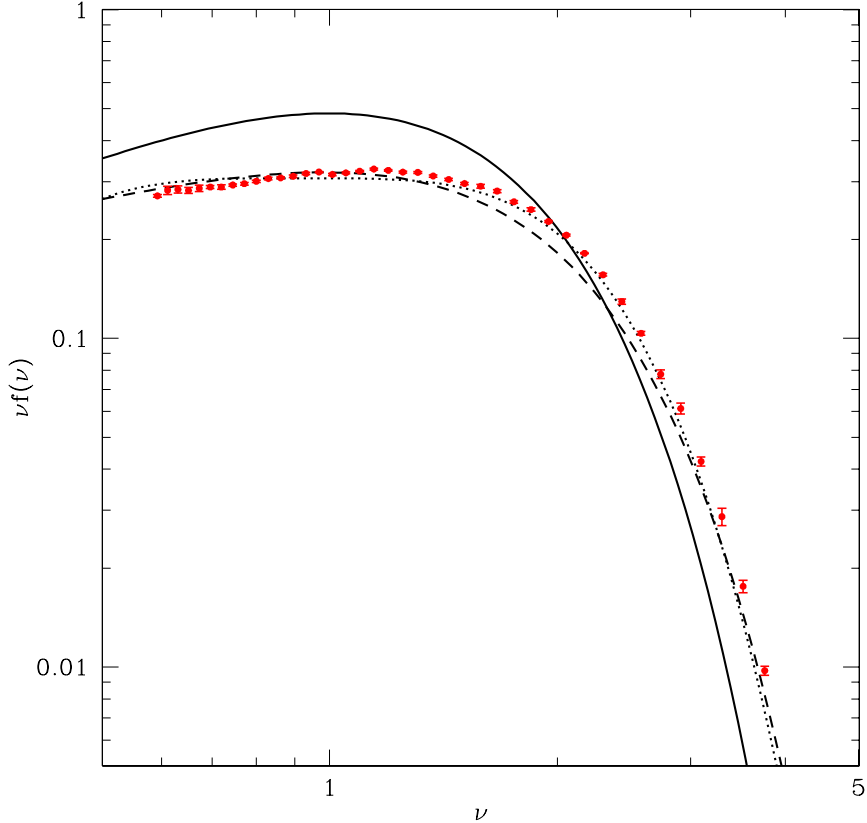}
\caption{Collapsed mass fractions.  The solid line represents 
the standard excursion set theory predictions.  The 
dashed and dotted lines represent the improved fits of 
Sheth \& Tormen~\cite{sheth_tormen99} and 
Jenkins et al.~\cite{jenkins_etal01} 
respectively.  The points represent numerical data from a 
suite of N-body simulations reluctantly provided by J. L. Tinker.
\label{fig:fnu}
}
\end{center}
\end{figure}

Analytic mass spectra are often compared to simulation 
data without reference to specific power spectra or 
cosmological models by 
comparing the fraction of mass in collapsed objects 
per logarithmic interval in $\nu$, 
$\nu f(\nu) \equiv \dd F/\dd \ln \nu$.  For the 
standard excursion set theory, 
\beq
\label{eq:fnu}
f(\nu) = \sqrt{\frac{2}{\pi}} \exp\Bigg(-\frac{\nu^2}{2}\Bigg).
\eeq
In Fig.~\ref{fig:fnu}, I compare the Press-Schechter/Excursion 
set predictions for the fraction of mass in collapsed objects 
with the results of a suite of cosmological numerical simulations.  
While excursion set theory may explain the gross features of the 
mass spectrum of dark matter halos, it fails in its detail.  
The Press-Schechter/Excursion set relations predict too many 
low-mass halos and too few high-mass halos.
Given the simplicity of the excursion set model described above, 
this is not surprising and the level of agreement provides 
encouragement that the excursion set model is a useful tool 
for understanding the gross features of halo abundance, 
formation, and clustering.  In particular, these gross features 
are set by the statistics of the initial density fluctuations.

One detail not discussed so far is the assignment of mass to 
a particular filter radius $R$ or variance $S$.  As I have 
already mentioned, the Press-Schechter relation 
follows directly by assuming a window function that is a 
tophat in Fourier space, but such a window does not have a 
well-defined volume to associate with it.  The method of 
circumventing this that is implemented most often is to 
simply use the formulas derived from the sharp k-space 
filtering assumption, but assign mass according to a 
configuration space tophat filter so that 
$M=4\pi R^3/3 \rhomean$ (though this is not the method 
used by Bond et al. in Ref.~\cite{bond_etal91}).

In the context of excursion set theory, 
the k-space tophat window function 
serves only to simplify the solution for $\Pi(\delta,S)$ 
[Eq.~(\ref{eq:heatsolved})] and the first barrier 
up-crossing distribution [Eq.~(\ref{eq:fdelta})].  In 
principle, other windows can be used but the mathematics 
become significantly more complicated.  This is because 
the steps are no longer independent so it is necessary 
to compute the entire trajectory at once to account for 
the correlations between the steps.  
Ref.~\cite{bond_etal91} demonstrates the general procedure 
which amounts to generating a large number of solutions 
to a Langevin equation of random motion 
(\eg, see Ref.~\cite{chandrasekhar43}) with a 
correlated stochastic force.  Each solution 
yields a single trajectory of density at each value 
of the smoothing scale.  The probability distributions 
can be computed from an ensemble of trajectories.  This 
procedure is potentially complicated and 
time consuming, and it has the drawback 
that it does not yield a closed-form solution 
for the $\Pi(\delta,S)$ or first up-crossing distributions.  
For these reasons, other filters are rarely discussed; however, 
it is important to keep in mind that the lack of correlations 
between different steps is not a prediction of 
excursion set theory.  Contrarily, it is a simplifying 
assumption used in the most common implementation of 
excursion set theory.

\section{Excursion Set Theory of the Spatial Bias of Dark Matter Halos}
\label{section:bias}

It has long been appreciated that within the context of 
cosmological structure formation theories, 
the clustering of dark matter halos differs 
from the overall clustering of 
matter~\cite{kaiser84,efstathiou_etal88,mo_white96,sheth_tormen99,seljak_warren04}.
The excursion set formalism provides a neat framework 
with which to understand the relative clustering of halos.  
The argument is much the same 
as the peak-background split approach developed in 
Refs.~\cite{kaiser84,efstathiou_etal88,cole_kaiser89}, 
and was expressed in the framework 
of the excursion set model by Mo \& White~\cite{mo_white96} 
as I outline below.

Consider the solution to the excursion set problem in  
Eq.~(\ref{eq:heatsolved}).  This gives the probability 
distribution of $\delta$ given that on a smoothing scale 
$S_0$, the smoothed density fluctuation is $\delta_0$.  
Notice that the important quantity is the relative height 
of the density threshold $\dc - \delta_0$ so that in regions 
with $\delta_0 > 0$ on large scales, trajectories are more 
likely to penetrate the barrier at $\dc$ and conversely 
for $\delta_0 < 0$.  This means that 
Eq.~(\ref{eq:Fexcursion}) and Eq.~(\ref{eq:fdelta}) 
allow for the calculation of the mass function in 
regions with specific large-scale density fluctuations.

The fraction of mass in collapsed halos of mass greater 
than $M$ in a region that has a smoothed density fluctuation 
$\delta_0$ on scale $S_0$ (corresponding to mass $M_0$ and 
volume $V_0$) is given by Eq.~(\ref{eq:Fexcursion}), 
\beq
\label{eq:Fcondition}
F(M|\delta_0,S_0) = \erfc\Bigg(\frac{\dc-\delta_0}{2\DS}\Bigg).
\eeq
Notice that as the density of the region increases, 
$F$ increases because smaller upward excursions are 
needed to cross the threshold.  When $\delta_0 \to \dc$, 
$F \to 1$ because the entire region will then be interpreted 
as a collapsed halo of mass $M_0$.  The fraction of mass 
in halos with mass in the range $M$ to $M+\dd M$ is 
\begin{eqnarray}
\label{eq:fcondition}
f(M|\delta_0,S_0)\Bigg|\frac{\dd S}{\dd M}\Bigg|\dd M & 
\equiv & \frac{\dd F(M|\delta_0,S_0)}{\dd M}\dd M \nonumber \\
 & = & \frac{1}{\sqrt{2\pi}}\frac{\dc-\delta_0}{\DS^{3/2}}
\Bigg|\frac{\dd S}{\dd M}\Bigg|\ \exp\Bigg[-\frac{(\dc-\delta_0)^2}{2\DS}\Bigg]\dd M,
\end{eqnarray}
so that regions with smoothed density $\delta_0$ on scale $S_0$ 
contain, on average, 
\beq
\Nscript(M|\delta_0,S_0)\dd M 
= \frac{M_0}{M}\ f(M|\delta_0,S_0)\Bigg|\frac{\dd S}{\dd M}\Bigg|\dd M
\eeq
halos in this mass range.

The quantity of interest is the relative over-abundance of halos in 
dense regions compared to the mean abundance of halos, 
\beq
\label{eq:dhldef}
\deltahl = \frac{\Nscript(M|\delta_0,S_0)}{(\dd n(M)/\dd M)V_0} - 1,
\eeq
where $\dd n(M)/\dd M$ is the mean number density of halos in a mass 
range of width $\dd M$ about $M$ from Eq.~(\ref{eq:mfexcursion}).  
The superscript $\mathrm{L}$ indicates that this is the overdensity in 
the initial Lagrangian space determined by the mass distribution at 
some very early time, ignoring the dynamical evolution 
of the overdense patch.

The relative overdensity of halos in large overdense and underdense 
patches is easy to compute.  In sufficiently large regions, 
$S_0 \ll S$, $\delta_0 \ll \dc$.  Expanding Eq.~(\ref{eq:dhldef}) to 
first order in the variables $S_0/S$ and $\delta_0/\dc$ gives a 
simple relation between halo abundance and dark matter density 
(see also Refs.~\cite{efstathiou_etal88,cole_kaiser89}) 
\beq
\label{eq:biasL}
\deltahl = \frac{\nu^2 - 1}{\dc}\delta_0,
\eeq
where $\nu = \dc/S^{1/2} = \dc/\sigma(M)$ as before.  The overdensity 
in the initial Lagrangian space is proportional to the dark matter 
overdensity and is a function of halo mass through $\nu$.  The final 
ingredient needed to relate the abundance of halos to the matter density 
is a model for the dynamics that can map the initial Lagrangian volume 
to the final Eulerian space.  Let $V$ and $\delta$ represent the 
Eulerian space variables corresponding to the Lagrangian space 
variables $V_0$ and $\delta_0$.  The final halo abundance is 
\beq
\label{eq:dh1}
\deltah = \frac{\Nscript(M|\delta_0,S_0)}{(\dd n(M)/\dd M)V}-1.
\eeq
%

\begin{figure}[t]
\begin{center}
\includegraphics[height=10.0cm]{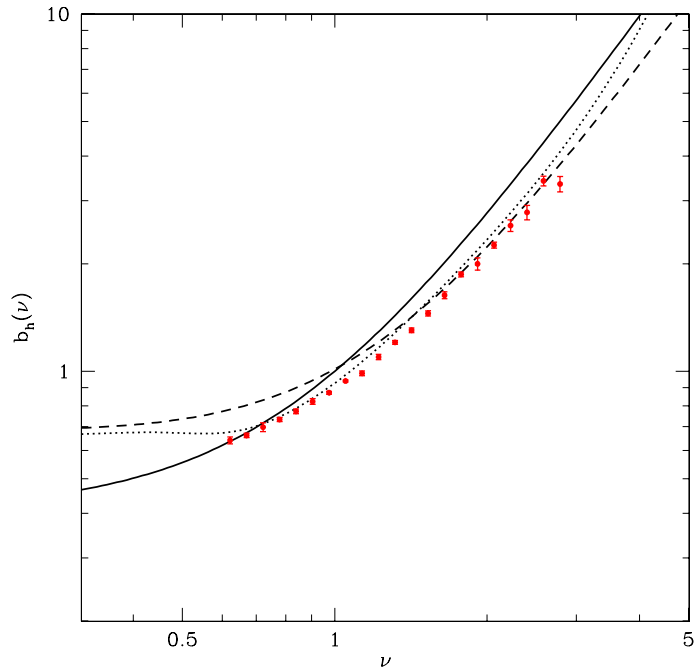}
\caption{
The large-scale bias of dark matter halos as a function 
of the scaled variable $\nu$.  The solid line gives the 
prediction of the standard Press-Schechter or excursion 
set theory in Eq.~(\ref{eq:dh}).  The dashed line is a 
modified form for the bias that follows from a modified 
form for the barrier criterion given by 
Sheth \& Tormen in Ref.~\cite{sheth_tormen99}.  
The dotted line represents an 
empirical fit to N-body simulation results given by 
Seljak \& Warren~\cite{seljak_warren04}.  
The points represent numerical data from 
a suite of N-body simulations kindly provided 
by J.~L. Tinker.  
\label{fig:bias}
}
\end{center}
\end{figure}

Mo \& White~\cite{mo_white96} give an extensive discussion of the 
mapping from Lagrangian to Eulerian coordinates and 
use a spherical collapse model to determine the 
appropriate mapping quantitatively.  
In the limit of a small overdensity 
$\delta_0 \ll 1$, $V \simeq V_0(1+\delta)$, 
$\delta \simeq \delta_0$, and 
\begin{eqnarray}
\label{eq:dh}
\deltah & = & \Bigg( 1 + \frac{\nu^2 - 1}{\dc}\Bigg)\delta \\
 & \equiv & \bh\delta.
\end{eqnarray}
The halo overabundance is proportional to the matter overdensity 
and always has the same sense.  This proportionality is commonly 
referred to has the halo bias $\bh = (1 + [\nu^2-1]/\dc)$.  
Interestingly, Eq.~(\ref{eq:dh}) gives a mass 
scale at which halos are not biased with respect to the dark matter.  
At a halo mass equal to the typical collapse mass, 
$M=\Mstar$, $\nu=1$ or $\sigma(M)=\dc$ so that the 
halo overdensity is the same as the total matter 
overdensity, $\deltah = \delta$.  
Larger halos will have larger values of 
$\nu = \dc/\sigma(M)$ because $\sigma(M)$ decreases with 
mass, so larger halos cluster considerably more strongly than 
the overall clustering of mass, while halos with $M<\Mstar$ cluster 
more weakly than the overall mass distribution.

The large-scale halo bias relation of Eq.~(\ref{eq:dh}) 
is plotted as a function of scaled peak height $\nu$, 
in Fig.~\ref{fig:bias}.  In that figure, I also plot 
numerical results for the relative clustering bias of 
halos as well as improved fits for halo bias given by 
Sheth \& Tormen~\cite{sheth_tormen99} and 
Seljak \& Warren~\cite{seljak_warren04}.  
The fit of Sheth \& Tormen in 
Ref.~\cite{sheth_tormen99} can be motivated by 
excursion set theory with a modified form for the 
shape of the barrier as a function of smoothing scale 
(\eg Refs.~\cite{sheth_etal01,sheth_tormen02}, 
and see \S~\ref{section:beyond}). 
These simple prescriptions for halo bias form one of the cornerstones 
of the analytic halo model that is often used to quantify the 
clustering of both dark matter and galaxies rapidly without 
resorting to a full numerical 
treatment~\cite{scherrer_bertschinger91,seljak00,ma_fry00,peacock_smith00,berlind_weinberg02}.  
As with the mass function, the standard excursion set approach 
gives an understanding of the general features of the 
halo bias, but fails to reproduce the halo bias in 
detail.

\section{An Excursion Set Model for Voids in the Galaxy Distribution}
\label{section:voids}

The excursion set model relates the initial density field on 
particular scales to specific structures in the evolved 
density field.  To compute the properties of halos, one 
considers overdensities in the initial smoothed density field 
but it seems natural that this logic can be extended to underdense 
regions, and in particular, to voids in the density distribution.  
This is a particularly interesting application of the 
excursion set theory because of the relatively simple 
dynamics of underdense patches.  Whereas 
overdense patches collapse, underdense patches expand more 
rapidly than the universal expansion rate.  Contrary to the 
complicated behavior of overdensities, general underdense 
patches that are not initially spherically-symmetric, tophat 
underdensities evolve toward spherically-symmetric tophat 
underdensities 
~\cite{icke84,bertschinger85,weygaert_kampen93,sheth_weygaert04}.  
This fact is encouraging.  It suggests 
that a density threshold criterion based on a spherical, tophat 
model may be a more realistic model of void evolution than 
it is of the collapse of overdense patches.

Sheth \& Van Weygaert~\cite{sheth_weygaert04} 
made precisely this realization and 
applied the excursion set model to predict the abundance and 
sizes of voids in the universe.  Most of the details of the 
excursion set model of voids can be found in their paper. 
The logic is quite similar to that of the previous two sections.  
First, it is necessary to assign a barrier height to designate 
a region that should be identified with a void, $\dv < 0$.  
(At this point, I reiterate that the overdensities that we 
consider are initial overdensities evaluated at some very 
early time, before any scales of interest have gone 
nonlinear, and extrapolated to $z=0$ 
using the linear theory of evolution 
so that overdensities are {\em not} 
constrained to $\delta \ge -1$.)  
The peculiar acceleration of a mass shell in a spherical 
underdense region is proportional to the integrated deficit in 
mass within the spherical region.  A consequence of this is 
that inner shells accelerate outward faster than shells of 
mass on the outer edge of an underdense region.  Mass shells 
start to accumulate in a narrow range of radii leading to a 
very high density ridge on the outer boundary of the underdense 
patch.  Eventually, the inner shells overtake the outer shells 
at the moment of {\em shell crossing}.  The relevant calculations 
are collected and reviewed in 
the appendices of Ref.~\cite{sheth_weygaert04}.

In the now defunct standard cold dark matter model with 
$\Omegam=1$, shell crossing for a spherical tophat 
overdensity occurs at a linearly-extrapolated  overdensity 
of $\dv = -2.81$.  This is analogous to the case of spherical 
tophat overdensity collapse, where collapse to a point occurs 
when the linearly-extrapolated overdensity is $\dc=1.69$.  As with 
the spherical tophat overdensity, there is a weak dependence on 
cosmological parameters that I will ignore for the purposes of 
this discussion.  Blumenthal et al.~\cite{blumenthal_etal92} 
argued that galaxy voids should be identified with 
regions that have just experienced shell crossing.  
Again, this is analogous to defining halos as regions that 
have just experienced complete collapse.  In this case, 
the galaxies in the void walls would lie on the overdense 
ridges that form at shell crossing.  The model of 
Sheth \& Van Weygaert~\cite{sheth_weygaert04} 
adopts this identification 
of voids with regions that just experience shell 
crossing and identify voids with the largest 
regions for which the smoothed density achieves 
the threshold $\dv=-2.81$.

In contrast to the case of halos, voids present an 
additional challenge to computing the first crossing 
distribution.  The additional difficulty is that 
an underdense region may be contained in a still 
larger overdense region which may collapse around 
the underdensity, engulfing and eliminating the void.  
Therefore, it is sensible that there be a second criterion that 
the smoothed density not exceed some overdensity 
threshold $\dcv > 0$, on some scale larger than the first 
crossing of $\dv$.  The appropriate height of the second 
barrier $\dcv$ is not obvious.  
It is clear that $\dcv \le \dc$ because a void cannot 
reside within a virialized halo.  Thus, the least 
restrictive choice that can be made is 
$\dcv=\dc=1.69$.  However, this does not account 
for the fact that some voids may be squeezed due 
to large-scale overdensities that have not yet 
collapsed and virialized and so it likely leads to an 
overestimate of 
the volume filling factor of voids in the universe.  
Sheth \& Van Weygaert~\cite{sheth_weygaert04} also 
explore $\dcv=1.06$ as this linearly-extrapolated 
overdensity corresponds to the moment of 
turnaround, or the transition from expansion to 
contraction, for a spherical overdensity.  
The drawback of this choice is that it completely 
erases voids in regions that are only just turning 
around so that this choice likely leads to an 
underestimate of the filling factor of voids in the universe.  
This discussion suggests that the appropriate value of 
the ``void-squashing'' overdensity should be in the range 
$1.06 < \dcv < 1.69$.

The first step in modeling the void population 
with the excursion set theory should now be clear.  
It is necessary to solve for the first crossing 
distribution of the barrier $\dv$ subject to the 
additional restriction that the barrier at 
$\dcv$ is not achieved at some larger scale either.  
As in \S~\ref{section:excursion}, it is convenient 
to solve for the probability distribution of 
trajectories that do not impact either barrier $\Pi(\delta,S)$, 
and to derive the first-crossing distribution from 
$\Pi(\delta,S)$.  As before, the distribution of 
$\delta$ at any value of $S$ obeys 
\beq
\label{eq:heateqvoid}
\frac{\partial \Pi}{\partial S} = \frac{1}{2}\frac{\partial^2 \Pi}{\partial \delta^2}.
\eeq
In the case of voids, trajectories are removed from consideration 
when they impinge upon either of the two barriers at 
$\dcv$ or $\dv$ so that the boundary conditions are 
$\Pi(\dv,S)=\Pi(\dcv,S)=0$.

This system of equations 
is the one-dimensional heat equation describing heat flow in 
a long, thin bar of finite length held at fixed temperature at 
both ends.  As such, it has a solution that is likely 
to be familiar.  To begin with, unlike the case of the 
single-barrier problem in \S~\ref{section:excursion}, 
the solution to this boundary value problem can only 
be a superposition of a discrete set of sinusoidal modes, 
namely those modes that individually obey the homogeneous 
boundary conditions.

As before, it is convenient to define a new variable 
$\gamma \equiv \delta -\dv$.  In that case, 
Eq.~(\ref{eq:heateqvoid}) still holds with 
$\delta$ replaced by $\gamma$ and the boundary conditions 
are $\Pi(\gamma=0,S)=\Pi(\gamma=\gcv=\dcv-dv,S)=0$.  
As before, trajectories start at $\delta=0$ at 
$S=0$ so that the initial condition is 
$\Pi(\gamma,S=0)=\dirac(\gamma+\dv)$.  
The solution follows straightforwardly using 
separation of variables.  Assuming that 
$\Pi(\gamma,S)=G(\gamma)P(S)$, the functions 
$G(\gamma)$ and $P(S)$ obey 
\beq
\label{eq:diffeq_sov}
\frac{1}{P}\frac{\dd P}{\dd S} = \frac{1}{2G}\frac{\dd^2 G}{\dd \gamma^2} = -k^2,
\eeq
where $k$ is some constant.  
Therefore $P(S) \propto \exp( -k^2 \ S)$.  
The boundary conditions on $G(\gamma)$ can only be satisfied if 
$k > 0$.  Thus the solution is a superposition of sines and 
cosines.  It is easy to show that the boundary conditions 
require that 
\beq
\label{eq:ggamma}
G(\gamma)=\sum_{n=1}^{\infty} \ c_n \  \sin\Bigg( \frac{n\pi}{\gcv}\gamma\Bigg).
\eeq
The initial condition can be used to determine the $c_n$ 
by using the orthogonality relation for sinusoids, 
$\int_0^{\pi} \ \sin(ny) \ \sin(my) \dd y = (\pi/2)\delta_{nm}$, 
where $\delta_{nm}$ is a Kronecker delta.  
The result is that $c_n = 2 \sin ( n\pi \abs{\dv}/\gcv )/\pi$.  
Consequently, the solution to the boundary value problem is 
\begin{eqnarray}
\Pi(\delta,S) & = & \frac{1}{\dcv-\dv} 
\sum_{n=1}^{\infty}\ 
\Bigg[ \cos\Bigg( \frac{n\pi}{\dcv-\dv}\delta\Bigg)\  
- \ \cos\Bigg( \frac{n\pi}{\dcv-\dv}[2\dv-\delta]\Bigg)\Bigg]\ \nonumber\\ 
 &  & \ \ \times \exp \Bigg( -\frac{n^2\pi^2}{2(\dcv-\dv)}S\Bigg).
\end{eqnarray}

In analogy with Eq.~(\ref{eq:Fexcursion}) and 
Eq.~(\ref{eq:fdelta}), trajectories are 
removed from the region between the two boundaries 
at a ``rate'' of 
\beq
\label{eq:fvoid}
\fv(S)\dd S = -\frac{1}{2} 
\frac{\partial \Pi}{\partial \delta}\Bigg\vert_{\dv}^{\dcv}\ \dd S.
\eeq
The lower limit of this divergence of trajectories indicates the 
trajectories that penetrate the lower boundary at $\delta=\dv$.  
These are the void regions that we aim to model, so the first 
crossing distribution is 
\begin{eqnarray}
\label{eq:fvoideval}
\fv(S)\dd S & = & \frac{1}{2} 
\frac{\partial \Pi}{\partial \delta}\Bigg\vert_{\dv} \nonumber \\
 & = & \sum_{n=1}^{\infty} \ 
\frac{n\pi\DD^2}{\dv^2}\ \sin ( n\pi \DD )\ 
\exp\Bigg( -\frac{n^2\pi^2\DD^2}{2}\frac{S}{\dv^2}\Bigg)\dd S,
\end{eqnarray}
where $\DD \equiv \abs{\dv}/(\dcv+\abs{\dv})$ is the 
{\em void-in-cloud parameter}, 
following the nomenclature of Ref.~\cite{sheth_weygaert04}.  
The parameter $\DD$ quantifies the relative importance of eliminating 
voids that lie within some larger overdense region in which the 
smoothed density contrast exceeds $\dcv$.

\begin{figure}[t]
\begin{center}
\includegraphics[height=7.0cm]{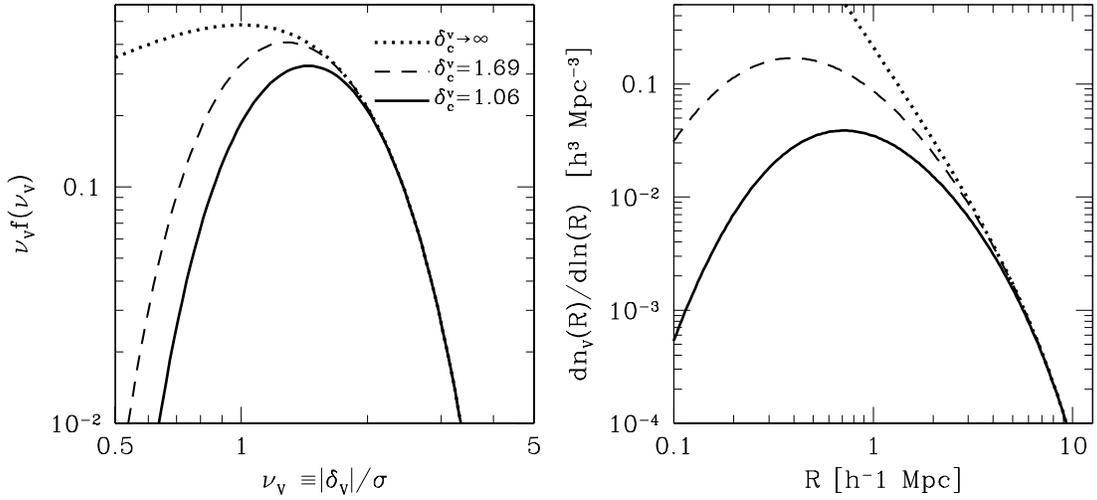}
\caption{
{\em Left}:  The mass distribution of voids predicted by the 
excursion set model of Sheth \& Van Weygaert~\cite{sheth_weygaert04}.  
The mass distribution is given in scaled units.  
The horizontal axis is the absolute value of the 
void threshold in units of the $\sigma(M)$ for 
a particular mass scale $\nuv = \abs{\dv}/S^{1/2}$.  
The vertical axis is the fraction of 
trajectories that first cross the threshold 
in a logarithmic interval of $\nuv$.  Physically, 
this is the fraction of mass in voids that have 
a mass corresponding to a logarithmic interval in $\nuv$.  
The three lines show three choices of the overdensity 
threshold $\dcv$.  
{\em Right}:  The physical size distribution of voids.  
The three lines represent the number densities of 
voids of radius $R$ for the same three choices of 
$\dcv$ as in the left panel.  
\label{fig:voidmass}
}
\end{center}
\end{figure}

From the first-crossing distribution of Eq.~(\ref{eq:fvoideval}), 
it is straightforward to compute the ``mass function'' of voids, 
meaning the number of voids containing a specific amount of mass.  
As before, it is most convenient to represent this distribution 
in terms of a scaled variable $\nuv \equiv \abs{\dv}/S^{1/2}$, 
so that the physical size distribution is determined by the 
size of the smoothing scale at a particular value of the 
variance $S$ and the parameter $\DD$.  Then $\nuv \fv(\nuv)$ defined as the 
fraction of mass in voids per logarithmic interval 
$\dd \ln \nuv$ then follows by changing variables in 
Eq.~(\ref{eq:fvoideval}) from $S$ to $\nuv$ as in \S~\ref{section:excursion}.  
Explicitly, 
\beq
\label{eq:nuvfnuv}
\nuv \fv(\nuv) = \sum_{n=1}^{\infty}\ 
\frac{2n\pi \DD^2}{\nu^2}\ \sin(n \pi \DD)\ 
\exp\Bigg(-\frac{n^2\pi^2\DD^2}{2\nu^2}\Bigg).
\eeq

This void mass spectrum is shown in the left panel of 
Fig.~\ref{fig:voidmass} for the 
choices of $\dv=-2.81$ and $\dcv=1.06$ advocated by 
Sheth \& Van Weygaert~\cite{sheth_weygaert04}.  
Two additional mass spectra derived from different values 
of $\dcv$ are shown in Fig.~\ref{fig:voidmass} to illustrate the 
importance of the void-in-cloud parameter.  
The choice $\dcv \to \infty$ corresponds to ignoring completely 
the possibility for a larger-scale overdensity to overwhelm a 
smaller void within it.  Decreasing $\dcv$ makes the probability 
for a larger overdensity to overwhelm a void within it higher 
and decreases the number of voids of all sizes.  Two things are 
worthy of note.  First, the number of large voids (high $\nuv$) 
is independent of $\dcv$.  This is because it is very unlikely for 
a trajectory to cross both the $\dcv$ and the $\dv$ thresholds at 
small values of $S$.  Second, regardless of the choice of $\dcv$, 
the fraction of mass in voids is peaked around voids with 
$\nuv \approx 1-1.5$.  This places the typical void mass in the 
broad range $M_{\mathrm{void}} \sim 10^{11}-10^{13} \hMsun$.  
Of course, as with the mass function, the number of voids of a 
particular size will be shifted toward smaller objects because 
a fixed fraction of mass can be partitioned into more small 
voids and fewer large voids.

The void size spectrum in Eq.~(\ref{eq:nuvfnuv}) gives the 
fraction of mass, or of volume in the initial Lagrangian space, 
in voids of a particular size.  The size variable is 
the scaled variable $\nuv$.  To convert this to a 
physical size spectrum where the sizes of objects are measured 
in units of length or volume requires relating the initial 
Lagrangian volume to an Eulerian volume.  The simplest 
conversion follows from the fact that in the spherical 
tophat model, the tophat radius expands by a factor or 
$\sim 1.72$ by the time shell crossing occurs.  Finally, 
one also needs to assume a power spectrum to relate 
particular values of $\nuv$ with volume in the initial 
Lagrangian coordinate system.  As before, I will assume the 
spectrum in Fig.~\ref{fig:ps}.  The logic is then the 
same as that of the mass function.  The number density 
of voids as a function of physical void radius is shown 
in the right panel of Fig.~\ref{fig:voidmass} and reflects 
the basic features of the voids model.  The number density 
of voids is independent of $\dcv$ for large voids, and 
the number of small voids is suppressed because small, 
underdense regions are likely to be contained within 
larger overdense regions.

So far, I have ignored the primary shortfall of the void model.  
The weakness of the approach of Ref.~\cite{sheth_weygaert04} 
is that it defines a void as a region that is underdense with 
respect to total mass whereas observationally voids are 
identified as underdensities in the galaxy distribution, as 
noted in Ref.~\cite{sheth_weygaert04}.  
Furlanetto \& Piran~\cite{furlanetto_piran05} have extended the 
Sheth \& Van Weygaert model to identify voids 
as regions that are underdense in the galaxy distribution.  
To do this, these authors added to the excursion set theory 
a method for assigning galaxies to halos and thereby 
established a mapping between the galaxy number density in a 
region and a linearly-extrapolated overdensity.  Methods for 
assigning galaxies to halos are beyond the scope of this 
article so I will not discuss the model of 
Ref.~\cite{furlanetto_piran05} 
any further except to say that it is a more practical approach 
to voids because it phrases predictions in terms of the 
observed galaxy overdensity.

\section{Halo Formation in the Excursion Set Theory}
\label{section:trees}

Excursion set theory is also a powerful formalism with 
which to study the formation history of halos.  It is most 
natural to picture evolution occuring through the linear 
growth of the density field.  In that way, the smoothed 
overdensity scales with time according to the linear 
growth factor $D(a)$.  Writing the smoothed density field as a 
function of both the smoothing scale $S$ and the 
cosmological expansion factor $a$, this evolution is 
$\delta(S,a) = D(a)\delta(S,a=1)$.  As a matter of convention, 
$D(a)$ is normalized so that $D(a=1)=1$.  In the framework of 
excursion sets, predictions depend upon the density field in 
the ratio $\nu = \dc/\sigma(M)$ and 
it is often easier to envision the density 
field as fixed at $\delta(S) \equiv \delta(S,a=1)$ and to 
consider the height of the critical density threshold as 
a function of time.  In that way, collapse at a scale 
factor $a' \ne 1$ corresponds to the $a=1$ density fluctuation 
penetrating a barrier of height $\dc(a') = \dc/D(a')$.

In their pioneering paper, Bond et al.~\cite{bond_etal91} 
briefly discussed the utility of 
the excursion set approach for understanding the 
formation histories of halos.  Subsequently, 
Lacey \& Cole~\cite{lacey_cole93} 
studied this problem in great detail in a rather 
elegant paper and I largely follow 
their treatment in what follows.  I will adopt the 
simplifying notation of Ref.~\cite{lacey_cole93} and refer 
to the barrier height for collapse at a specific time as 
\beq
\omega(a) \equiv \dc/D(a).
\eeq
The variable $\omega$ indicates the time of the 
collapse through only the linear growth function $D(a)$.  
In the ``old standard'' CDM model with $\Omegam=1$, 
the growth function is simply $D(a)=a$ and the conversion 
between $\omega$ and $a$ is trivial.  However, in the 
currently-favored $\Lambda$CDM model, the growth function 
is given by an incomplete beta function 
integral~\cite{bildhauer_etal92} and the conversion is 
more complicated.

Many of the insights into halo formation can be gleaned 
by considering the two-barrier problem, just as 
with halo bias and void regions 
in the previous two sections.  Again 
following the work of Lacey \& Cole~\cite{lacey_cole93}, 
I also adopt a simplifying notation 
for the conditional probabilities 
of barrier crossings.  I illustrate this notation by 
re-writing the two-barrier probability distribution 
already given in Eq.~(\ref{eq:fdelta}) and 
Eq.~(\ref{eq:fcondition}) as 
\beq
\label{eq:fdef}
f(S_2,\w_2|S_1,\w_1)\dd S_2 
 = \frac{1}{\sqrt{2\pi}} \frac{\Delta\w}{\DS^{3/2}}\ 
\exp\Bigg[-\frac{(\Delta\w)^2}{2\DS}\Bigg] \dd S_2,
\eeq
where $\Delta\w \equiv \w_2 - \w_1$ is the difference between the two 
barrier heights (assumed to be arbitrary with $\w_2 > \w_1$), 
$\DS \equiv S_2 - S_1$.  As always, some mass $M_1$ and 
$M_2$ are associated with the variances $S_1$ and $S_2$ 
respectively.  The quantity $f(S_2,\w_2|S_1,\w_1)\dd S_2$ is 
the conditional probability that a trajectory pierces the barrier $\w_2$ 
in an interval of width $\dd S_2$ about $S_2$ on the condition that 
the trajectory first pierces $\w_1$ at $S_1$.  As a further shorthand, 
I take $f(S_2,\w_2)\dd S_2$ to denote the probability starting from 
a very large smoothing scale $S_1=0$, and $\w_1=0$.  

In this section and those that follow, the interpretation 
of Eq.~(\ref{eq:fdef}) is different from that in 
\S~\ref{section:excursion} and \S~\ref{section:bias}.  
In those sections, the initial condition for the random 
walk was fixed by the large scale environment on some larger 
filtering scale $S_1 \ll S_2$.  Here, the two barriers represent the 
critical density for collapse at two different times so the 
change in barrier height $\Delta\w$ represents a shift in 
time.

\subsection{The Conditional Mass Function}
\label{subsection:cmf}

\begin{figure}[t]
\begin{center}
\includegraphics[height=12.0cm]{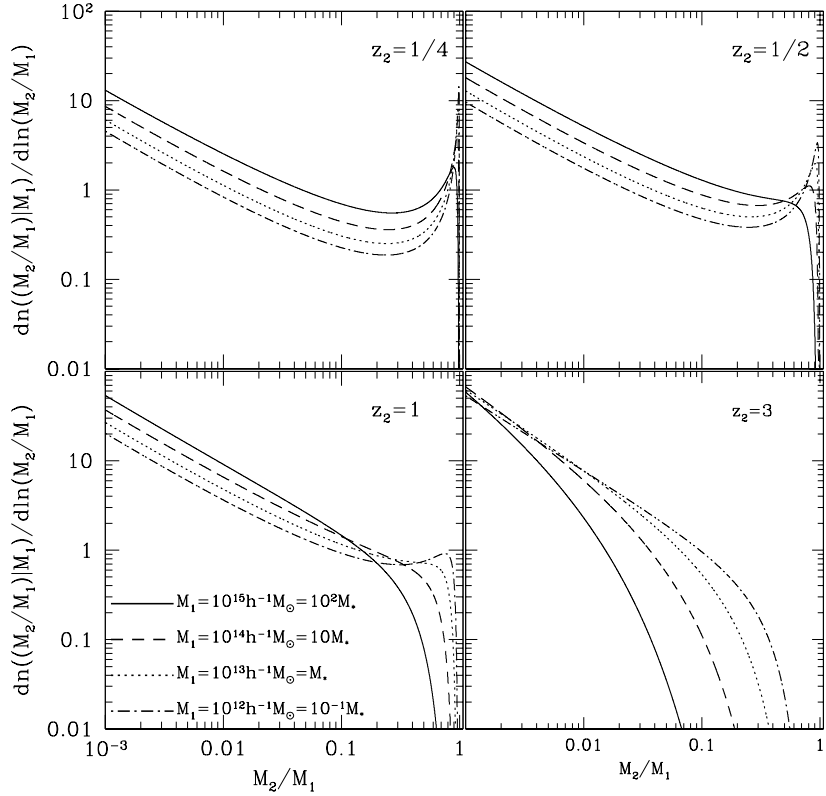}
\caption{
The conditional mass function.  All panels refer to 
the conditional mass functions of halos fixed at $z=0$ 
looking back toward higher redshifts as labeled in each 
panel.  Four halo masses of interest are shown.  
\label{fig:condmf}
}
\end{center}
\end{figure}

The most immediate quantity of interest is the conditional 
mass function.  Given a halo of mass $M_1$ at time $t_1$, 
one can compute the average manner in which this mass was partitioned 
among smaller halos at some earlier time $t_2 < t_1$ ($\w_2 > \w_1$).  
The conditional mass function is simply the average number of halos 
of mass $M_2$ at time $t_2$ that are incorporated into a mass 
$M_1$ object at time $t_1$.  
This is precisely the two-barrier problem.  A halo of mass $M_1$ 
has its mass partitioned, on average, among a spectrum of halos 
at time $w_2$ given almost directly from Eq.~(\ref{eq:fdef}) 
(\eg see Refs.~\cite{lacey_cole93,somerville_kolatt99}), 
\beq
\label{eq:mfconditional}
\frac{\dd n(M_2|M_1)}{\dd M_2} = 
\frac{M_1}{M_2}\ f(S_2,\w_2|S_1,\w_1)\ 
\Bigg|\frac{\dd S_2}{\dd M_2}\Bigg| \dd M_2.
\eeq
The function $f(S_2,\w_2|S_1,\w_1)$ gives the probability 
of the second barrier crossing at a particular value of 
$S_2$, while the factor $M_1/M_2$ converts this from a 
probability per unit mass of halo $M_1$ into the number of 
halos of mass $M_2$.  

Figure~\ref{fig:condmf} shows conditional mass functions for 
standard CDM power spectra normalized such that 
$\Mstar=10^{13}\hMsun$ ($\sigma_8 \simeq 0.93$) for 
four halo masses at four different lookback redshifts, 
$z=1/4$, $z=1/2$, $z=1$, and $z=3$ 
(all panels have final redshift at $z=0$).  
The evolution toward increased fragmentation 
as the second epoch moves to higher redshifts is 
evident.  Notice that more massive halos fragment into small sub-units 
more quickly.  Also, notice that each conditional mass 
function goes toward an approximate power law as 
$M_2/M_1 \ll 1$.  This can be understood directly from 
Eq.~(\ref{fig:condmf}).  Assume that the power 
spectrum is an approximate power law 
$S(M) \simeq \dc^2(M/\Mstar)^{-\beta}$ 
in the mass range of interest.  In the limit of a large mass difference, 
$\Delta\w/\DS \ll 1$ and $M_2/M_1 \ll 1$ and 
$\dd n/\dd M_2 \propto M_2^{-(4-3\beta)/2}$.  
Near $\Mstar$, CDM-like power spectra vary slowly, with 
an effective power-law index near $\beta \approx 0.35$ giving 
$\dd n/\dd M_2 \propto M_2^{-3/2}$, which is approximately 
the power law shown in Figure~\ref{fig:condmf}.

\subsection{Halo Accretion Rates}
\label{subsection:accretion}

The excursion set theory also provides a framework 
for computing mass accretion rates as well.  The 
two-barrier problem of \S~\ref{subsection:cmf} can be 
manipulated to yield an average mass accretion rate.  
Consider the probability of transitioning forward in time, 
from $M_2$ at barrier $\w_2$ to 
$M_1$ at $\w_1$.  Bayes' rule for combining probabilities gives 
the forward probability as~\cite{lacey_cole93}
\begin{eqnarray}
\label{eq:fbayes}
f(S_1,\w_1|S_2,\w_2)\dd S_1 &
= & \frac{f(S_2,\w_2|S_1,\w_1)f(S_1,\w_1)}{f(S_2,\w_2)}\dd S_1 \\
 & = & \frac{1}{\sqrt{2\pi}}\frac{\w_1(\w_2-\w_1)}{\w_2}
\Bigg[\frac{S_2}{S_1(S_2-S_1)}\Bigg]^{3/2} \\
 & & \times 
\exp\Bigg(-\frac{(\w_1 S_2 - \w_2 S_1)}{2 S_1 S_2 (S_2-S_1)}\Bigg)\ \dd S_1.
\end{eqnarray}
In language familiar to cosmology, $f(S_1,\w_1|S_2,\w_2)$ is the 
posterior probability, $f(S_1,\w_1)$ is the prior, and 
$f(S_2,\w_2|S_1,\w_1)$ is the likelihood.

\begin{figure}[t]
\begin{center}
\includegraphics[height=7.0cm]{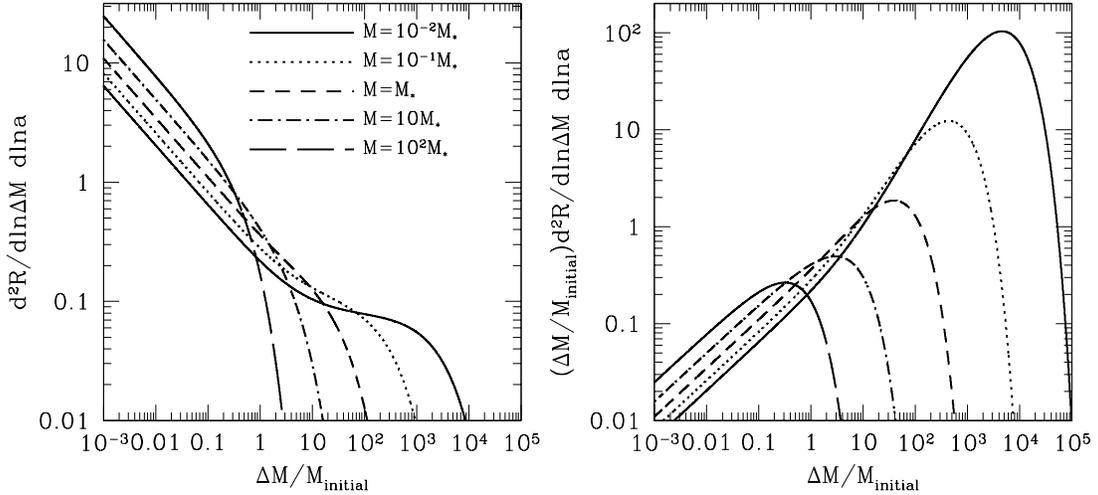}
\caption{
Halo accretion rates per logarithmic interval of mass change.  
This is based on a very similar plot in Ref.~\cite{lacey_cole93}.  
The plots show the accretion rates per unit mass 
change for halos of five different masses (shown in 
the left panel) at $z=0$.  The left panel shows the 
probability for a change $\Delta M$ as a function of 
$\Delta M$ [Eq.~(\ref{eq:accrate})], 
while the right panel shows the fractional 
mass accretion rate given by 
$(\Delta M/M_2) \times \dd^2 R/\dd \ln \Delta M \dd \ln a$.  
\label{fig:accrate}
}
\end{center}
\end{figure}

The instantaneous probability per unit shift in 
barrier height follows by taking 
$\Delta\w = (\w_2-\w_1) \to 0$ in Eq.~(\ref{eq:fbayes}), 
giving 
\beq
\frac{\dd^2 R}{\dd \w \ \dd S_1} = 
\frac{1}{\sqrt{2\pi}}\Bigg[\frac{S_2}{S_1(S_2-S_1)}\Bigg]^{3/2} 
\exp\Bigg[-\frac{\w^2(S_2-S_1)}{2 S_2 S_1}\Bigg],
\eeq
from which the probability per unit change in mass 
$\Delta M = M_1 - M_2$, per unit 
time (or $\dd \ln a = H(a)\dd t$) is 
\begin{eqnarray}
\label{eq:accrate}
\frac{\dd^2 R}{\dd \ln \Delta M \dd \ln a} & = &
\sqrt{\frac{2}{\pi}} \frac{\Delta M}{M_1}
\frac{\w/\sigma(M_1)}{(1-S_1/S_2)^{3/2}} \nonumber \\
 & & \times \exp\Bigg[-\frac{\w^2 (S_2-S_1)}{2 S_1 S_2}\Bigg]
\Bigg|\frac{\dd \ln \w}{\dd \ln a}\Bigg|
\Bigg|\frac{\dd \ln \sigma}{\dd \ln M_1}\Bigg|.
\end{eqnarray}
The average rate of mass accretion per unit change 
in mass can be obtained by multiplying Eq.~(\ref{eq:accrate}) 
by $\Delta M$, while the average total mass 
accretion rate follows from multiplying 
Eq.~(\ref{eq:accrate}) by $\Delta M$ and integrating 
over $\ln \Delta M$.  
%
%

\begin{figure}[t]
\begin{center}
\includegraphics[height=8.0cm]{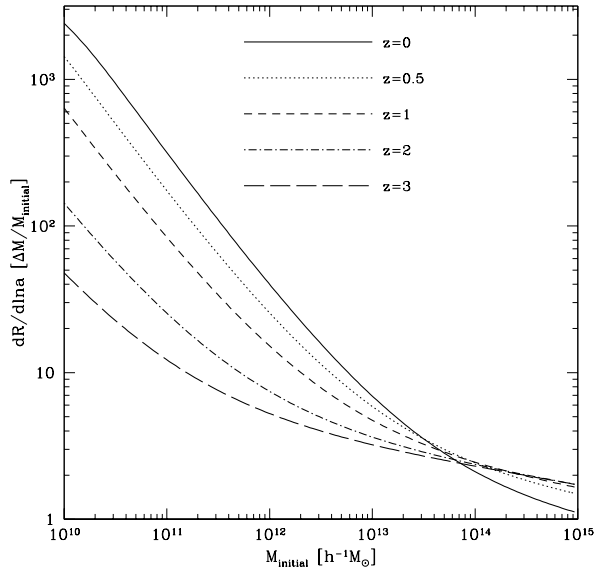}
\caption{
Total halo mass accretion rates in units of halo mass and 
as a function of halo mass at four different redshifts.  
\label{fig:totrate}
}
\end{center}
\end{figure}

The excursion set accretion rates of halos in the 
standard $\Lambda$CDM cosmology are summarized in 
Fig.~\ref{fig:accrate} and Fig.~\ref{fig:totrate}.  
The left panel Fig.~\ref{fig:accrate} shows 
the probability for a particular fractional mass 
change as a function of the fractional mass change.  
The rate of accretion by small mass increments 
diverges. However, the average rate of mass increase 
(in the right panel of Fig.~\ref{fig:accrate}) 
converges and is dominated by the less frequent, 
high-mass mergers.  Notice in particular that 
low-mass halos increase their masses very quickly 
via mergers with halos larger then themselves while 
halos with $M \gsim \Mstar$ experience only infrequent 
mergers with halos larger than themselves.

The total mass accretion rate given by integrating 
Eq.~(\ref{eq:accrate}) over $\ln \Delta M$ is shown 
in Fig.~\ref{fig:totrate} for various redshifts as 
a function of halo masses.  Fig.~\ref{fig:totrate} 
shows explicitly that low-mass halos should be expected 
increase there mass many times over in a single Hubble time.  
The standard interpretation of the collection of results 
shown in Fig.~\ref{fig:accrate} and Fig.~\ref{fig:totrate} is that 
low-mass halos ($M<\Mstar$) should be expected to increase their 
masses by mergers with halos larger than themselves, while 
high-mass halos ($M>\Mstar$) typically absorb smaller halos 
and increase their masses relatively slowly.

\subsection{Halo Formation Times}
\label{subsection:formtimes}

It is often useful to have a sense of a time by which a 
halo acquired most of its mass.  For example, one may be 
modeling galaxies and may try to estimate stellar evolution 
histories within a particular halo.  Several definitions of 
halo formation time exist.  An intuitive 
definition is the time that a halo first acquires 
half of its final mass because once a progenitor of a 
halo has achieved this threshold it can uniquely be 
identified as the main progenitor of the final object.  
Consider the formation time given by this definition 
$\wf$ or, in terms of expansion factor, $\af$.

It is not trivial to predict halo formation 
times from excursion set theory by considering 
the two-barrier problem as before.  The reason 
for this difficulty is simple to understand.  
Consider the formation of a halo of mass $M_1$ at time $\w_1$.  
The probability of crossing a higher threshold $\w_2$ 
(corresponding to earlier time) at some 
value of the variance greater than $S_2 = S(M_2=M_1/2) > S_1$ 
gives only the probability per unit mass 
that the halo has some progenitor with mass 
$M_2 \le M_1/2$ at time $\w_2$.  It does not guarantee 
that this is the mass of the most massive of all of 
the subunits that merged to form the halo of mass 
$M_1$ at $\w_1$.  Lacey \& Cole~\cite{lacey_cole93} 
circumvent this difficulty by giving an approximate counting 
argument for the formation time.  I reproduce the 
argument of Ref.~\cite{lacey_cole93} in what follows.

First, the results of \S~\ref{subsection:accretion} 
give the number of halos of mass $M_2$ at time 
$t_2$ that are incorporated into halos of mass 
$M_1$ at $t_1 > t_2$ as 
\beq
\frac{\dd^2 n}{\dd M_2 \dd S_1} = 
\frac{\dd n(M_2)}{\dd M_2}\ f(S_1,\w_1|S_2,\w_2).
\eeq
So long as $M_2 > M_1/2$, each trajectory 
must connect two unique halos because there cannot be 
two paths each of which contain more than half of 
the final halo mass.  However, it is possible that a halo 
of mass $M_1$ at $t_1$ has {\em no} progenitor of mass 
greater than $M_1/2$ at time $t_2$.  The counting argument 
accounts for this.  The probability that a 
halo of mass $M_1$ at $t_1$ has a progenitor in the 
mass range $M_1 \ge M_2 > M_1/2$ at time $t_2$ is 
then given by the ratio of halos that evolve to 
a halo of mass $M_1$ relative to the total number 
of halos of mass $M_1$, 
\beq
\frac{\dd P}{\dd M_2} = 
\frac{[\dd n(M_2)/\dd M_2]f(S_1,\w_1|S_2,\w_2)}
{[\dd n(M_1)/\dd M_1] \vert|\dd S_1/\dd M_1\vert^{-1}}.
\eeq
Using Bayes' rule again and the definitions of the 
mass functions in terms of the first-crossing 
distributions gives 
\beq
\label{eq:dpdm}
\frac{\dd P}{\dd M_2} = 
\Bigg(\frac{M_1}{M_2}\Bigg)\ f(S_2,\w_2|S_1,\w_1) 
\Bigg|\frac{\dd S_2}{\dd M_2}\Bigg|.
\eeq
The cumulative probability of a formation time prior to time $t_2$ is 
then the integral 
\beq
\label{eq:pcum}
P(t_{\mathrm{F}}<t_2) = P(\wf>\w_2) = 
\int_{S_1}^{S_2=S(M_1/2)}\ 
\Bigg(\frac{M_1}{M_2}\Bigg) f(S_2',\w_2|S_1,\w_1) \dd S_2',
\eeq
and the differential probability can be obtained by 
differentiating Eq.~(\ref{eq:pcum}) with respect to either 
$\wf$ or $t_{\mathrm{F}}$.

\begin{figure}[t]
\begin{center}
\includegraphics[height=8.0cm]{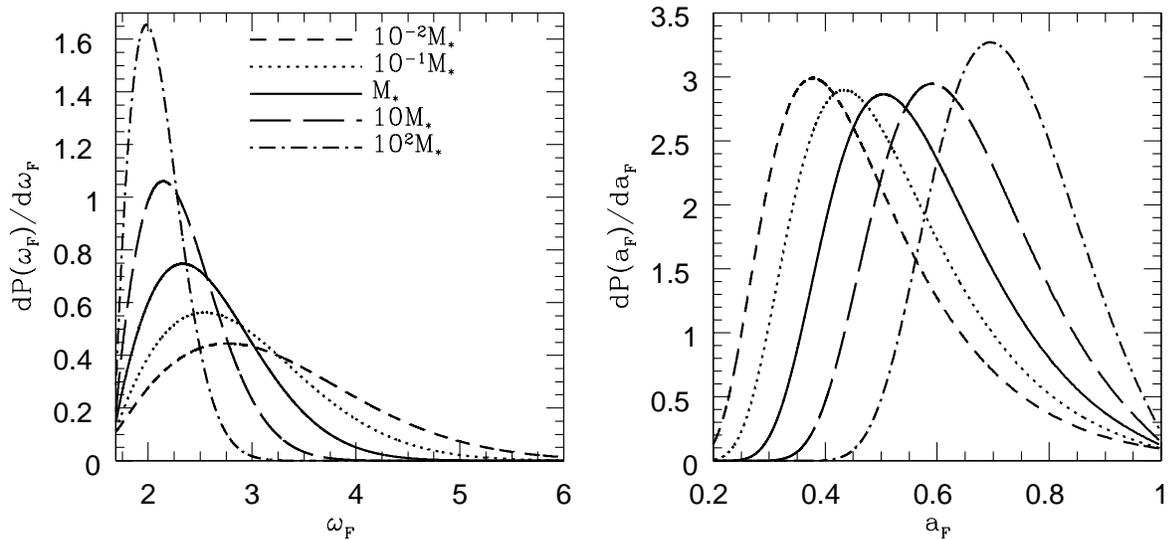}
\caption{
The probability distributions for formation times 
in the standard $\Lambda$CDM cosmology with a scale-invariant 
primordial power spectrum normalized to $\sigma_8=0.93$, so 
that $\Mstar=10^{13}\hMsun$.  The left panel shows the 
distribution as a function of threshold height $\dd P(\wf)/\dd \wf$ 
while the right panel shows the probability distribution of 
formation scale factors $\dd P(\af)/\dd \af$.  
\label{fig:wform}
}
\end{center}
\end{figure}

Formation time distributions computed from Eq.~(\ref{eq:pcum}) 
are shown in Fig.~\ref{fig:wform} for a variety of halo 
masses at $z=0$ ($a=1$, $\wf=\dc\simeq 1.69$).  The most 
obvious feature of formation times is that small halos 
form early and large halos form relatively late in accordance 
with our general expectation in hierarchical models of 
structure formation.  Note also that in the simplest implementation 
of the excursion set formalism, in particular making use of 
k-space tophat filtering of the density field, the conditional 
probability $f(S_2,\w_2|S_1,\w_1)$ is a function only of 
the differences $S_2-S_1$ and $\w_2-\w_1$.  In this case, this 
means that the formation times in particular and, more generally, 
the entire formation histories of halos are independent of the 
larger-scale environments that halos reside in.  
In the next section, I discuss halo formation 
times in the context of halo merger trees.

\section{Halo Merger Trees}
\label{section:mt}

The probability of first up-crossing at a second barrier 
given a particular starting point is given by the solution 
to the two-barrier problem in Eq.~(\ref{eq:fdef}).  This 
relation has been repeated several times throughout these notes 
in several forms because it is the fundamental relation 
on which most physical results are based.  In the preceding 
sections, I have applied this formula to derive relations 
about various aspects of halo fragmentation and formation 
over time.  So far, these quantities have been averages 
(as in \S~\ref{subsection:cmf} and \S~\ref{subsection:accretion}) 
or probability distributions (as in \S~\ref{subsection:formtimes}).  
However, as Ref.~\cite{lacey_cole93} describes, 
the interpretation of Eq.~(\ref{eq:fdef}) 
as a probability for a transition 
from one barrier to another means that one should be able 
to draw specific fragmentation probabilities from this 
distribution repeatedly for a number of time intervals 
$\Delta \w_i$, and thereby reconstruct individual 
trajectories of halo mass versus cosmic expansion factor.  
This enables one to follow the fragmentation 
of halos on an object-by-object basis.  This is the logic 
behind the algorithm for the generation 
of Monte Carlo merger trees described by 
(see also earlier methods in 
Refs.~\cite{cole_kaiser88,cole91,kauffmann_etal93}).

The trajectories $\delta(S)$ have structure on 
arbitrarily small scales corresponding to mergers with very 
small halos.  This is a restatement of the divergence of 
the mean number of transitions with small mass changes 
shown in the left panel of Fig.~\ref{fig:accrate}.  
Consequently, it is necessary to examine 
trajectories with a particular resolution in order to 
``smooth over'' the mergers with numerous tiny halos.  
In practice, each application to the prediction of a 
specific physical quantity has a 
minimum mass scale of interest $\Mmin$, 
so it is natural to set the resolution 
with which trajectories are computed according to 
this minimum scale.  Consider Eq.~(\ref{eq:fdef}) in the limit 
$\Delta\w/\sqrt{\DS} \ll 1$.  This gives a transition 
probability $f = \Delta\w/(\sqrt{2\pi}\DS^{3/2})$, that 
is directly proportional to the time interval $\Delta \w$.  
Lacey \& Cole~\cite{lacey_cole93} interpreted this as indicative of a 
probability that is due to a single, improbable merger 
event.  For this reason, Lacey \& Cole refer to the limit 
$\Delta\w/\sqrt{\DS} \ll 1$ as the ``binary regime.''  
This can be used to motivate a particular 
choice of step-size $\Delta\w$ that sets the resolution 
of the trajectory.

First, consider making the change of variable to 
$u \equiv \Delta \w/\sqrt{\DS}$ in Eq.~(\ref{eq:fdef}).  
This converts the probability distribution to a 
Gaussian in the variable $u$ with a mean of zero and 
a unit variance.  Of course, the probability distribution 
is one-sided in that it is subject to the constraint $u \ge 0$.  
To resolve individual, binary mergers of a parent halo 
of mass $M$ with a merging halo at the threshold 
$\Mmin$ the step-size should obey 
\beq
\label{eq:dw}
\Delta \w \lsim \sqrt{\Bigg| \frac{\dd S(M)}{\dd M}\Bigg| \Mmin},
\eeq
so that these mergers are in the binary regime.  Authors 
often parameterize this guideline and set 
$\Delta \w = \fstep \sqrt{\vert \dd S/\dd M\vert \Mmin}$.  
The guideline in Eq.~(\ref{eq:dw}) 
makes a transition with $\Delta M \ge \Mmin$ a rare 
event.  The motivation behind this choice is that it is 
very unlikely that the transition could be due to two 
rare events, so that most mergers with objects of mass 
$> \Mmin$ are resolved as binary encounters.  
In particular, the probability for a transition with 
$\Delta M \ge \Mmin$ is 
\begin{eqnarray}
\label{eq:fp}
P(>\Delta M) & = & \frac{2}{\sqrt{\pi}}\int_0^{\fstep/\sqrt{2}} 
\exp( -y^2) \dd y \nonumber \\
 & = & \erf (\fstep/\sqrt{\pi}) \\
 & \approx & \sqrt{\frac{2}{\pi}}\fstep, \nonumber
\end{eqnarray}
where the last step assumes $\fstep \ll 1$.  
The choice of $\fstep$ is motivated by the 
need to resolve all mergers of objects of mass $\Mmin$,  
driving $\fstep$ toward small values, and 
the desire to spend as little effort as possible 
computing the probabilities for transitions with 
$\Delta M < \Mmin$, driving $\fstep$ toward large 
values.  As I discuss below 
Ref.~\cite{somerville_kolatt99} advocates $\fstep \lsim 0.3$ 
in order to guarantee accuracy down to $\Mmin$.  
Typical values are $\fstep \lsim 0.1$.

With these preliminaries out of the way, 
the Lacey \& Cole~\cite{lacey_cole93} 
algorithm for generating 
merger histories is as follows.  First, determine 
the appropriate timestep using Eq.~(\ref{eq:dw}).  
Second, select a transition 
$\DS$ from the probability distribution 
of Eq.~(\ref{eq:fdef}) and invert the $S(M)$ relation 
to obtain both the change in mass $\Delta M$ and the 
new main progenitor mass $M'=M-\Delta M$.  One 
then repeats this procedure at the new values of 
$\omega$ and $S(M')$ in order to determine the 
next fragmentation of the main progenitor.  This 
process continues until the remaining mass is 
less than $\Mmin$.

The trajectory for the main progenitor, defined as 
the most massive progenitor at each timestep 
according to this algorithm, may form the trunk 
of a ``merger tree.''  All of the merging halos 
of mass $\Delta M$ have prior fragmentation 
histories of their own.  For each mass 
$\Delta M$ above the threshold $\Mmin$, one can 
generate an independent history for the infalling ``branch'' 
on the tree using the same algorithm.

The process of constructing a mass accretion history 
or merger tree in this way must be repeated numerous 
times in order to sample the variety of ways in which a 
halo of a fixed mass at a fixed time might 
build up its mass.  This is a Monte Carlo method 
for exploring the various mass accretion histories.  
It is conventional to refer 
to each individual tree generated in 
this way as a particular {\em realization} in 
an ensemble of merger histories.

\begin{figure}[t]
\begin{center}
\includegraphics[height=5.75cm]{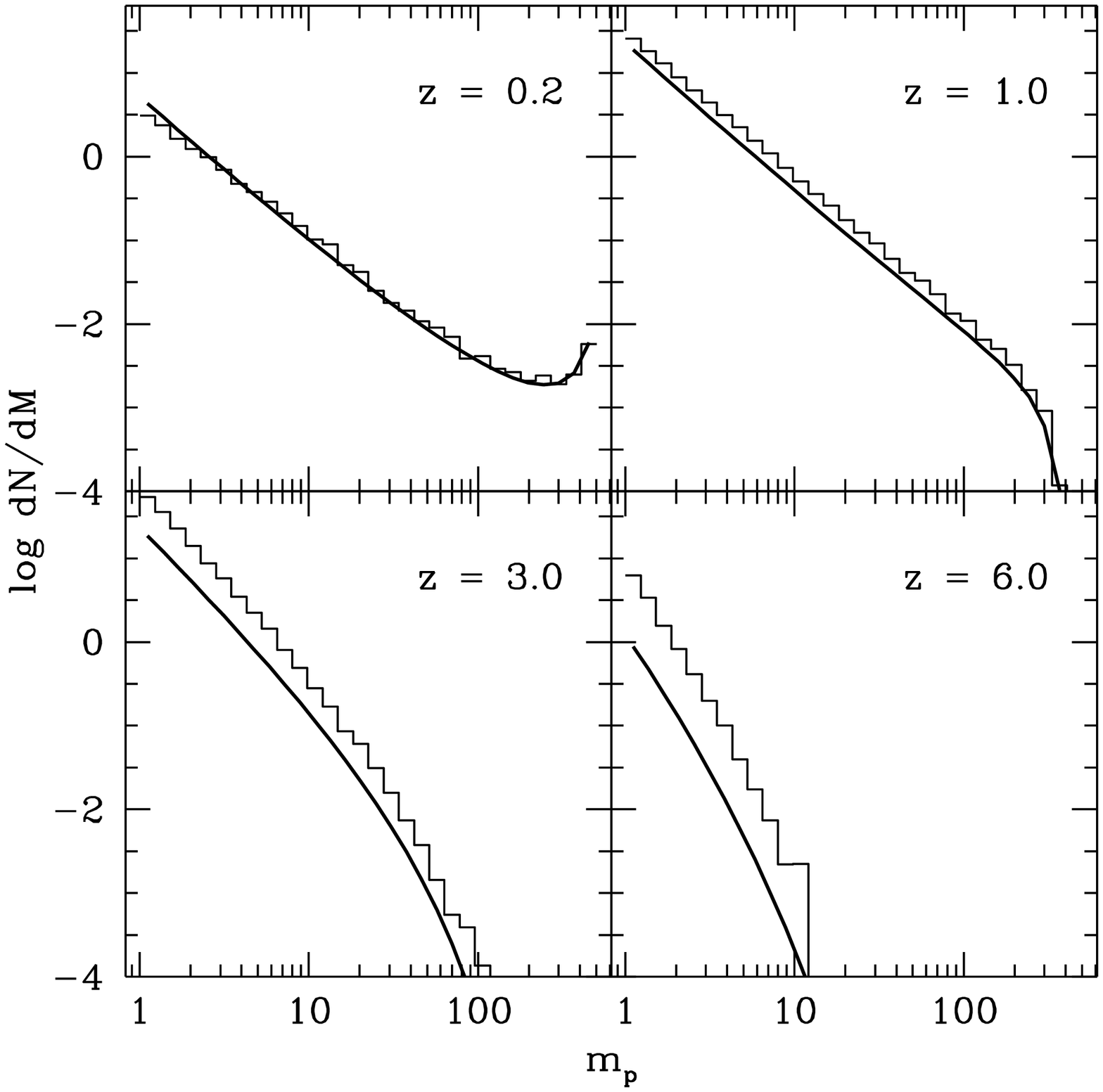}
\includegraphics[height=5.75cm]{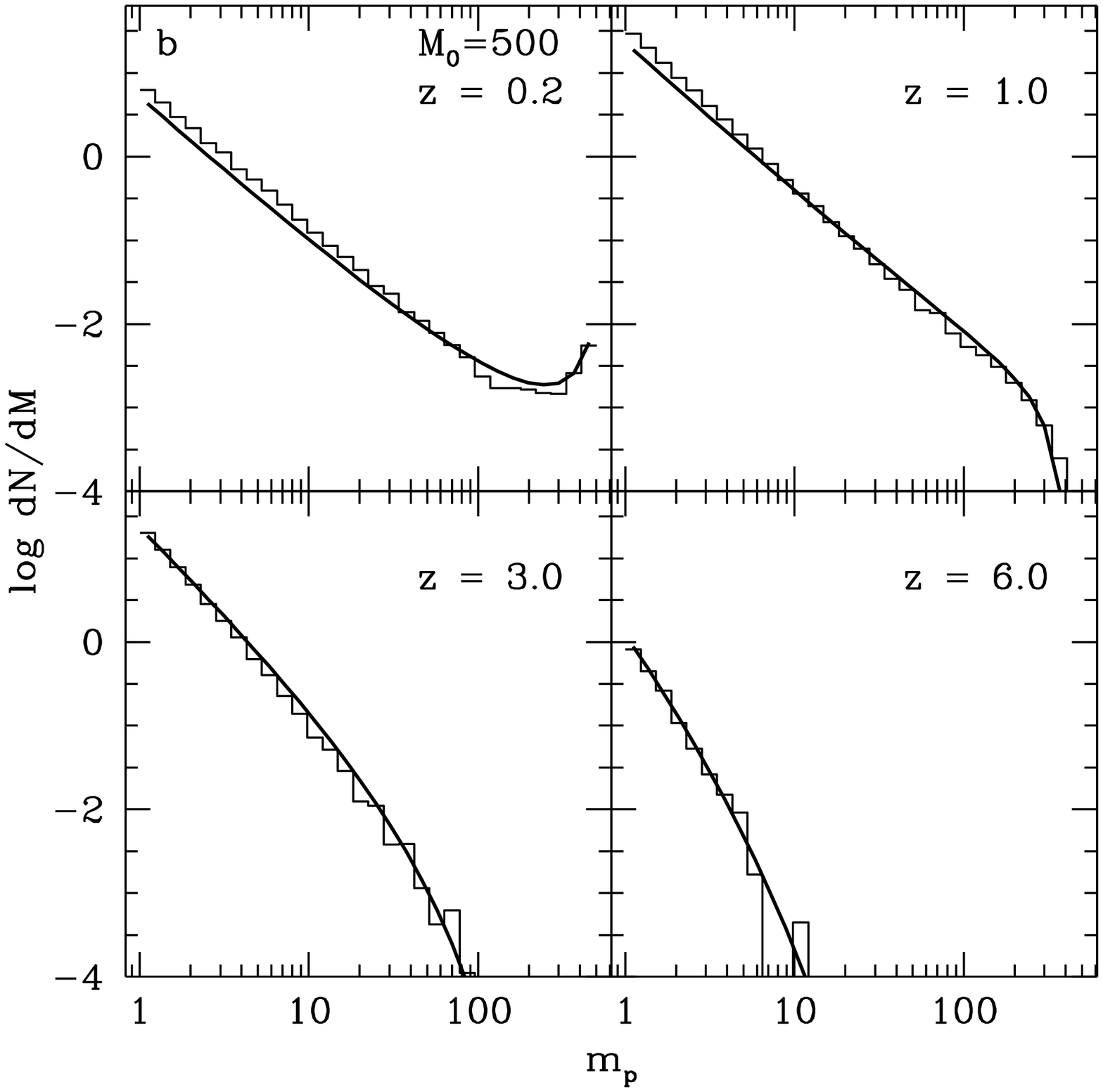}
\caption{
These figures are reproduced from Ref.~\cite{somerville_kolatt99}.  
The left panels show the distribution of progenitor masses 
$\log \dd N/\dd M$ for a halo of initial mass $M_i=500$ at 
four different redshifts.  The histograms were 
computed using the binary merger 
tree method of Lacey \& Cole in Ref.~\cite{lacey_cole93} 
and the {\em smooth} lines are 
the conditional mass functions of excursion set theory.  
The mass units are arbitrary.  The discrepancy between the 
conditional mass functions and the Monte Carlo merger tree 
results is clear and becomes larger with increasing redshift.  
The right panels show the same physical quantities.  
In this case the {\em smooth} lines are the 
same conditional mass functions as 
before.  The histograms show the results of the improved 
Monte Carlo merger tree algorithm proposed by 
Somerville \& Kolatt~\cite{somerville_kolatt99}.  
The discrepancy between the Monte Carlo 
method and the analytic conditional mass functions 
is clearly reduced using the Somerville \& Kolatt 
modification to the merger tree algorithm.
\label{fig:mtcmf}
}
\end{center}
\end{figure}

The algorithm for generating 
Monte Carlo merger trees described above 
is convenient because of its simplicity.  
Unfortunately, while this algorithm conserves 
mass by construction, it overpredicts the number 
of progenitor halos with masses $M > \Mmin$ at 
previous timesteps relative to the analytic distribution 
of Eq.~(\ref{eq:mfconditional}).  This point has been emphasized 
by Somerville \& Kolatt~\cite{somerville_kolatt99} 
(see also Ref.~\cite{sheth_lemson99}).  
In other words, the Monte Carlo 
procedure does not lead to a mean 
population of halos at high redshift that is 
consistent with the excursion set relations of 
the previous section.  In the left panel of 
Figure~\ref{fig:mtcmf}, I reproduce 
Figure~2 of Ref.~\cite{somerville_kolatt99} 
in order to emphasize this point.  The smooth 
lines in the left panel of this figure 
represent the analytic conditional 
mass functions described in \S~\ref{subsection:cmf} and 
the histograms represent conditional mass functions  
from an ensemble of merger histories constructed according 
to the method of Lacey \& Cole~\cite{lacey_cole93}.  That the 
merger tree method of Lacey \& Cole becomes more and 
more inconsistent with the analytic conditional mass function 
with increasing redshift is evident.

%

Somerville \& Kolatt~\cite{somerville_kolatt99} suggested two 
very plausible sources for this discrepancy.  
The first is that after choosing the first mass 
$\Delta M$ from Eq.~(\ref{eq:fdef}), the mass of the 
second body in the merger is forced to be 
$M'=M-\Delta M$.  This choice is made 
in order to conserve mass, but without 
regard to the probability of a transition to 
the mass $M'$.  This drives an overabundance of 
high-mass halos because all of the remaining mass 
is forced into a halo of mass $M'$.  The second 
source stems from the fact that the simple 
algorithm above largely neglects accretion with 
transitions $\Delta M < \Mmin$ when the probability 
for such a transition becomes infinitely large as 
$\Delta M \to 0$.  While Ref.~\cite{somerville_kolatt99} 
does not present a rigorous demonstration of a self-consistent 
algorithm, they give a qualitatively well-motivated 
algorithm that reproduces analytic halo 
conditional mass functions at high redshift 
very well.  This new algorithm is now the most 
commonly used method and I summarize it below.  
A successful algorithm that is, 
in practice, very similar to that of Ref.~\cite{somerville_kolatt99} 
is developed through the papers of 
Sheth~\cite{sheth96}, Sheth \& Pitman~\cite{sheth_pitman97}, 
and Sheth \& Lemson~\cite{sheth_lemson99}, 
though as far as I am aware, this approach also lacks rigorous 
support other than that it is exact for clumps in 
an uncorrelated density field.  Cole et al.~\cite{cole_etal00} 
provide another variation of the Lacey \& Cole~\cite{lacey_cole93} 
merger tree algorithm that is similar in spirit and effect to 
the that of Ref.~\cite{somerville_kolatt99}.

The Somerville \& Kolatt~\cite{somerville_kolatt99} 
algorithm begins in the 
same manner as the original Lacey \& Cole~\cite{lacey_cole93} procedure.  
First, choose a timestep corresponding to the desired 
mass resolution in accord with Eq.~(\ref{eq:dw}).  
At each timestep choose a transition 
$\DS$ from Eq.~(\ref{eq:fdef}) and convert it 
to a mass change $\Delta M$.  The algorithm now deviates  
from that of Lacey \& Cole 
because the Somerville \& Kolatt algorithm allows 
for both multiple mergers and diffuse mass accretion.  
At this point, I note that the term ``diffuse mass'' 
is used simply to refer to accretion of objects with 
mass below the threshold $\Mmin$.  In the standard 
excursion set approach, there is no truly diffuse mass 
as all mass is bound into halos of some size.

The Somerville \& Kolatt algorithm continues 
as follows.  If $\Delta M < \Mmin$, 
remove it from further consideration and regard it as 
diffuse accretion.  If $\Delta M \ge \Mmin$ treat it as a 
progenitor halo that has merged.  Next, 
update the remaining mass to $\Mrem = M-\Delta M$.  
Instead of requiring that the second member of the merger 
be a halo of mass $M'=\Mrem$, the Somerville \& Kolatt 
algorithm allows for an arbitrary number of progenitors, 
each of which is chosen from Eq.~(\ref{eq:fdef}), as well as an 
arbitrary amount of accreted mass.  
The next step is as follows.  If $\Mrem > \Mmin$, 
choose another transition and get the new mass change 
$\Delta M'$.  If $\Delta M' > \Mmin$, treat it as a merger 
and if $\Delta M' < \Mmin$, treat it as accretion.  Update 
$\Mrem$ by subtracting $\Delta M'$.  This procedure is then 
repeated until $\Mrem \le \Mmin$ at which point the remaining 
mass is regarded as diffuse accretion.  This produces a list 
of progenitors at a single timestep, though the list typically 
consists of two members and a small quantity of accreted mass.  
To fill out the tree, one applies 
this procedure with new timesteps to all the 
progenitors in the list and so on, recursively building a 
merger tree.  The tree is of finite extent because each 
branch off of the tree terminates when all mass enters 
in units that are $< \Mmin$, so that each unit is regarded 
as accreted material.

The right panel of Figure~\ref{fig:mtcmf} is reproduced from 
Figure~6 of Ref.~\cite{somerville_kolatt99}.  This panel 
shows the conditional mass functions of progenitors 
from an ensemble of merger trees generated using their algorithm 
compared to the analytic excursion set conditional mass functions.  
The suggestions of Somerville \&  Kolatt~\cite{somerville_kolatt99} 
have succeeded in drastically reducing the discrepancy 
between the analytic progenitor mass 
functions and the results of the Monte Carlo merger trees.

\begin{figure}[t]
\begin{center}
\includegraphics[height=10.0cm]{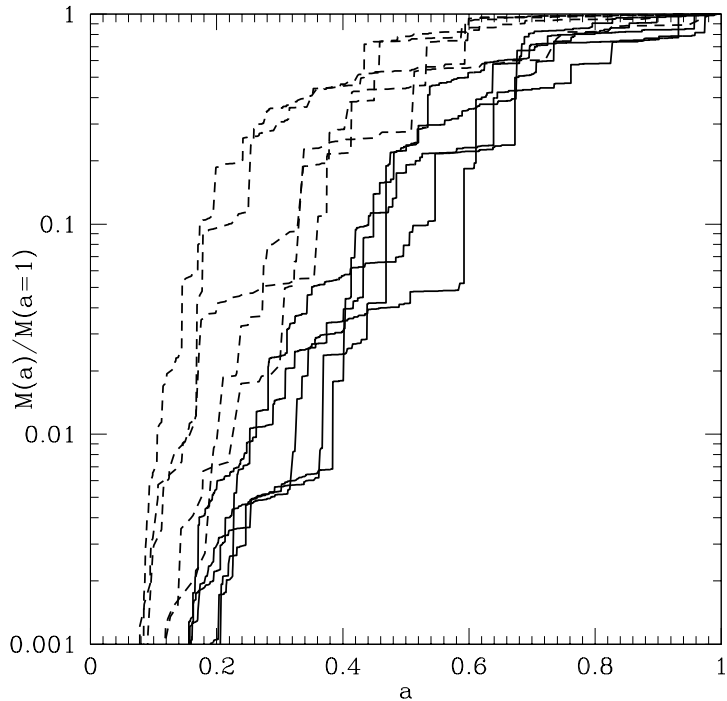}
\caption{
Ten examples of mass accretion histories as a function 
of cosmological scale factor for main progenitors.  
These histories were generated using the Monte Carlo 
merger history method based on excursion set 
theory as described in the text.  The solid lines 
correspond to a large cluster halo with a mass of 
$10^{15}\hMsun$ at $z=0$ ($a=1$).  The dashed lines 
correspond to a Milky Way-sized halo with mass 
$M(a=1) = 10^{12} \hMsun$.  
\label{fig:mhist}
}
\end{center}
\end{figure}

\begin{figure}[t]
\begin{center}
\includegraphics[height=10.0cm]{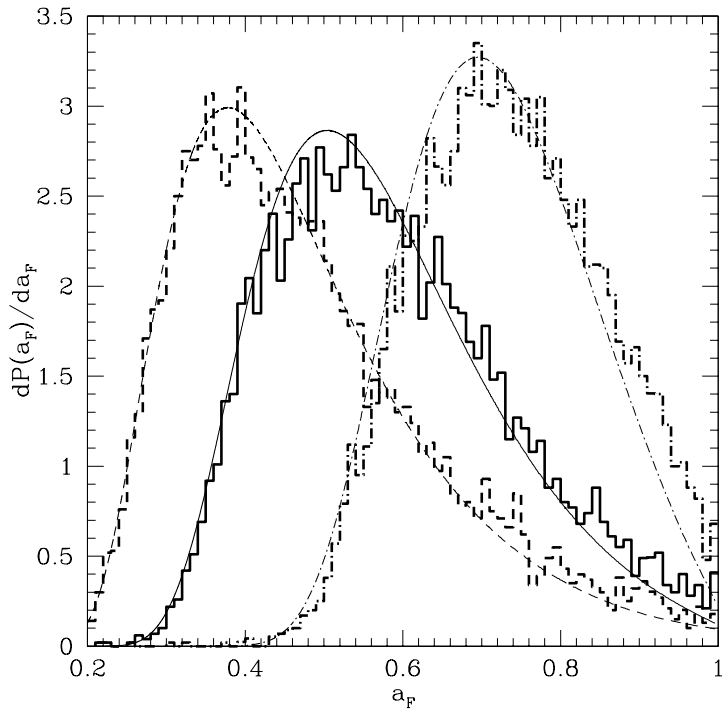}
\caption{
Halo formation times computed from the 
Monte Carlo merger tree algorithm compared 
to the closed form distribution of formation 
times given in \S~\ref{subsection:formtimes}.  
The dashed lines correspond to halos of mass 
$10^{11} \hMsun$, the solid lines correspond to 
halos of mass $10^{13} \hMsun = \Mstar$, and the 
dash-dot lines correspond to halos of mass 
$10^{15} \hMsun$.  The thin, smooth lines are the 
analytic predictions while the thick lines are the 
results $1000$ realizations of the 
Monte Carlo procedure.
\label{fig:mcaf}
}
\end{center}
\end{figure}

Examples of mass accretion histories for  
main progenitors (the most massive progenitor 
at each timestep) computed according to this 
merger tree prescription are shown in Fig.~\ref{fig:mhist}.  
The large vertical jumps correspond to very sudden 
increases in mass due to (perhaps multiple) major 
mergers.  The variety of different paths to the 
same final mass is evident.  Also clear in Fig.~\ref{fig:mhist} 
is the fact that smaller halos acquire their masses 
relatively earlier than larger halos, a picture that 
is consistent with the results shown in Fig.~\ref{fig:accrate}.  
Formation times are also easy to compute directly 
from these merger histories.  The distribution of 
formation times among Monte Carlo realizations 
is shown in Fig.~\ref{fig:mcaf}.  Also shown in 
Fig.~\ref{fig:mcaf} are the formation time distributions 
from \S~\ref{subsection:formtimes}, which appear to be 
in reasonable agreement (Note that the agreement here 
is better than that shown in the paper by 
Lacey \& Cole~\cite{lacey_cole93} due 
to the use of the Somerville \& Kolatt~\cite{somerville_kolatt99} 
algorithm for generating merger trees 
rather than a binary-split algorithm).

\begin{figure}[t]
\begin{center}
\includegraphics[height=10.0cm]{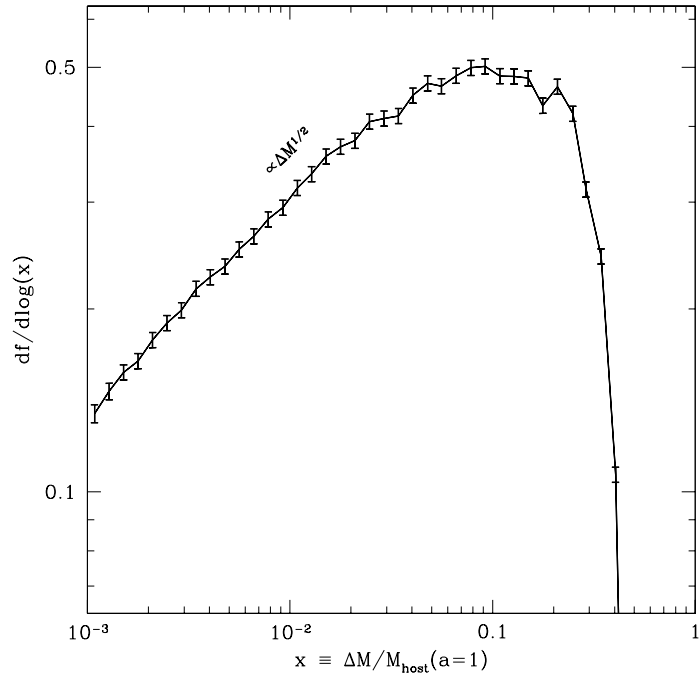}
\caption{
The total fraction of a host halo mass at $a=1$ 
accreted in units of size $\Delta M/M$ per logarithmic 
interval $\dd \ln (\Delta M)$, $\dd F/\dd \ln (\Delta M)$.  
The result shown is from $1000$ merger histories of a halo 
with mass $10^{13}\hMsun = \Mstar$ at $a=1$.  
\label{fig:dfdx}
}
\end{center}
\end{figure}

Merger trees also give a wealth of information about 
the manner in which the mass is acquired by the 
parent halos.  Some properties of the 
distribution of all mergers are extremely simple.  
Consider the transition probability in the binary 
regime and in the regime $\Delta M \ll M$ so that the 
merging object is much smaller than the main progenitor 
with which it is merging.  In this case, the transition 
probability is 
$f \sim [\Delta\w /(\sqrt{2\pi} \abs{\dd S(M)/\dd M})] \Delta M^{-3/2}$.  
As such, the average mass fraction that is accreted in 
units in units of mass in a 
logarithmic interval $\ln (\Delta M)$ varies as 
$\dd F/\dd \ln (\Delta M) \sim [\Delta \w/(\sqrt{2\pi}\abs{\dd S/\dd M}^{1/2})] \Delta M^{1/2}$ 
for all timesteps.  This requires only that the 
probability be in the binary regime and that the mass change be small.  
Integrating over all time, the full tree has this property as well
with a deviation at $\Delta M \sim 0.1 M$ necessitated by 
mass conservation.  The total mass accreted in units of mass 
$\Delta M$ per logarithmic interval $\dd \ln (\Delta M)$ integrated 
over the entire history of $1000$ realizations of the merger 
history of an $\Mstar$ halo is shown in Fig.~\ref{fig:dfdx}.  
The scaling $\dd F/\dd \ln (\Delta M) \propto \Delta M^{1/2}$ at 
$\Delta M \ll M$ is evident.  Integrating over $\ln (\Delta M)$ 
gives $F=1$ because all mass is acquired in the form of 
halos of a particular mass in the excursion set model.  

Before moving on to extensions of the simple excursion set 
approach it is worth drawing attention to a simple but 
profound weakness of the excursion set theory that is 
not often discussed in the literature, though one 
noteworthy exception is the lecture by White~\cite{white96}.  
In this and the preceding sections, I described how the excursion set 
theory is used to predict the merger rates between distinct 
dark matter halos.  However, these distinct halos are never 
clearly identified in the excursion set approach.  
As an example, consider 
some patch of the initial Lagrangian space.  
We can draw a sphere about each point 
in that patch that has a smoothed density that 
is above threshold at some scale and thereby assign to each 
point a halo mass.  In general, each point will be assigned a 
mass that is different from its neighboring points.  
The result of this exercise is 
a continuous ``halo mass field,'' rather than the 
definite assignment of mass in certain patches to 
distinct halos of particular masses.  This makes tests of 
excursion set theory on an object-by-object basis a 
poorly-defined exercise because it is unclear how to assign 
individual particles to halos in the initial Lagrangian space. 
In fact, such tests have been made with 
different choices for assigning mass elements to halos 
(compare Ref.~\cite{white96} and Ref.~\cite{sheth_etal01}). 
This fundamental shortcoming of the excursion set theory 
may partially explain the difficulty of the theory in 
producing self-consistent merger trees and progenitor mass 
functions and its shortcomings compared to full numerical 
treatments.

\section{Beyond the Simple Excursion Set Approach}
\label{section:beyond}

In the proceeding sections, I have reviewed the 
simplest implementation of the excursion set approach 
to halo formation.  While this approach provides 
valuable insight into the influence of the statistical 
properties of the density field on the distribution and 
formation of halos, in its detail it is known to have 
a number of shortcomings, a number of which we have 
encountered already, that render it a less than ideal 
tool for applications where precision is required 
(references~\cite{lacey_cole94,gelb_bertschinger94,white96,jing98,tormen98,sheth_tormen99,somerville_kolatt99,jenkins_etal01,wechsler_etal02,sheth_tormen02,sheth_tormen04,benson_etal05,li_etal05} represent an incomplete sample of 
numerous studies that point out a variety issues).  
I will discuss briefly some of these shortcomings and 
some insights into the physical origin of these 
shortcomings in what follows.  One point that is 
important to emphasize is that the excursion set approach 
is itself an approximation that neglects the details of 
nonlinear dynamics aside from the collapse threshold; 
however, the predictions discussed 
in the previous sections largely result from making 
several further assumptions that are motivated primarily 
by simplicity and may not be the basis of the most 
faithful implementation of excursion set theory.

\subsection{Nontrivial Barrier Shapes and Ellipsoidal Collapse}
\label{subsection:ellipsoid}

One way in which the excursion set predictions are simplified 
is to assume that collapse and virialization of a halo 
occurs whenever the smoothed density exceeds a critical 
value $\dc$, that is independent of any further details 
of the density field.  Even some of the gross qualitative 
aspects of excursion set theory no longer hold when this 
simple form for the barrier is no longer required.

Following Sheth~\cite{sheth98}, it is instructive to begin by 
studying a simple toy example such as the linear barrier where the 
threshold increases with the variance $B(S) = \dc + \beta S$ 
(constructed so that it can be easily solved).  
In this case the barrier recedes from the line $\delta=0$ 
at a ``rate''  $\beta$.  Solving for the first-crossing 
distribution of this barrier is equivalent to solving 
for the first-crossing distribution of a constant barrier 
where the transition probability causes trajectories to ``drift'' 
away from the barrier at a rate $\beta$.  In this case, we 
can revisit the logic that led to Eq.~(\ref{eq:inter}) but 
note that the mean transition should be $\avg{\Ddelta} = -\beta \DS$ 
so that Eq.~(\ref{eq:heateq}) is replaced by 
\beq
\label{eq:heatdrift}
\frac{\partial \Pi}{\partial S} = 
\frac{1}{2}\frac{\partial^2 \Pi}{\partial \delta^2} 
+ \beta \frac{\partial \Pi}{\partial \delta}.
\eeq
The boundary conditions and the initial condition are the 
same as before.  Note that in this case, the solution by 
using an image source of trajectories cannot be used because 
the image trajectories would not satisfy the same differential 
equation [Eq.~(\ref{eq:heatdrift})].

The solution to Eq.~(\ref{eq:heatdrift}) can be greatly simplified 
by making the transformation 
$\Pi = U \exp(\beta[\gamma-\beta S/2])$ where 
$\gamma \equiv \dc - \delta$ as in \S~\ref{section:excursion}.  
In that case $U(S,\gamma)$ satisfies 
\beq
\label{eq:heateqU}
\frac{\partial U}{\partial S} = \frac{1}{2}\frac{\partial^2 U}{\partial \gamma^2},
\eeq
which is the diffusion equation with no drift term, and the 
conditions
\beq
U(S=0,\gamma) = \dirac(\gamma-\dc) \exp(-\beta \gamma)
\eeq
and
\beq
U(S,\gamma=0) = 0
\eeq
Following \S~\ref{section:excursion}, this gives the result 
\beq
\label{eq:pilinear}
\Pi(S,\delta) = \frac{\exp(-\beta[\beta S/2 + \delta])}{\sqrt{2\pi S}}
\bigg[ \exp \Bigg(-\frac{\delta^2}{2S}\Bigg) 
- \exp\Bigg(-\frac{[2\dc - \delta]^2}{2S}\Bigg) \Bigg].
\eeq
From Eq.~(\ref{eq:pilinear}), the fundamental relation for the 
probability for a mass element to be above threshold at a 
variance between $S$ and $S+\dd S$ is 
\beq
\label{eq:flinear}
f_{\mathrm{lin}}(S|\delta_0=0,S_0=0)\dd S = 
\frac{\dc}{\sqrt{2\pi}S^{3/2}} \exp \Bigg(-\frac{[\beta S + \dc]^2}{2S}\Bigg).
\eeq
From this solution to the simple linear barrier a mass 
function can be derived in the usual way.

It is useful to note a few things from this example.  First, 
consider a mass function arising from Eq.~(\ref{eq:flinear}).  
For $S \ll \dc/\beta$, the mass function will have an 
exponential cut-off for high-mass (low-$S$) objects as 
before.  Additionally, for $S \gg \dc/\beta$ the mass function 
will feature an exponential suppression of low-mass object 
as well [compared to $f(S)\sim S^{-3/2}$ for $S>>\dc^2$ in 
the constant barrier case, Eq.~(\ref{eq:fdelta})].  
This is due to the fact that the barrier increases 
in height as $S$ while the standard deviation of the walk increases 
only as $\sqrt{S}$ so that the barrier becomes harder and harder 
to pierce at high $S$.  In contrast, the constant barrier 
will always be crossed because the standard deviation of the 
walk $\sqrt{S}$, always becomes much larger than the barrier 
height as $S$ increases.  This leads to a second interesting 
feature.  Integrating Eq.~(\ref{eq:flinear}) over all $S$ 
gives the total fraction of mass in bound halos as 
$F = \exp(-2\beta \dc) < 1$.  The statement that all matter 
is bound into halos is not generally true in the excursion 
set approach, but is a result of specific barrier shapes.

In fact, nontrivial barriers may play an important practical 
role in understanding things like the mass function and bias 
of halos.  A number of papers  
have, essentially, argued that the spherical 
collapse model that leads to the constant barrier at $\dc$ 
is an inadequate description of the collapse of 
overdense patches and that the barrier shape can be 
more complex effectively to reflect both the more 
complicated collapse process and the relative differences 
between large and small overdense 
patches~\cite{monaco95,bond_myers96,monaco97a,monaco97b,audit_etal97,sheth_etal01} .  
Sheth \& Tormen~\cite{sheth_tormen99} noted that the GIF simulations 
obeyed a mass function where the fraction of mass in 
collapsed objects is modified from Eq.~(\ref{eq:fnu}) to 
\beq
\label{eq:fst}
f_{\mathrm{ST}}(\nu) = A\sqrt{\frac{2a}{\pi}}
\Bigg(1+\frac{1}{(\sqrt{a}\nu)^{2p}}\Bigg)\nu \ \exp(-a\nu^2/2),
\eeq
with $a=0.707$, $p=0.3$, and $A=0.322$ in order to guarantee 
that all mass is bound into halos of some mass.
This is the form shown in Fig.~\ref{fig:fnu} and it clearly 
represents a major improvement over the simple excursion 
set mass function.  Reference~\cite{sheth_tormen99} also 
showed that this form translated into a much improved model 
for halo bias (using the same logic as \S~\ref{section:bias}) 
as shown in Fig.~\ref{fig:bias}.

Subsequently, Sheth et al.~\cite{sheth_etal01} and 
Sheth \& Tormen~\cite{sheth_tormen02} discussed 
how the collapse fraction in Eq.~(\ref{eq:fst}) is very similar 
to the collapse fraction one would derive from of an excursion 
set model with a nontrivial barrier, the form of which is 
motivated by considering the ellipsoidal, rather than 
spherical, collapse of overdense patches.  In general, integrating  
the details of ellipsoidal collapse into the excursion 
set formalism is quite complicated 
(some examples include 
Refs.~\cite{monaco95,audit_etal97,lee_shandarin98}).  
Sheth et al.~\cite{sheth_etal01} greatly simplified this 
problem by considering the barrier that results 
form the mode of the distributions for ellipticity and 
prolateness of overdense patches 
(see Refs.~\cite{bond_myers96,sheth_etal01,sheth_tormen02} 
for details).  In short, the 
simplification stems from assigning overdense patches only 
a single ellipsoidal shape rather than the full distribution of 
possible shapes that would be realized in the cosmological 
density contrast field.  This yields a single barrier shape 
to describe ellipsoidal collapse rather than, at least effectively, 
considering a barrier with a shape that should depend upon 
the specific properties of each overdense patch.  
As it turns out, this approximation works fairly well.  
The resulting, simplified barrier is~\cite{sheth_etal01,sheth_tormen02}
\beq
\label{eq:ellbarrier}
\dell = \dc (1+\beta \nu^{-\gamma}),
\eeq
where the parameters 
$\beta \approx 0.47$ and $\gamma \approx 0.62$ are 
determined by considering the evolution of 
an ellipsoidal density perturbation rather than a 
perfectly spherical density perturbation.

Notice that the ellipsoidal collapse barrier of 
Eq.~(\ref{eq:ellbarrier}) is higher for low-mass (low-$\nu$) objects.  
The physical reasoning for this is that low-mass objects 
typically exhibit greater ellipticity 
(\eg Refs.~\cite{bond_myers96,sheth_etal01}) and are more prone 
to disruption by tidal interactions so that their densities 
must be higher in order to overcome this countervailing effect.  
Notice that one of the primary deficiencies of the constant-barrier 
excursion set mass function is that it predicts considerably 
too many low-mass halos (see Fig.~\ref{fig:fnu}).  As the 
ellipsoidal barrier is higher for higher $S$ (lower mass), 
one should expect a reduction in low-mass halos relative to the 
standard excursion set mass function and this is precisely 
what is shown in Fig.~\ref{fig:fnu}.

Finally, Sheth \& Tormen~\cite{sheth_tormen02} 
showed that the ellipsoidal model still failed to model 
accurately conditional mass functions at small lookback 
times and large mass ratios.  Moreover, these authors 
also pointed out that the model did a poor job with predictions of 
the number densities of halos in very overdense regions.  
This failure leads to the next 
simplification of the excursion set approach which 
I will address in \S~\ref{subsection:correlations}.  However, 
before moving on, I will briefly note the methods used to solve 
the problem of a walk toward a generic barrier, as this problem 
seems to be increasingly interesting.

\subsection{General Barriers}
\label{subsection:generalbarriers}

\begin{figure}[t]
\begin{center}
\includegraphics[height=7.00cm]{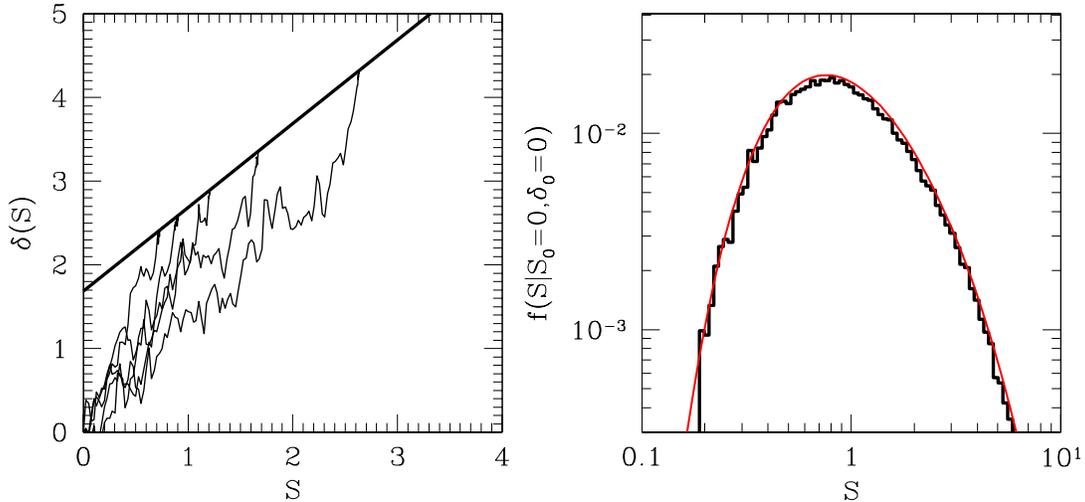}
\caption{
Monte Carlo solution for the first-crossing distribution of a 
linear barrier with $\beta=1$, $B(S)=\dc+S$.  The {\em left} panel shows a 
sample of five $\delta(S)$ trajectories that cross the 
linear threshold.  The {\em thin} lines are the five trajectories 
and the {\em thick} line is the linear barrier.  
The histogram in the {\em right} panel shows the tabulated 
first-crossing distribution constructed from $10^5$ trajectories.  
The {\em thin} line shows the analytic solution to the first-crossing 
distribution from Eq.~(\ref{eq:flinear}).  It is clear that the 
Monte Carlo procedure reproduces the analytic result in the 
case of the linear barrier.
\label{fig:linmc}
}
\end{center}
\end{figure}

It can sometimes be useful to consider more general barrier shapes.  
For example, Furlanetto et al.~\cite{furlanetto_etal04} used a barrier with 
a more complicated shape to model the sizes of ionized regions 
during the epoch of reionization, while Sheth \& Tormen~\cite{sheth_tormen02} 
suggested that moving barriers could effectively encapsulate a 
wide variety of phenomena such as suppression of the collapse 
of small (low-mass) overdense patches in models where the 
dark matter is warm.  In the latter case, the suppression is 
due to the non-negligible primordial velocity 
dispersions of the dark matter particles and the barrier 
shape could encode the importance of these velocities relative 
to the depths of the potentials.  
In this section, I describe two numerical methods 
to construct the first-crossing distribution for 
more general barrier shapes.

The most commonly used method to solve for 
the first-crossing distribution of a barrier 
of more general shape is direct Monte Carlo construction of 
the distribution.  This method is straightforward.  One only 
needs to compute a large number of $\delta(S)$ trajectories.  
To construct a trajectory, one computes values of 
$\delta(S)$ at a discrete set of values of $S$.  
Maintaining the assumption that the smoothing of the 
density fluctuation field is performed with a k-space 
tophat filter, each step is independent and is 
drawn from Eq.~(\ref{eq:Ptransition}) for a 
step size $\DS$.  At this point, one only needs 
to keep track of the value of $S$ at first crossing of the 
barrier.  As with generating halo merger trees, the choice of 
step size $\DS$ is important because all trajectories will 
be resolved only to within $\DS$.  The step size must 
be small enough that a transition above the 
barrier is rare on any given step.  This ensures that multiple 
barrier crossings are unlikely to be missed.

I give an explicit example of the Monte Carlo 
method for the linear barrier with $B(S) = \dc + S$ in 
Fig.~\ref{fig:linmc}.  The left panel shows a sample of five 
trajectories that cross the linear threshold.  Of course, most 
trajectories do not cross the threshold in the range of $S$ that 
I show.  I have selected five that do and so these trajectories 
are a biased subsample of the ensemble of all trajectories.  
The right panel shows the tabulated 
first-crossing distribution constructed 
from $10^5$ trajectories.  It is evident that the Monte Carlo 
method reproduces the analytic result of the previous section 
quite well.

Recently, Zhang and Hui~\cite{zhang_hui06} described a very 
useful and elegant method for solving the first-crossing problem 
for barriers of a very wide range of possible shapes.  
Zhang \& Hui pointed out that the problem can 
be recast in the form of an integral equation that 
can be solved by well-known methods.

The first step is to recognize that all trajectories either 
are included in $\Pi(\delta,S)$ or penetrate the barrier so that 
\beq
\label{eq:alltraj}
\int_0^S\ f(S') \dd S' + \int_{-\infty}^{B(S)}\ \Pi(\delta,S) \dd \delta = 1,
\eeq
where $f(S)$ is the first-crossing distribution and $B(S)$ is the 
scale-dependent barrier height.  If there 
were no barrier, the distribution of $\Pi(\delta,S)$ would 
be given by $\Psi(\delta,S)$ from Eq.~(\ref{eq:Ptransition}).  Therefore, 
$\Pi(\delta,S)$ should be given by $\Psi(\delta,S)$ minus the 
trajectories that pierced the barrier at some $S'<S$ and migrated 
downward to $\delta$ by a variance $S$.  Thus 
\beq
\label{eq:Pi2}
\Pi(\delta,S) = \Psi(\delta,S) - \int_0^S\ f(S')\Psi(\delta-B(S),S-S')\dd S'.
\eeq
Taking the derivative of Eq.~(\ref{eq:alltraj}) with respect to $S$ gives 
an integral relation for $f(S)$ in terms of $\Pi(\delta,S)$ and 
the barrier shape.  
Substituting $\Pi(\delta,S)$ from Eq.~(\ref{eq:Pi2}) into this 
integral relation for $f(S)$ gives 
\beq
\label{eq:volterra}
f(S) = G(S) + \int_0^S\ f(S')H(S,S')\dd S',
\eeq
where $G(S) \equiv [B(S)/S-2\dd B(S)/\dd S]\Psi(B(S),S)$, 
$H(S,S') \equiv [2\dd B(S)/\dd S - (B(S)-B(S'))/(S-S')]\Psi(B(S)-B(S'),S-S')$, 
and $H(S,S')=0$ if $S'>S$.

Equation~(\ref{eq:volterra}) is an integral equation for the 
first-crossing distribution of $f(S)$.  Equation~(\ref{eq:volterra}) has 
the variable $S$ as a limit of integration, and the unknown 
function of interest $f(S)$ both under and outside the 
integral.  This is the form of a 
{\em Volterra Equation of the second kind}.  
This equation can be solved numerically 
using standard methods (\eg Ref.~\cite{arfken_weber95}).  
The first step is to discretize the integral 
on an interval from $0$ to $S$.  The simplest approach 
is to choose a uniform step in $S$.  
Let $n=1,2,...,N$, $\Delta S = S/N$, $S_n=n\Delta S$, 
$f_n=f(S_n)$, $G_n=G(S_n)$, and $H_nm=H(S_n,S_m-\DS/2)$.  Then, the 
integral can be rewritten as 
\beq
\label{eq:volterra_integral}
\int_0^{S_n} \ f(S')\Psi(\delta-B(S_n),S_n-S')\dd S' \simeq 
\frac{\DS}{2} \sum_{m=1}^{N}\ H_{nm} [f_m + f_{m-1}],
\eeq 
so that
\beq
\label{eq:volterra_discrete}
f_n = \frac{G_n \ 
+ \ \DS/2 (\sum_{m=1}^{n-1}\ H_{nm}[f_m+f_{m-1}]\ \ + \ H(n,n)f_{n-1})}
{1-(\DS/2)H(n,n)}
\eeq
with the exceptions
\begin{eqnarray}
\label{eq:volterra_exc}
f_0 & = & G_0 \quad \quad \mathrm{and} \nonumber \\
f_1 & = & 
\frac{G_1 + (\DS/2)H(2,1)[f_1+f_0]}
{1-(\DS/2)H(2,2)}\nonumber.
\end{eqnarray}
The values $f_n$ can be constructed iteratively based on 
Eq.~(\ref{eq:volterra_discrete}).  Utilizing 
Eq.~(\ref{eq:volterra_discrete}) as a solution for 
Eq.~(\ref{eq:volterra}) is generally  a more efficient method 
to obtain a precise estimate for $f(S)$ than 
computing a large ensemble of trajectories.

\begin{figure}[t]
\begin{center}
\includegraphics[height=10.00cm]{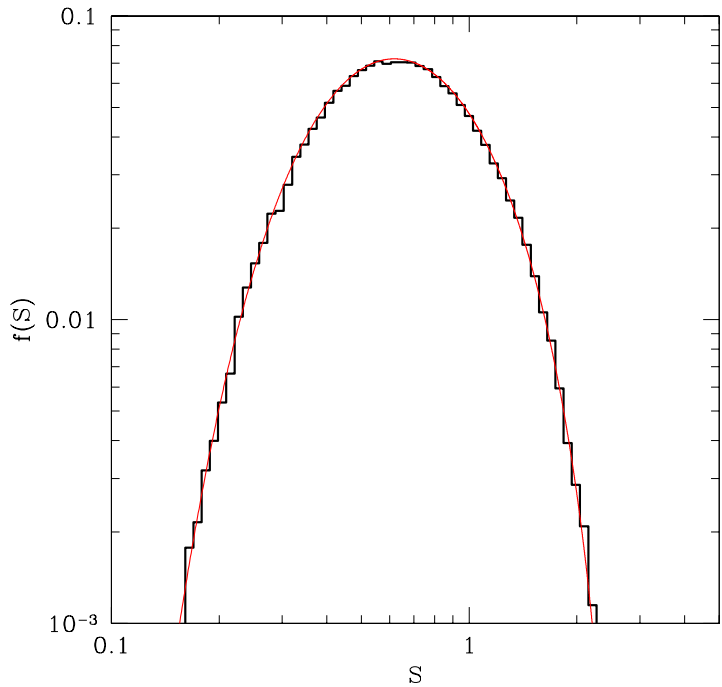}
\caption{
The first-crossing distribution for the quadratic barrier 
$B(S) = \dc + 0.5S^2$.  The histogram is the result of 
a Monte Carlo construction of the distribution while the 
{\em smooth}, {\em solid} line is the iterative solution to the 
Volterra equation [Eq.~(\ref{eq:volterra})] described by 
Zhang \& Hui~\cite{zhang_hui06}.  
\label{fig:fquad}
}
\end{center}
\end{figure}

As a simple example of this approach, 
I show the first-crossing distribution of the barrier 
$B(S)=\dc + 0.5S^2$ in Figure~\ref{fig:fquad}.  The histogram 
is the first-crossing distribution constructed from $10^5$ 
Monte Carlo trajectories, while the smooth line is the 
iterative solution to Eq.~(\ref{eq:volterra}) in 
Eq.~(\ref{eq:volterra_discrete}).  The two solutions are 
clearly in good agreement.  Note also that the first-crossing 
distribution declines more rapidly at high $S$ (small mass) 
than the first-crossing distribution for the linear barrier.  
This is a natural consequence of the fact that the barrier 
height grows more rapidly with respect to the variance 
than in the linear case so that the barrier is even more 
difficult to penetrate at high $S$.

\subsection{Correlations Between Scales}
\label{subsection:correlations}

\begin{figure}[t]
\begin{center}
\includegraphics[height=7.00cm]{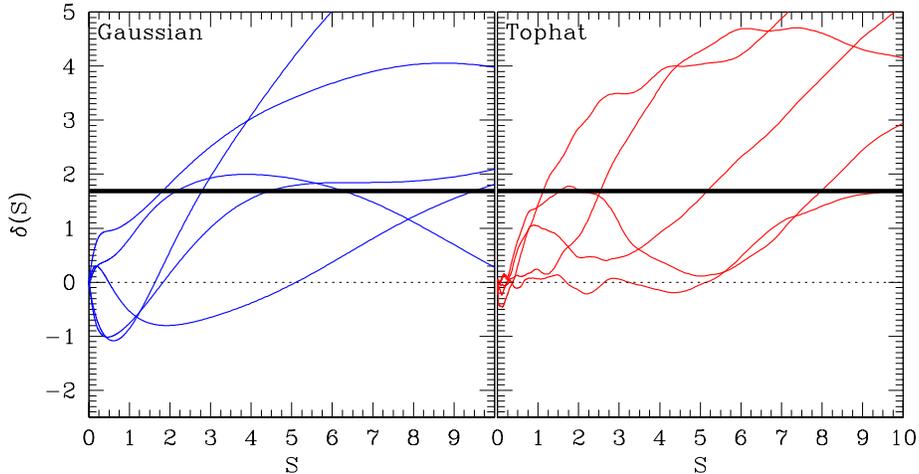}
\caption{
Correlated random walks.  The {\em thin, solid} lines 
show five examples of correlated random walks of $\delta(S)$ defined 
with a Gaussian filter in the left panel and a tophat filter in 
the right panel.  These trajectories were constructed 
from five integrations of the Langevin equation 
[Eq.~(\ref{eq:langevin})] in each case.
The {\em thick} line represents a threshold at $\delta=\dc$.  
The trajectories in this figure have been selected to 
pierce the barrier at $S<10$ which is why they all tend to 
lie in the upper half plane.  Notice that these trajectories 
vary smoothly, unlike the uncorrelated case.
\label{fig:corrtraj}
}
\end{center}
\end{figure}

As I described in \S~\ref{section:excursion}, the 
standard implementation of the excursion set theory 
makes use of a density contrast field filtered with 
a k-space tophat window function.  This choice of a 
filter that is local in Fourier space 
requires that the mass that collapses 
onto a halo is not local in configuration space.  
In fact, the previous sentence understates the 
issue.  In \S~\ref{section:excursion}, I noted that the use 
of the sharp k-space filter represents something 
of a problem because the sharp k-space window is 
difficult to associate with any particular mass because 
its volume diverges.  However, once some 
method for assigning mass to a filter scale 
is decided upon, the use of the sharp k-space filter 
also has an important implication for the properties 
of halos.  
In the case of a sharp k-space window function, the 
transitions between different smoothed densities are 
independent Gaussian random variables as in 
Eq.~(\ref{eq:Ptransition}).  This implies that the 
formation histories of halos are completely 
unrelated to their local environments because 
in the excursion set formalism a halo of mass $M$ 
has $\delta(S) = \dc$ at $S=S(M)$ while all steps 
away from this value are independent of each other.

Though the effects are small, several studies have 
recently pointed out that the formation histories of 
halos in cosmological numerical simulations are correlated with their 
environments~\cite{sheth_tormen04,gao_etal05,wechsler_etal06,wetzel_etal06} 
as are internal properties of halos~\cite{wechsler_etal06,wetzel_etal06,maccio_etal06}.  
The implication of these studies is that  
correlations between scales cannot be neglected completely.   
Though the impact of these effects on observable quantities 
has not been thoroughly explored, the first observation of such 
an effect was recently presented by Berlind et al.~\cite{berlind_etal06}.

\begin{figure}[t]
\begin{center}
\includegraphics[height=7.00cm]{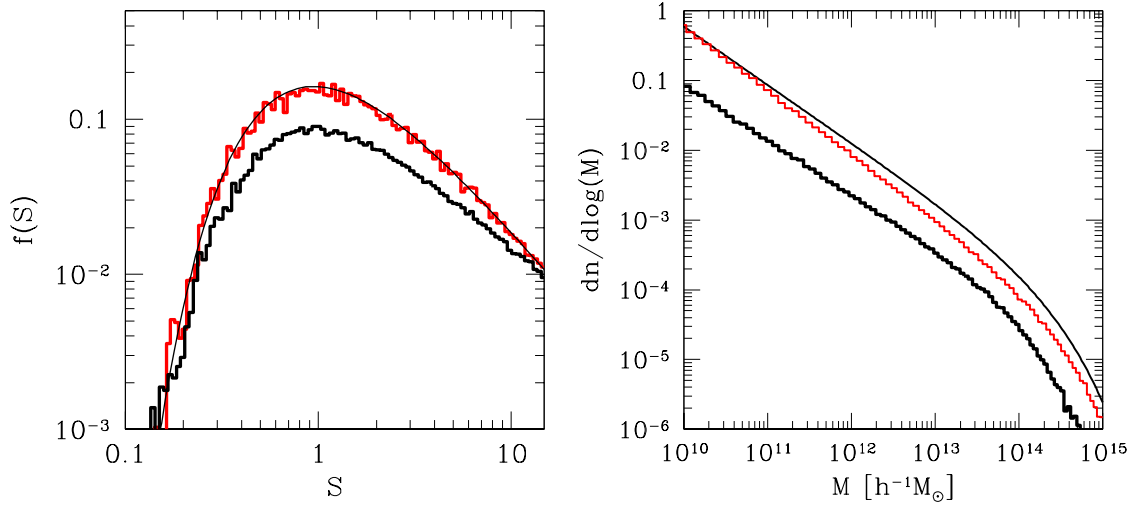}
\caption{
The first-crossing distribution of the smoothed density field 
defined with a Gaussian window function.  The left panel shows 
the first-crossing distribution explicitly.  First the {\em smooth, upper} 
line is the analytic distribution for the sharp k-space window function 
given in Eq.~(\ref{eq:fdelta}).  The {\em upper} histogram shows the 
first-crossing distribution constructed by integrating the Langevin 
equation for this case and demonstrates that the two solutions are 
in good agreement.  The lower histogram is the first-crossing distribution 
for the smoothed density field defined with a Gaussian window function.  
The right panel shows mass functions.  The {\em smooth} line is the 
standard excursion set mass function.  The {\em lower} histogram shows the mass 
function of halos using the Gaussian first-crossing distribution from the 
left panel and the $S(M)$ relationship for a Gaussian filter function.  
The {\em upper} histogram shows the mass function constructed from the 
Gaussian first-crossing distribution of the left panel but using the 
$S(M)$ relation for a real-space tophat window function.  The difference 
between the two histograms represents the effect of the different mass 
definitions with a fixed first-crossing distribution.
\label{fig:fofsgauss}
}
\end{center}
\end{figure}

Along other lines Sheth \& Tormen~\cite{sheth_tormen02}, attempted 
to determine whether or not their conditional mass 
functions were in error because of an incomplete 
treatment of ellipsoidal collapse and concluded that 
a more detailed treatment of ellipsoidal collapse 
could not explain the failures of their model to 
predict conditional mass functions and density-dependent 
mass functions.  They reasoned that because the 
worst failures occurred in high-density regions and 
for conditional mass functions at small lookback 
times and large mass ratios, the deficiency may 
be due to the neglect of correlations between different 
scales.  The reasoning is as follows.  Consider the 
specific example of the conditional mass function.  
To compute the conditional mass function, 
two smoothing scales are necessary.  
One scale describes the parent object while the 
other scale describes the progenitor halo at some 
earlier time.  At high mass ratios and small 
lookback times, the two smoothing scales are 
quite similar so one might expect correlations 
to be important.  However, in the low-mass-ratio 
regime where the progenitors are much smaller than 
the parent halo, the two smoothing scales are 
very different so that it might not be surprising 
for correlations between these very different 
scales to be relatively less important.

\begin{figure}[t]
\begin{center}
\includegraphics[height=7.00cm]{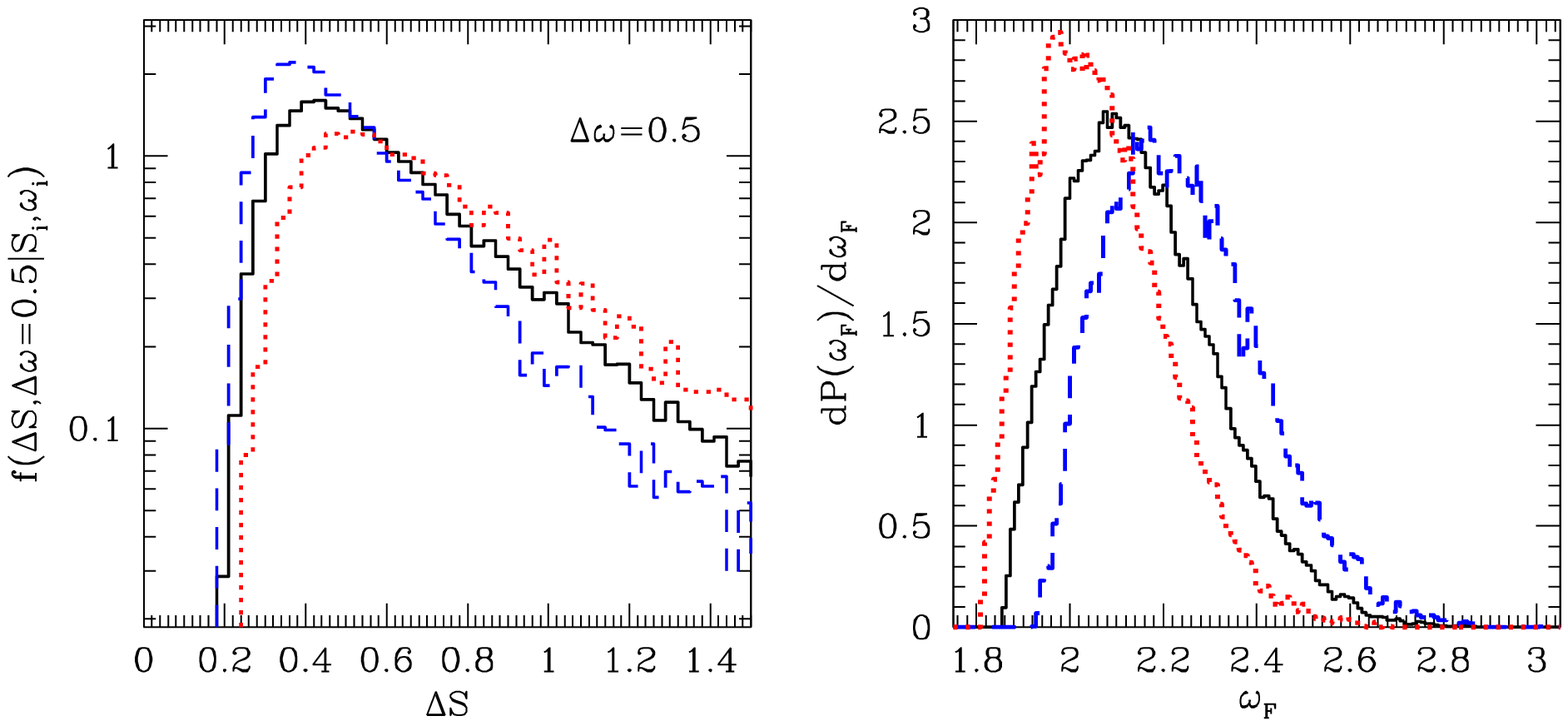}
\caption{
The dependence of halo formation times on halo environment in 
the excursion set theory.  The figure shows results derived from 
trajectories of $\delta(S)$ defined using a Gaussian window function.  
The left panel shows the conditional probability 
for making a first crossing of a barrier of height 
$\w = \dc + \Delta \w$ with $\Delta \w=0.5$, given that the first crossing 
of the barrier $\w_{\mathrm{i}}=\dc$ occurred in the interval 
$1.3 \le S_{\mathrm{i}} \le 1.4$.  This corresponds to a mass 
range $1.8 \le M/[10^{13}\hMsun] \le 2.2$.  The {\em solid} histogram 
shows the full conditional probability distribution derived from 
$10^5$ such trajectories.  The {\em dotted} histogram shows the 
conditional probability derived from the quartile of trajectories 
that have the highest smoothed densities on a scale of 
$R_{\mathrm{env}}=10\hMpc$.  For the purposes of this exercise, the 
smoothed density on a scale of $R_{\mathrm{env}}$ serves to define the 
environment.  The {\em dashed} histogram shows the conditional probability 
from the quartile of trajectories with the smallest smoothed densities on 
a scale of $R_{\mathrm{env}}$.  The right panel shows the distributions of 
formation times (in terms of the variable $\wf$) derived from these trajectories.  
Again, the {\em solid} histogram shows the mean distribution of formation 
times for all halos, the {\em dotted} histogram shows the distribution of 
formation times for the quartile of halos with the highest large-scale 
densities, and the {\em dashed} histogram shows the formation time distribution 
for the quartile of halos in the lowest density environments.  Notice that 
at fixed halo mass, halos with low large-scale densities typically 
form earlier (at higher $\wf$) than halos that reside in regions that 
are overdense on large scales. 
\label{fig:pform}
}
\end{center}
\end{figure}

Correlations between scales can be incorporated into 
the excursion set approach simply by modifying the 
window function.  Any window function other than a 
sharp k-space window will result in walks with steps that 
are correlated to some degree.  In the case of correlated walks, 
Eq.~(\ref{eq:markoff}) no longer applies and so the 
results of \S~\ref{section:excursion} are not valid.  
For correlated walks, the $f(S)\dd S$ distribution 
must be computed by integrating numerically 
a Langevin equation as described in the 
seminal work of Bond et al. in Ref.~\cite{bond_etal91}.

The Brownian motion of a particle can 
be treated by considering the acceleration 
of the particle as due to a fluctuating stochastic force 
$\vec{F}(t)$.  The Langevin equation for the particle motion 
is $\dd \vec{v}/\dd t = \vec{F}(t)$, where $\vec{v}$ is the 
particle velocity (for example, return to the review 
of Chandrasekhar in Ref.~\cite{chandrasekhar43}).  
In a similar way, the trajectory 
$\delta(S)$ experiences a series of stochastic inputs from 
specific realizations of the different Fourier modes of the 
density field.  The trajectory for the smoothed density 
field obeys 
\beq
\label{eq:langevin}
\frac{\partial \delta(S)}{\partial \ln k} = Q(\ln k)W(k;\Rw),
\eeq
where the variance $S$ and the smoothing scale $\Rw$ are 
related through Eq.~(\ref{eq:smoothvariance}) as usual.  
The ``stochastic force'' in this case is $Q( \ln k )$ which 
is a Gaussian random variable with zero mean and a 
variance of 
\beq
\label{eq:varq}
\avg{Q^2(\ln k)} = \frac{\dd S}{\dd \ln k} = \Delta^2(k).
\eeq
A large number of trajectories can be built up numerically 
by summing a large number of realizations of the stochastic 
forces.  The algorithm is to select $Q( \ln k )$ on a grid 
of wavenumbers, to apply the smoothing to each $Q(\ln k)$ 
on a grid of smoothing scales, and to sum the contributions 
of each step in wavenumber to construct each trajectory.  
As before, the steps in smoothing scale should be sufficiently 
small that crossing the barrier of interest on any particular 
interval is unlikely, so it is often most efficient to use a 
non-uniform grid in resolution scale 
(\eg see Ref.~\cite{bond_etal91}).

Figure~\ref{fig:corrtraj} gives examples of trajectories with 
correlated steps computed by integrating numerically the 
Langevin equation of Eq.~(\ref{eq:langevin}).  The left 
panel shows trajectories generated using a Gaussian 
window function and the right panel shows trajectories 
generated using a tophat window function.  The most 
obvious feature of these trajectories is that they do not 
exhibit the frenetic variations of the uncorrelated random 
walks shown in Fig.~\ref{fig:traj} and Fig.~\ref{fig:linmc}.  
On the contrary, because all of the steps in the trajectory 
are correlated to some degree, the trajectories are smooth 
functions of scale.

The first-crossing distribution for a set of trajectories with 
correlated steps can be constructed numerically by generating a 
large number of trajectories like those in Figure~\ref{fig:corrtraj}.  
Figure~\ref{fig:fofsgauss} shows the first-crossing distribution 
for trajectories defined using a Gaussian window function compared 
to the standard first-crossing distribution for sharp k-space filtering.  
In the range of $S$ corresponding to halo masses of interest, the 
first crossing distribution of trajectories defined relative to a 
Gaussian window is lower than the standard, sharp k-space result.  
At large values of the variance ($S \gsim 25$), the first-crossing 
distribution of the trajectories defined relative to a Gaussian 
filter overtakes the sharp k-space distribution 
(these features have been discussed in Ref.~\cite{bond_etal91}).  
Recall that translating $f(S)$ into a mass function 
requires knowledge of the relationship between 
the variance $S$ and the mass scale $M$.  
Therefore, comparing $f(S)$ for two different filtering functions 
does not reflect the differences in the mass function 
directly because the two different $S(M)$ relations must be accounted for.  
In the case of the sharp k-space filter, the definition of mass is 
far from obvious because the filter has a divergent volume associated 
with it.  Standard practice is to use the $S(M)$ relation set by 
a real-space tophat window function.  The right panel of 
Figure~\ref{fig:fofsgauss} shows the mass function for the 
Gaussian first-crossing distribution using both the Gaussian 
window relation for $S(M)$ and the real-space tophat relation 
for $S(M)$ so that the effect of changing the $S(M)$ relation at 
fixed $f(S)$ on the mass function is evident.

\subsection{Halo Formation Times and Environment:  An Example}
\label{subsection:environment}

The excursion set theory using filters other than the sharp k-space 
filter has been considered only relatively rarely in the literature, 
but in light of recent results it is interesting to ask what excursion 
set theory might say about the relationship between halo properties, 
such as formation time, and halo environments.  One can ask what 
predictions implementing a window that is localized in configuration 
space, rather than Fourier space, makes for the relation between, 
in particular, halo formation histories and large-scale halo environments.

Consider as a concrete example the dependence 
of halo formation time on environment at fixed halo mass.  
A simple and physically reasonable way 
to introduce correlations between the past formation 
histories of halos and the environments in which halos reside is to 
consider trajectories of $\delta(S)$ defined using a window function 
that is localized in configuration space.  The logic for determining 
formation times is precisely the same as that in \S~\ref{subsection:formtimes}, 
but the probability distributions must be constructed numerically.  
I show the result of a simple toy example 
in Figure~\ref{fig:pform}.  The argument is fairly general, 
but for concreteness, take the specific 
example of the formation times of $10^{13} \hMsun$ halos 
with environment defined as the smoothed density field on scale of 
$R_{\mathrm{env}} = 10\hMpc$, 
$\delta_{\mathrm{env}} \equiv \delta(\Rw=R_{\mathrm{env}})$.  
The calculation must be constructed numerically 
from a large ensemble of trajectories so 
it is necessary to take a bin of halo mass that has finite extent.  
For this exercise, I consider halos that 
make a first crossing of the barrier $\dc$ between 
$1.3 \le S_{\mathrm{i}} \le 1.4$.  This corresponds to a mass range 
of approximately $1.8 \lsim M/[10^{13} \hMsun] \lsim 2.2$.  This 
toy model is based on $10^5$ trajectories of $\delta(S)$ defined 
using Gaussian filtering so that the steps of the random walks 
are correlated.

\begin{figure}[t]
\begin{center}
\includegraphics[height=7.00cm]{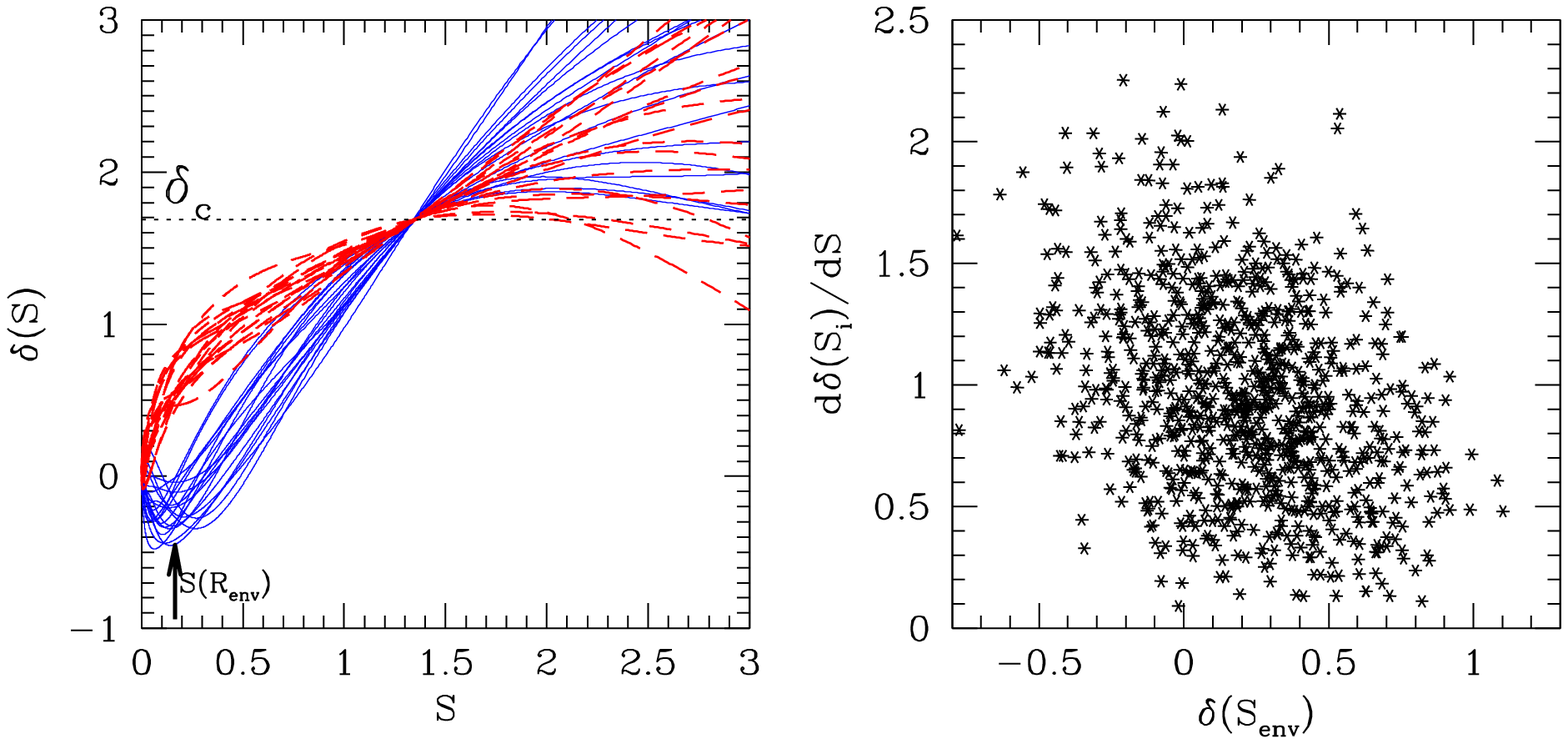}
\caption{
Trajectories exhibiting low and high large-scale 
densities.  The left panel shows twenty randomly-selected 
trajectories with values of $\delta_{\mathrm{env}}$ in 
the lowest quartile ({\em solid lines}) and $\delta_{\mathrm{env}}$ 
in the highest quartile ({\em dashed lines}).  
The {\em horizontal, dotted} line marks $\delta=\dc$ indicating 
collapse at $z=0$.  All trajectories represent halos of nearly 
identical mass so they cross $\dc$ at nearly identical points.  
The arrow indicates the variance on the length scale 
$R_{\mathrm{env}}=10 \hMpc$ which is where the environment measure 
$\delta_{\mathrm{env}}$ is defined.  Notice that trajectories with 
low large-scale densities tend to rise more rapidly after crossing 
the threshold at $\dc$.  The right panel shows a 
scatter plot of large-scale density contrast $\delta_{\mathrm{env}}$ 
and the rate of change of $\delta(S)$ at the first-crossing of 
$\dc$ at $S_{\mathrm{i}}$, $\dd \delta (S_{\mathrm{i}})/\dd S$.  
The inverse correlation between large-scale density and local 
rate of increase is evident.
\label{fig:condtraj}
}
\end{center}
\end{figure}

The left panel of Figure~\ref{fig:pform} shows an explicit 
example of a conditional probability for crossing a second barrier 
of height $\w = \dc + \Delta \w$ with $\Delta \w = 0.5$ in an interval 
of width $\dd S$ about $S$ given a first crossing of the barrier at 
$\dc$ in the range $1.3 \le S_{\mathrm{i}} \le 1.4$, as specified in 
the previous paragraph.  Also shown in the left panel of 
Figure~\ref{fig:pform} is the conditional probability for the 
quartiles of trajectories with the highest and lowest values of 
large-scale density $\delta_{\mathrm{env}}$.  
Notice that the conditional probability 
$f(\DS,\Delta \w|S_{\mathrm{i}},\w_{\mathrm{i}})$ is shifted toward 
lower $\DS$ in regions of low density.  Trajectories in 
low-density regions rise more rapidly in the vicinity of 
the threshold.  According to the logic of 
\S~\ref{subsection:formtimes}, the consequence of this 
is that trajectories in low-density regions generally will 
achieve higher values by the time the smoothing scale decreases 
to include half of the initial mass.  As a direct result, 
these trajectories will generally correspond to halos with 
earlier formation times.  I show this explicitly in the 
right panel of Figure~\ref{fig:pform}.  This panel shows the 
distribution of formation times computed from the same set of 
trajectories for all halos, as well as for the high and low quartiles of 
$\delta_{\mathrm{env}}$.

Figure~\ref{fig:pform} shows clearly that introducing correlations 
between different scales into the excursion set formalism generally 
leads to halos in dense environments forming later and halos in 
diffuse environments forming earlier at fixed halo mass.  The example 
of Figure~\ref{fig:pform} described in the previous paragraphs is 
fairly specific; however, while the magnitude of this effect 
varies, the sense of this effect is general and 
holds for different filter shapes and for $\delta_{\mathrm{env}}$ 
defined on widely different scales.  Interestingly, I have computed 
the environment-dependent formation time distribution for halos of 
numerous masses, and the sense of this shift is 
independent of the mass of halo under consideration.

The generality of this result implies that the excursion 
set theory with a density field smoothed by a filter that 
is localized in configuration space will lead to dense regions 
that preferentially host late-forming halos and vice versa 
for regions of low density.  The result stems from the general fact that the 
conditional probability $f(\DS,\Delta \w|S_{\mathrm{i}},\w_{\mathrm{i}})$ 
is shifted toward low $\DS$ in low density regions.  There is a 
simple qualitative explanation for this behavior.  
All trajectories are required to pass through 
the point $\w_{\mathrm{i}}$ at $S_{\mathrm{i}}$ in order to 
be considered a halo of a particular mass at all.  
Consider beginning a walk at $S=0$ 
and $\delta=0$.  Trajectories that start so 
that the large-scale smoothed density is relatively low at 
some small value of the variance $S$ must rise rapidly in order 
to cross $\w_{\mathrm{i}}$ at $S_{\mathrm{i}}$.  The steps 
in the walk or correlated, so that these trajectories typically 
continue to rise after crossing this threshold.  The opposite 
is true in high-density regions.  On average, trajectories 
that start at low large-scale densities, proceed to pierce 
subsequent thresholds earlier in the walk than counterpart 
trajectories with high large-scale densities.  This general 
trend is illustrated in Figure~\ref{fig:condtraj} where I 
show both examples of trajectories in low- and high-density 
regions and a scatter plot of $\delta_{\mathrm{env}}$ 
against the rate of change of the trajectories about 
the crossing of $\dc$ at $S_{\mathrm{i}}$, 
$\dd \delta(S_{\mathrm{i}})/\dd S$.

The results of this toy model for the environmental 
dependence of halo formation indicate that the natural 
expectation of excursion set theory should be that halos 
in dense environments tend to form relatively later than average, 
while halos in diffuse environments tend to form relatively 
earlier than average.  Excursion set theory with a constant 
barrier height predicts this relative trend 
independent of halo mass.  The studies in 
Refs.~\cite{gao_etal05,wechsler_etal06,wetzel_etal06,maccio_etal06} 
indicate that low-mass halos ($M \ll \Mstar$) that 
form early should be found preferentially in high-density 
environments and conversely for late-forming halos.  This is 
opposite the excursion set expectation described in the 
preceding paragraphs.  
On the contrary, the recent results of 
Wechsler et al.~\cite{wechsler_etal06} and 
Wetzel et al.~\cite{wetzel_etal06} 
indicate that in the high-mass regime ($M \gsim \Mstar$), 
it is indeed the late-forming halos that are preferentially 
located in dense environments.

Wang et al.~\cite{wang_etal06} have 
recently studied the origin of the link between formation 
time and environment using a cosmological numerical simulation.  
These authors found that low-mass halos that form early 
are preferentially in dense environments, in agreement with 
previous work.  More interestingly, Wang et al. continued 
to show that the accretion of matter onto small halos in dense 
environments is cut off prematurely 
(relative to excursion set expectations) due to tidal 
interactions and heating of the ambient environment by 
significantly more massive neighbors.  The 
low-mass halos are forced to be early formers because 
their late accretion is eliminated in a competition with massive 
neighbors.  Wang et al. propose this as the origin 
of the environment-dependent formation times measured in previous 
numerical studies.  Finally, Ref.~\cite{wang_etal06} went on 
to show that this competition-limited 
growth is not important for high-mass halos.

It is interesting to conjecture that 
the prediction of the excursion set theory 
and the study by Wang et al.~\cite{wang_etal06} can be combined into a 
global picture that explains environment-dependent formation 
times at all mass scales.  In such a scenario, the statistics 
of overdense patches in the primordial density contrast field 
would favor a bias toward late halo formation in dense regions.  
In the high-mass regime, halo growth would not be limited by 
competition in any meaningful way, so this bias would persist in 
the sense seen by Ref.~\cite{wechsler_etal06} and 
Ref.~\cite{wetzel_etal06} for high-mass halos in numerical 
simulations.  However, in the low-mass regime, 
competition would stifle halo growth, 
forcing small halos in dense environments to be early 
formers.  This gives rise to the bias of small halos in 
dense regions to be preferentially early formers.  

To my knowledge, the issues mentioned in this section have 
not been addressed in any adequate way at present.  The model 
I have given here is an incomplete toy model intended to illustrate 
the method for generalizing the excursion set predictions and 
to indicate the sense of the effect that excursion set theory predicts. 
A more detailed exploration of these points is an 
interesting undertaking, but is well beyond the scope of 
this review.

\subsection{The \pinocchio Algorithm}

In closing, I would like to point toward another tool 
for estimating halo mass functions and formation histories 
quickly and accurately.  This is the publicly-available 
\pinocchio code developed by P. Monaco and collaborators 
~\cite{monaco_etal02,monaco_etal02b,taffoni_etal02}$^\mathrm{,}$\endnote{The \pinocchio code can be obtained from 
{\tt URL} http://adlibitum.oats.inaf.it/monaco/Homepage/Pinocchio/}. 
(A similar tool that can be used to generate mock catalogs of 
halos rapidly is {\tt PTHalos} developed by Scoccimarro and 
Sheth~\cite{scoccimarro98,scoccimarro_sheth02}) 
The great value of the \pinocchio code is that it is 
considerably less computationally expensive than a 
cosmological N-body simulation but predicts halo properties 
and formation histories that are in better agreement 
with N-body results relative to the simple excursion set 
approach (for a recent example see Ref.~\cite{li_etal05}).

Briefly, the \pinocchio code uses an algorithm for 
identifying halos and halo properties such as mass, velocity, 
angular momentum, in particular realizations of the linear 
density field.  The \pinocchio code also contains 
a parameterized algorithm for building filamentary structures 
in the density field and for generating the merging histories 
of halos.  \pinocchio creates a particle-based realization 
of the linear density field in a manner similar to standard 
N-body simulations.  \pinocchio then computes the approximate 
evolution of mass elements using a particular truncation of 
third order Lagrangian perturbation theory (for example, 
first order Lagrangian perturbation theory constitutes the 
well-known Zel'dovich Approximation used to initialize most 
N-body simulations).  As \pinocchio is perturbative, the 
computation progresses at a fraction of the computational 
effort of an N-body simulation.  \pinocchio groups mass 
elements together into halos using an orbit-crossing 
condition.  When the orbits of two mass elements coincide, 
Lagrangian perturbation theory breaks down because the 
mapping from Eulerian position to Lagrangian position becomes 
multi-valued and the density is infinite.  At this point, 
the mass elements are grouped together into a halo (or 
perhaps a filament) in analogy with collapse to a point in 
the evolution of a spherical tophat overdensity.

The development of the algorithm and the 
\pinocchio code involves a great deal of 
detail and is well-beyond the scope of this review.  
However, this development is well-documented and the 
references above provide a good starting point for 
understanding the calculation.

\vskip 0.5truecm

\noindent{\bf Acknowledgments}
\vskip 0.25truecm

I would like to thank Jeremy Tinker for making is numerical 
data available to me for this review.  I thank James Bullock, 
Neal Dalal, Anatoly Klypin, Andrey Kravtsov, Savvas Koushiappas, 
Brant Robertson, Eduardo Rozo, Douglas Rudd, Frank van den Bosch, Risa Wechsler, 
and David Weinberg for useful discussions and/or comments on the manuscript.  
This review is based on notes for lectures given at the Sixth Summer School 
of the Helmholtz Institute for Supercomputational Physics on 
Supercomputational Cosmology.  The Summer School is funded by 
the Ministry for Science, Research, and Culture Brandenburg.  
I would like to thank the 
organizer of the Summer School on Supercomputational Cosmology, 
Anatoly Klypin, for inviting me to give these lectures.  I thank 
Anatoly Klypin, Andrey Kravtsov and Francisco Prada 
for encouraging me to expand them to the current review.  
I thank the Astrophysical Institute Potsdam for hospitality 
during my stay.  This work was funded by the 
Kavli Institute for Cosmological Physics 
at the University of Chicago, by the National 
Science Foundation under grant NSF PHY 0114422, 
and by the National Science Foundation through the 
Astronomy and Astrophysics Postdoctoral Fellowship 
Program under grant AST 0602122.

\bibliography{eps}

\begin{thebibliography}{84}
\expandafter\ifx\csname natexlab\endcsname\relax\def\natexlab#1{#1}\fi
\expandafter\ifx\csname bibnamefont\endcsname\relax
  \def\bibnamefont#1{#1}\fi
\expandafter\ifx\csname bibfnamefont\endcsname\relax
  \def\bibfnamefont#1{#1}\fi
\expandafter\ifx\csname citenamefont\endcsname\relax
  \def\citenamefont#1{#1}\fi
\expandafter\ifx\csname url\endcsname\relax
  \def\url#1{\texttt{#1}}\fi
\expandafter\ifx\csname urlprefix\endcsname\relax\def\urlprefix{URL }\fi
\providecommand{\bibinfo}[2]{#2}
\providecommand{\eprint}[2][]{\url{#2}}

\bibitem[{\citenamefont{{Press} and {Schechter}}(1974)}]{press_schechter74}
\bibinfo{author}{\bibfnamefont{W.~H.} \bibnamefont{{Press}}} \bibnamefont{and}
  \bibinfo{author}{\bibfnamefont{P.}~\bibnamefont{{Schechter}}},
  \bibinfo{journal}{\apj} \textbf{\bibinfo{volume}{187}}, \bibinfo{pages}{425}
  (\bibinfo{year}{1974}).

\bibitem[{\citenamefont{{Somerville} and {Kolatt}}(1999)}]{somerville_kolatt99}
\bibinfo{author}{\bibfnamefont{R.~S.} \bibnamefont{{Somerville}}}
  \bibnamefont{and} \bibinfo{author}{\bibfnamefont{T.~S.}
  \bibnamefont{{Kolatt}}}, \bibinfo{journal}{\mnras}
  \textbf{\bibinfo{volume}{305}}, \bibinfo{pages}{1} (\bibinfo{year}{1999}).

\bibitem[{\citenamefont{{Cole} et~al.}(2000)\citenamefont{{Cole}, {Lacey},
  {Baugh}, and {Frenk}}}]{cole_etal00}
\bibinfo{author}{\bibfnamefont{S.}~\bibnamefont{{Cole}}},
  \bibinfo{author}{\bibfnamefont{C.~G.} \bibnamefont{{Lacey}}},
  \bibinfo{author}{\bibfnamefont{C.~M.} \bibnamefont{{Baugh}}},
  \bibnamefont{and} \bibinfo{author}{\bibfnamefont{C.~S.}
  \bibnamefont{{Frenk}}}, \bibinfo{journal}{\mnras}
  \textbf{\bibinfo{volume}{319}}, \bibinfo{pages}{168} (\bibinfo{year}{2000}).

\bibitem[{\citenamefont{{Somerville} and
  {Primack}}(1999)}]{somerville_primack99}
\bibinfo{author}{\bibfnamefont{R.~S.} \bibnamefont{{Somerville}}}
  \bibnamefont{and} \bibinfo{author}{\bibfnamefont{J.~R.}
  \bibnamefont{{Primack}}}, \bibinfo{journal}{\mnras}
  \textbf{\bibinfo{volume}{310}}, \bibinfo{pages}{1087} (\bibinfo{year}{1999}),
  \eprint{astro-ph/9802268}.

\bibitem[{\citenamefont{{Kauffmann} and
  {Haehnelt}}(2000)}]{kauffmann_haehnelt00}
\bibinfo{author}{\bibfnamefont{G.}~\bibnamefont{{Kauffmann}}} \bibnamefont{and}
  \bibinfo{author}{\bibfnamefont{M.}~\bibnamefont{{Haehnelt}}},
  \bibinfo{journal}{\mnras} \textbf{\bibinfo{volume}{311}},
  \bibinfo{pages}{576} (\bibinfo{year}{2000}), \eprint{astro-ph/9906493}.

\bibitem[{\citenamefont{{Neistein} et~al.}(2006)\citenamefont{{Neistein}, {van
  den Bosch}, and {Dekel}}}]{neistein_etal06}
\bibinfo{author}{\bibfnamefont{E.}~\bibnamefont{{Neistein}}},
  \bibinfo{author}{\bibfnamefont{F.~C.} \bibnamefont{{van den Bosch}}},
  \bibnamefont{and} \bibinfo{author}{\bibfnamefont{A.}~\bibnamefont{{Dekel}}},
  \bibinfo{journal}{ArXiv Astrophysics e-prints}  (\bibinfo{year}{2006}),
  \eprint{astro-ph/0605045}.

\bibitem[{\citenamefont{{Zentner} and {Bullock}}(2003)}]{zentner_bullock03}
\bibinfo{author}{\bibfnamefont{A.~R.} \bibnamefont{{Zentner}}}
  \bibnamefont{and} \bibinfo{author}{\bibfnamefont{J.~S.}
  \bibnamefont{{Bullock}}}, \bibinfo{journal}{ApJ}
  \textbf{\bibinfo{volume}{598}}, \bibinfo{pages}{49} (\bibinfo{year}{2003}).

\bibitem[{\citenamefont{{Benson} et~al.}(2004)\citenamefont{{Benson}, {Lacey},
  {Frenk}, {Baugh}, and {Cole}}}]{benson_etal04}
\bibinfo{author}{\bibfnamefont{A.~J.} \bibnamefont{{Benson}}},
  \bibinfo{author}{\bibfnamefont{C.~G.} \bibnamefont{{Lacey}}},
  \bibinfo{author}{\bibfnamefont{C.~S.} \bibnamefont{{Frenk}}},
  \bibinfo{author}{\bibfnamefont{C.~M.} \bibnamefont{{Baugh}}},
  \bibnamefont{and} \bibinfo{author}{\bibfnamefont{S.}~\bibnamefont{{Cole}}},
  \bibinfo{journal}{\mnras} \textbf{\bibinfo{volume}{351}},
  \bibinfo{pages}{1215} (\bibinfo{year}{2004}), \eprint{astro-ph/0307298}.

\bibitem[{\citenamefont{{Koushiappas} et~al.}(2004)\citenamefont{{Koushiappas},
  {Zentner}, and {Walker}}}]{koushiappas_etal04}
\bibinfo{author}{\bibfnamefont{S.~M.} \bibnamefont{{Koushiappas}}},
  \bibinfo{author}{\bibfnamefont{A.~R.} \bibnamefont{{Zentner}}},
  \bibnamefont{and} \bibinfo{author}{\bibfnamefont{T.~P.}
  \bibnamefont{{Walker}}}, \bibinfo{journal}{\prd}
  \textbf{\bibinfo{volume}{69}}, \bibinfo{pages}{043501}
  (\bibinfo{year}{2004}), \eprint{astro-ph/0309464}.

\bibitem[{\citenamefont{{Zentner} et~al.}(2005)\citenamefont{{Zentner},
  {Berlind}, {Bullock}, {Kravtsov}, and {Wechsler}}}]{zentner_etal05}
\bibinfo{author}{\bibfnamefont{A.~R.} \bibnamefont{{Zentner}}},
  \bibinfo{author}{\bibfnamefont{A.}~\bibnamefont{{Berlind}}},
  \bibinfo{author}{\bibfnamefont{J.~S.} \bibnamefont{{Bullock}}},
  \bibinfo{author}{\bibfnamefont{A.}~\bibnamefont{{Kravtsov}}},
  \bibnamefont{and} \bibinfo{author}{\bibfnamefont{R.~H.}
  \bibnamefont{{Wechsler}}}, \bibinfo{journal}{\apj, 624, 505}
  (\bibinfo{year}{2005}).

\bibitem[{\citenamefont{{Pe{\~n}arrubia} and
  {Benson}}(2005)}]{penarrubia_benson05}
\bibinfo{author}{\bibfnamefont{J.}~\bibnamefont{{Pe{\~n}arrubia}}}
  \bibnamefont{and} \bibinfo{author}{\bibfnamefont{A.~J.}
  \bibnamefont{{Benson}}}, \bibinfo{journal}{\mnras}
  \textbf{\bibinfo{volume}{364}}, \bibinfo{pages}{977} (\bibinfo{year}{2005}),
  \eprint{astro-ph/0412370}.

\bibitem[{\citenamefont{{Taylor} and
  {Babul}}(2005{\natexlab{a}})}]{taylor_babul05a}
\bibinfo{author}{\bibfnamefont{J.~E.} \bibnamefont{{Taylor}}} \bibnamefont{and}
  \bibinfo{author}{\bibfnamefont{A.}~\bibnamefont{{Babul}}},
  \bibinfo{journal}{\mnras} \textbf{\bibinfo{volume}{364}},
  \bibinfo{pages}{515} (\bibinfo{year}{2005}{\natexlab{a}}).

\bibitem[{\citenamefont{{Taylor} and
  {Babul}}(2005{\natexlab{b}})}]{taylor_babul05b}
\bibinfo{author}{\bibfnamefont{J.~E.} \bibnamefont{{Taylor}}} \bibnamefont{and}
  \bibinfo{author}{\bibfnamefont{A.}~\bibnamefont{{Babul}}},
  \bibinfo{journal}{\mnras} \textbf{\bibinfo{volume}{364}},
  \bibinfo{pages}{535} (\bibinfo{year}{2005}{\natexlab{b}}),
  \eprint{astro-ph/0410049}.

\bibitem[{\citenamefont{{Monaco}
  et~al.}(2002{\natexlab{a}})\citenamefont{{Monaco}, {Theuns}, and
  {Taffoni}}}]{monaco_etal02}
\bibinfo{author}{\bibfnamefont{P.}~\bibnamefont{{Monaco}}},
  \bibinfo{author}{\bibfnamefont{T.}~\bibnamefont{{Theuns}}}, \bibnamefont{and}
  \bibinfo{author}{\bibfnamefont{G.}~\bibnamefont{{Taffoni}}},
  \bibinfo{journal}{\mnras} \textbf{\bibinfo{volume}{331}},
  \bibinfo{pages}{587} (\bibinfo{year}{2002}{\natexlab{a}}).

\bibitem[{\citenamefont{{Monaco}
  et~al.}(2002{\natexlab{b}})\citenamefont{{Monaco}, {Theuns}, {Taffoni},
  {Governato}, {Quinn}, and {Stadel}}}]{monaco_etal02b}
\bibinfo{author}{\bibfnamefont{P.}~\bibnamefont{{Monaco}}},
  \bibinfo{author}{\bibfnamefont{T.}~\bibnamefont{{Theuns}}},
  \bibinfo{author}{\bibfnamefont{G.}~\bibnamefont{{Taffoni}}},
  \bibinfo{author}{\bibfnamefont{F.}~\bibnamefont{{Governato}}},
  \bibinfo{author}{\bibfnamefont{T.}~\bibnamefont{{Quinn}}}, \bibnamefont{and}
  \bibinfo{author}{\bibfnamefont{J.}~\bibnamefont{{Stadel}}},
  \bibinfo{journal}{\mnras} \textbf{\bibinfo{volume}{564}}, \bibinfo{pages}{8}
  (\bibinfo{year}{2002}{\natexlab{b}}).

\bibitem[{\citenamefont{{Taffoni} et~al.}(2002)\citenamefont{{Taffoni},
  {Monaco}, and {Theuns}}}]{taffoni_etal02}
\bibinfo{author}{\bibfnamefont{G.}~\bibnamefont{{Taffoni}}},
  \bibinfo{author}{\bibfnamefont{P.}~\bibnamefont{{Monaco}}}, \bibnamefont{and}
  \bibinfo{author}{\bibfnamefont{T.}~\bibnamefont{{Theuns}}},
  \bibinfo{journal}{\mnras} \textbf{\bibinfo{volume}{333}},
  \bibinfo{pages}{623} (\bibinfo{year}{2002}).

\bibitem[{\citenamefont{{Bardeen} et~al.}(1986)\citenamefont{{Bardeen}, {Bond},
  {Kaiser}, and {Szalay}}}]{bardeen_etal86}
\bibinfo{author}{\bibfnamefont{J.~M.} \bibnamefont{{Bardeen}}},
  \bibinfo{author}{\bibfnamefont{J.~R.} \bibnamefont{{Bond}}},
  \bibinfo{author}{\bibfnamefont{N.}~\bibnamefont{{Kaiser}}}, \bibnamefont{and}
  \bibinfo{author}{\bibfnamefont{A.~S.} \bibnamefont{{Szalay}}},
  \bibinfo{journal}{\apj} \textbf{\bibinfo{volume}{304}}, \bibinfo{pages}{15}
  (\bibinfo{year}{1986}).

\bibitem[{\citenamefont{{Bond} et~al.}(1991)\citenamefont{{Bond}, {Cole},
  {Efstathiou}, and {Kaiser}}}]{bond_etal91}
\bibinfo{author}{\bibfnamefont{J.~R.} \bibnamefont{{Bond}}},
  \bibinfo{author}{\bibfnamefont{S.}~\bibnamefont{{Cole}}},
  \bibinfo{author}{\bibfnamefont{G.}~\bibnamefont{{Efstathiou}}},
  \bibnamefont{and} \bibinfo{author}{\bibfnamefont{N.}~\bibnamefont{{Kaiser}}},
  \bibinfo{journal}{\apj} \textbf{\bibinfo{volume}{379}}, \bibinfo{pages}{440}
  (\bibinfo{year}{1991}).

\bibitem[{\citenamefont{{Lacey} and {Cole}}(1993)}]{lacey_cole93}
\bibinfo{author}{\bibfnamefont{C.}~\bibnamefont{{Lacey}}} \bibnamefont{and}
  \bibinfo{author}{\bibfnamefont{S.}~\bibnamefont{{Cole}}},
  \bibinfo{journal}{\mnras} \textbf{\bibinfo{volume}{262}},
  \bibinfo{pages}{627} (\bibinfo{year}{1993}).

\bibitem[{\citenamefont{{Lacey} and {Cole}}(1994)}]{lacey_cole94}
\bibinfo{author}{\bibfnamefont{C.}~\bibnamefont{{Lacey}}} \bibnamefont{and}
  \bibinfo{author}{\bibfnamefont{S.}~\bibnamefont{{Cole}}},
  \bibinfo{journal}{\mnras} \textbf{\bibinfo{volume}{271}},
  \bibinfo{pages}{676} (\bibinfo{year}{1994}), \eprint{astro-ph/9402069}.

\bibitem[{\citenamefont{{Sheth} and {Lemson}}(1999)}]{sheth_lemson99}
\bibinfo{author}{\bibfnamefont{R.~K.} \bibnamefont{{Sheth}}} \bibnamefont{and}
  \bibinfo{author}{\bibfnamefont{G.}~\bibnamefont{{Lemson}}},
  \bibinfo{journal}{\mnras} \textbf{\bibinfo{volume}{305}},
  \bibinfo{pages}{946} (\bibinfo{year}{1999}).

\bibitem[{\citenamefont{{Wax}}(1954)}]{wax_54}
\bibinfo{author}{\bibfnamefont{N.}~\bibnamefont{{Wax}}},
  \emph{\bibinfo{title}{{Selected Papers on Noise and Stochastic Processes}}}
  (\bibinfo{publisher}{New York: Dover Publication, 1954, edited by Wax,
  Nelson}, \bibinfo{year}{1954}).

\bibitem[{\citenamefont{{Chandrasekhar}}(1943)}]{chandrasekhar43}
\bibinfo{author}{\bibfnamefont{S.}~\bibnamefont{{Chandrasekhar}}},
  \bibinfo{journal}{\rmp} \textbf{\bibinfo{volume}{15}}, \bibinfo{pages}{2}
  (\bibinfo{year}{1943}).

\bibitem[{\citenamefont{{Rice}}(1944)}]{rice_44}
\bibinfo{author}{\bibfnamefont{S.~O.} \bibnamefont{{Rice}}},
  \bibinfo{journal}{Bell Systems Tech.~J., Volume 23, p.~282-332}
  \textbf{\bibinfo{volume}{23}}, \bibinfo{pages}{282} (\bibinfo{year}{1944}).

\bibitem[{\citenamefont{{Rice}}(1945)}]{rice_45}
\bibinfo{author}{\bibfnamefont{S.~O.} \bibnamefont{{Rice}}},
  \bibinfo{journal}{Bell Systems Tech.~J., Volume 24, p.~46-156}
  \textbf{\bibinfo{volume}{24}}, \bibinfo{pages}{46} (\bibinfo{year}{1945}).

\bibitem[{\citenamefont{{Eisenstein} and {Hu}}(1999)}]{eisenstein_hu99}
\bibinfo{author}{\bibfnamefont{D.~J.} \bibnamefont{{Eisenstein}}}
  \bibnamefont{and} \bibinfo{author}{\bibfnamefont{W.}~\bibnamefont{{Hu}}},
  \bibinfo{journal}{\apj} \textbf{\bibinfo{volume}{511}}, \bibinfo{pages}{5}
  (\bibinfo{year}{1999}), \eprint{astro-ph/9710252}.

\bibitem[{\citenamefont{{Williams} et~al.}(1991)\citenamefont{{Williams},
  {Heavens}, {Peacock}, and {Shandarin}}}]{williams_etal91}
\bibinfo{author}{\bibfnamefont{B.~G.} \bibnamefont{{Williams}}},
  \bibinfo{author}{\bibfnamefont{A.~F.} \bibnamefont{{Heavens}}},
  \bibinfo{author}{\bibfnamefont{J.~A.} \bibnamefont{{Peacock}}},
  \bibnamefont{and} \bibinfo{author}{\bibfnamefont{S.~F.}
  \bibnamefont{{Shandarin}}}, \bibinfo{journal}{\mnras}
  \textbf{\bibinfo{volume}{250}}, \bibinfo{pages}{458} (\bibinfo{year}{1991}).

\bibitem[{\citenamefont{{Carroll} et~al.}(1992)\citenamefont{{Carroll},
  {Press}, and {Turner}}}]{carroll_etal92}
\bibinfo{author}{\bibfnamefont{S.~M.} \bibnamefont{{Carroll}}},
  \bibinfo{author}{\bibfnamefont{W.~H.} \bibnamefont{{Press}}},
  \bibnamefont{and} \bibinfo{author}{\bibfnamefont{E.~L.}
  \bibnamefont{{Turner}}}, \bibinfo{journal}{\araa}
  \textbf{\bibinfo{volume}{30}}, \bibinfo{pages}{499} (\bibinfo{year}{1992}).

\bibitem[{\citenamefont{{Bildhauer} et~al.}(1992)\citenamefont{{Bildhauer},
  {Buchert}, and {Kasai}}}]{bildhauer_etal92}
\bibinfo{author}{\bibfnamefont{S.}~\bibnamefont{{Bildhauer}}},
  \bibinfo{author}{\bibfnamefont{T.}~\bibnamefont{{Buchert}}},
  \bibnamefont{and} \bibinfo{author}{\bibfnamefont{M.}~\bibnamefont{{Kasai}}},
  \bibinfo{journal}{A \& A} \textbf{\bibinfo{volume}{263}}, \bibinfo{pages}{23}
  (\bibinfo{year}{1992}).

\bibitem[{\citenamefont{{Epstein}}(1983)}]{epstein83}
\bibinfo{author}{\bibfnamefont{R.~A.} \bibnamefont{{Epstein}}},
  \bibinfo{journal}{\mnras} \textbf{\bibinfo{volume}{205}},
  \bibinfo{pages}{207} (\bibinfo{year}{1983}).

\bibitem[{\citenamefont{{Peacock} and {Heavens}}(1990)}]{peacock_heavens90}
\bibinfo{author}{\bibfnamefont{J.~A.} \bibnamefont{{Peacock}}}
  \bibnamefont{and} \bibinfo{author}{\bibfnamefont{A.~F.}
  \bibnamefont{{Heavens}}}, \bibinfo{journal}{\mnras}
  \textbf{\bibinfo{volume}{243}}, \bibinfo{pages}{133} (\bibinfo{year}{1990}).

\bibitem[{\citenamefont{{Bower}}(1991)}]{bower91}
\bibinfo{author}{\bibfnamefont{R.~J.} \bibnamefont{{Bower}}},
  \bibinfo{journal}{\mnras} \textbf{\bibinfo{volume}{248}},
  \bibinfo{pages}{332} (\bibinfo{year}{1991}).

\bibitem[{\citenamefont{{Adler}}(1981)}]{adler81}
\bibinfo{author}{\bibfnamefont{R.~J.} \bibnamefont{{Adler}}},
  \emph{\bibinfo{title}{{The Geometry of Random Fields}}}
  (\bibinfo{publisher}{The Geometry of Random Fields, Chichester: Wiley, 1981},
  \bibinfo{year}{1981}).

\bibitem[{\citenamefont{{Sheth} and {Tormen}}(1999)}]{sheth_tormen99}
\bibinfo{author}{\bibfnamefont{R.~K.} \bibnamefont{{Sheth}}} \bibnamefont{and}
  \bibinfo{author}{\bibfnamefont{G.}~\bibnamefont{{Tormen}}},
  \bibinfo{journal}{\mnras} \textbf{\bibinfo{volume}{308}},
  \bibinfo{pages}{119} (\bibinfo{year}{1999}).

\bibitem[{\citenamefont{{Jenkins} et~al.}(2001)\citenamefont{{Jenkins},
  {Frenk}, {White}, {Colberg}, {Cole}, {Evrard}, {Couchman}, and
  {Yoshida}}}]{jenkins_etal01}
\bibinfo{author}{\bibfnamefont{A.}~\bibnamefont{{Jenkins}}},
  \bibinfo{author}{\bibfnamefont{C.~S.} \bibnamefont{{Frenk}}},
  \bibinfo{author}{\bibfnamefont{S.~D.~M.} \bibnamefont{{White}}},
  \bibinfo{author}{\bibfnamefont{J.~M.} \bibnamefont{{Colberg}}},
  \bibinfo{author}{\bibfnamefont{S.}~\bibnamefont{{Cole}}},
  \bibinfo{author}{\bibfnamefont{A.~E.} \bibnamefont{{Evrard}}},
  \bibinfo{author}{\bibfnamefont{H.~M.~P.} \bibnamefont{{Couchman}}},
  \bibnamefont{and}
  \bibinfo{author}{\bibfnamefont{N.}~\bibnamefont{{Yoshida}}},
  \bibinfo{journal}{\mnras} \textbf{\bibinfo{volume}{321}},
  \bibinfo{pages}{372} (\bibinfo{year}{2001}).

\bibitem[{\citenamefont{{Kaiser}}(1984)}]{kaiser84}
\bibinfo{author}{\bibfnamefont{N.}~\bibnamefont{{Kaiser}}},
  \bibinfo{journal}{\apjl} \textbf{\bibinfo{volume}{284}}, \bibinfo{pages}{L9}
  (\bibinfo{year}{1984}).

\bibitem[{\citenamefont{{Efstathiou} et~al.}(1988)\citenamefont{{Efstathiou},
  {Frenk}, {White}, and {Davis}}}]{efstathiou_etal88}
\bibinfo{author}{\bibfnamefont{G.}~\bibnamefont{{Efstathiou}}},
  \bibinfo{author}{\bibfnamefont{C.~S.} \bibnamefont{{Frenk}}},
  \bibinfo{author}{\bibfnamefont{S.~D.~M.} \bibnamefont{{White}}},
  \bibnamefont{and} \bibinfo{author}{\bibfnamefont{M.}~\bibnamefont{{Davis}}},
  \bibinfo{journal}{\mnras} \textbf{\bibinfo{volume}{235}},
  \bibinfo{pages}{715} (\bibinfo{year}{1988}).

\bibitem[{\citenamefont{{Mo} and {White}}(1996)}]{mo_white96}
\bibinfo{author}{\bibfnamefont{H.~J.} \bibnamefont{{Mo}}} \bibnamefont{and}
  \bibinfo{author}{\bibfnamefont{S.~D.~M.} \bibnamefont{{White}}},
  \bibinfo{journal}{\mnras} \textbf{\bibinfo{volume}{282}},
  \bibinfo{pages}{347} (\bibinfo{year}{1996}).

\bibitem[{\citenamefont{{Seljak} and {Warren}}(2004)}]{seljak_warren04}
\bibinfo{author}{\bibfnamefont{U.}~\bibnamefont{{Seljak}}} \bibnamefont{and}
  \bibinfo{author}{\bibfnamefont{M.~S.} \bibnamefont{{Warren}}},
  \bibinfo{journal}{\mnras} \textbf{\bibinfo{volume}{355}},
  \bibinfo{pages}{129} (\bibinfo{year}{2004}), \eprint{astro-ph/0403698}.

\bibitem[{\citenamefont{{Cole} and {Kaiser}}(1989)}]{cole_kaiser89}
\bibinfo{author}{\bibfnamefont{S.}~\bibnamefont{{Cole}}} \bibnamefont{and}
  \bibinfo{author}{\bibfnamefont{N.}~\bibnamefont{{Kaiser}}},
  \bibinfo{journal}{\mnras} \textbf{\bibinfo{volume}{237}},
  \bibinfo{pages}{1127} (\bibinfo{year}{1989}).

\bibitem[{\citenamefont{{Sheth} et~al.}(2001)\citenamefont{{Sheth}, {Mo}, and
  {Tormen}}}]{sheth_etal01}
\bibinfo{author}{\bibfnamefont{R.~K.} \bibnamefont{{Sheth}}},
  \bibinfo{author}{\bibfnamefont{H.~J.} \bibnamefont{{Mo}}}, \bibnamefont{and}
  \bibinfo{author}{\bibfnamefont{G.}~\bibnamefont{{Tormen}}},
  \bibinfo{journal}{\mnras} \textbf{\bibinfo{volume}{323}}, \bibinfo{pages}{1}
  (\bibinfo{year}{2001}).

\bibitem[{\citenamefont{{Sheth} and {Tormen}}(2002)}]{sheth_tormen02}
\bibinfo{author}{\bibfnamefont{R.~K.} \bibnamefont{{Sheth}}} \bibnamefont{and}
  \bibinfo{author}{\bibfnamefont{G.}~\bibnamefont{{Tormen}}},
  \bibinfo{journal}{\mnras} \textbf{\bibinfo{volume}{329}}, \bibinfo{pages}{61}
  (\bibinfo{year}{2002}).

\bibitem[{\citenamefont{{Scherrer} and
  {Bertschinger}}(1991)}]{scherrer_bertschinger91}
\bibinfo{author}{\bibfnamefont{R.~J.} \bibnamefont{{Scherrer}}}
  \bibnamefont{and}
  \bibinfo{author}{\bibfnamefont{E.}~\bibnamefont{{Bertschinger}}},
  \bibinfo{journal}{\apj} \textbf{\bibinfo{volume}{381}}, \bibinfo{pages}{349}
  (\bibinfo{year}{1991}).

\bibitem[{\citenamefont{{Seljak}}(2000)}]{seljak00}
\bibinfo{author}{\bibfnamefont{U.}~\bibnamefont{{Seljak}}},
  \bibinfo{journal}{\mnras} \textbf{\bibinfo{volume}{318}},
  \bibinfo{pages}{203} (\bibinfo{year}{2000}).

\bibitem[{\citenamefont{{Ma} and {Fry}}(2000)}]{ma_fry00}
\bibinfo{author}{\bibfnamefont{C.-P.} \bibnamefont{{Ma}}} \bibnamefont{and}
  \bibinfo{author}{\bibfnamefont{J.~N.} \bibnamefont{{Fry}}},
  \bibinfo{journal}{\apj} \textbf{\bibinfo{volume}{543}}, \bibinfo{pages}{503}
  (\bibinfo{year}{2000}).

\bibitem[{\citenamefont{{Peacock} and {Smith}}(2000)}]{peacock_smith00}
\bibinfo{author}{\bibfnamefont{J.~A.} \bibnamefont{{Peacock}}}
  \bibnamefont{and} \bibinfo{author}{\bibfnamefont{R.~E.}
  \bibnamefont{{Smith}}}, \bibinfo{journal}{\mnras}
  \textbf{\bibinfo{volume}{318}}, \bibinfo{pages}{1144} (\bibinfo{year}{2000}).

\bibitem[{\citenamefont{{Berlind} and {Weinberg}}(2002)}]{berlind_weinberg02}
\bibinfo{author}{\bibfnamefont{A.~A.} \bibnamefont{{Berlind}}}
  \bibnamefont{and} \bibinfo{author}{\bibfnamefont{D.~H.}
  \bibnamefont{{Weinberg}}}, \bibinfo{journal}{\apj}
  \textbf{\bibinfo{volume}{575}}, \bibinfo{pages}{587} (\bibinfo{year}{2002}).

\bibitem[{\citenamefont{{Icke}}(1984)}]{icke84}
\bibinfo{author}{\bibfnamefont{V.}~\bibnamefont{{Icke}}},
  \bibinfo{journal}{\mnras} \textbf{\bibinfo{volume}{206}}, \bibinfo{pages}{1}
  (\bibinfo{year}{1984}).

\bibitem[{\citenamefont{{Bertschinger}}(1985)}]{bertschinger85}
\bibinfo{author}{\bibfnamefont{E.}~\bibnamefont{{Bertschinger}}},
  \bibinfo{journal}{\apjs} \textbf{\bibinfo{volume}{58}}, \bibinfo{pages}{1}
  (\bibinfo{year}{1985}).

\bibitem[{\citenamefont{{van de Weygaert} and {Van
  Kampen}}(1993)}]{weygaert_kampen93}
\bibinfo{author}{\bibfnamefont{R.}~\bibnamefont{{van de Weygaert}}}
  \bibnamefont{and} \bibinfo{author}{\bibfnamefont{E.}~\bibnamefont{{Van
  Kampen}}}, \bibinfo{journal}{\mnras} \textbf{\bibinfo{volume}{263}},
  \bibinfo{pages}{481} (\bibinfo{year}{1993}).

\bibitem[{\citenamefont{{Sheth} and {van de
  Weygaert}}(2004)}]{sheth_weygaert04}
\bibinfo{author}{\bibfnamefont{R.~K.} \bibnamefont{{Sheth}}} \bibnamefont{and}
  \bibinfo{author}{\bibfnamefont{R.}~\bibnamefont{{van de Weygaert}}},
  \bibinfo{journal}{\mnras} \textbf{\bibinfo{volume}{350}},
  \bibinfo{pages}{517} (\bibinfo{year}{2004}).

\bibitem[{\citenamefont{{Blumenthal} et~al.}(1992)\citenamefont{{Blumenthal},
  {Da Costa}, {Goldwirth}, {Lecar}, and {Piran}}}]{blumenthal_etal92}
\bibinfo{author}{\bibfnamefont{G.~R.} \bibnamefont{{Blumenthal}}},
  \bibinfo{author}{\bibfnamefont{L.}~\bibnamefont{{Da Costa}}},
  \bibinfo{author}{\bibfnamefont{D.~S.} \bibnamefont{{Goldwirth}}},
  \bibinfo{author}{\bibfnamefont{M.}~\bibnamefont{{Lecar}}}, \bibnamefont{and}
  \bibinfo{author}{\bibfnamefont{T.}~\bibnamefont{{Piran}}},
  \bibinfo{journal}{\apj} \textbf{\bibinfo{volume}{388}}, \bibinfo{pages}{234}
  (\bibinfo{year}{1992}).

\bibitem[{\citenamefont{{Furlanetto} and {Piran}}(2005)}]{furlanetto_piran05}
\bibinfo{author}{\bibfnamefont{S.}~\bibnamefont{{Furlanetto}}}
  \bibnamefont{and} \bibinfo{author}{\bibfnamefont{T.}~\bibnamefont{{Piran}}},
  \bibinfo{journal}{\mnras, in press}  (\bibinfo{year}{2005}),
  \eprint{arXiv:astro-ph/0509148}.

\bibitem[{\citenamefont{{Cole} and {Kaiser}}(1988)}]{cole_kaiser88}
\bibinfo{author}{\bibfnamefont{S.}~\bibnamefont{{Cole}}} \bibnamefont{and}
  \bibinfo{author}{\bibfnamefont{N.}~\bibnamefont{{Kaiser}}},
  \bibinfo{journal}{\mnras} \textbf{\bibinfo{volume}{233}},
  \bibinfo{pages}{637} (\bibinfo{year}{1988}).

\bibitem[{\citenamefont{{Cole}}(1991)}]{cole91}
\bibinfo{author}{\bibfnamefont{S.}~\bibnamefont{{Cole}}},
  \bibinfo{journal}{\apj} \textbf{\bibinfo{volume}{367}}, \bibinfo{pages}{45}
  (\bibinfo{year}{1991}).

\bibitem[{\citenamefont{{Kauffmann} et~al.}(1993)\citenamefont{{Kauffmann},
  {White}, and {Guiderdoni}}}]{kauffmann_etal93}
\bibinfo{author}{\bibfnamefont{G.}~\bibnamefont{{Kauffmann}}},
  \bibinfo{author}{\bibfnamefont{S.~D.~M.} \bibnamefont{{White}}},
  \bibnamefont{and}
  \bibinfo{author}{\bibfnamefont{B.}~\bibnamefont{{Guiderdoni}}},
  \bibinfo{journal}{\mnras} \textbf{\bibinfo{volume}{264}},
  \bibinfo{pages}{201} (\bibinfo{year}{1993}).

\bibitem[{\citenamefont{{Sheth}}(1996)}]{sheth96}
\bibinfo{author}{\bibfnamefont{R.~K.} \bibnamefont{{Sheth}}},
  \bibinfo{journal}{\mnras} \textbf{\bibinfo{volume}{281}},
  \bibinfo{pages}{1277} (\bibinfo{year}{1996}).

\bibitem[{\citenamefont{{Sheth} and {Pitman}}(1997)}]{sheth_pitman97}
\bibinfo{author}{\bibfnamefont{R.~K.} \bibnamefont{{Sheth}}} \bibnamefont{and}
  \bibinfo{author}{\bibfnamefont{J.}~\bibnamefont{{Pitman}}},
  \bibinfo{journal}{\mnras} \textbf{\bibinfo{volume}{289}}, \bibinfo{pages}{66}
  (\bibinfo{year}{1997}).

\bibitem[{\citenamefont{{White}}(1996)}]{white96}
\bibinfo{author}{\bibfnamefont{S.~D.~M.} \bibnamefont{{White}}}, in
  \emph{\bibinfo{booktitle}{Cosmology and Large Scale Structure}}, edited by
  \bibinfo{editor}{\bibfnamefont{R.}~\bibnamefont{{Schaeffer}}},
  \bibinfo{editor}{\bibfnamefont{J.}~\bibnamefont{{Silk}}},
  \bibinfo{editor}{\bibfnamefont{M.}~\bibnamefont{{Spiro}}}, \bibnamefont{and}
  \bibinfo{editor}{\bibfnamefont{J.}~\bibnamefont{{Zinn-Justin}}}
  (\bibinfo{year}{1996}), pp. \bibinfo{pages}{349--+}.

\bibitem[{\citenamefont{{Gelb} and {Bertschinger}}(1994)}]{gelb_bertschinger94}
\bibinfo{author}{\bibfnamefont{J.~M.} \bibnamefont{{Gelb}}} \bibnamefont{and}
  \bibinfo{author}{\bibfnamefont{E.}~\bibnamefont{{Bertschinger}}},
  \bibinfo{journal}{\apj} \textbf{\bibinfo{volume}{436}}, \bibinfo{pages}{467}
  (\bibinfo{year}{1994}), \eprint{astro-ph/9408028}.

\bibitem[{\citenamefont{{Jing}}(1998)}]{jing98}
\bibinfo{author}{\bibfnamefont{Y.~P.} \bibnamefont{{Jing}}},
  \bibinfo{journal}{\apjl} \textbf{\bibinfo{volume}{503}}, \bibinfo{pages}{L9+}
  (\bibinfo{year}{1998}), \eprint{astro-ph/9805202}.

\bibitem[{\citenamefont{{Tormen}}(1998)}]{tormen98}
\bibinfo{author}{\bibfnamefont{G.}~\bibnamefont{{Tormen}}},
  \bibinfo{journal}{\mnras} \textbf{\bibinfo{volume}{297}},
  \bibinfo{pages}{648} (\bibinfo{year}{1998}), \eprint{astro-ph/9802290}.

\bibitem[{\citenamefont{{Wechsler} et~al.}(2002)\citenamefont{{Wechsler},
  {Bullock}, {Primack}, {Kravtsov}, and {Dekel}}}]{wechsler_etal02}
\bibinfo{author}{\bibfnamefont{R.~H.} \bibnamefont{{Wechsler}}},
  \bibinfo{author}{\bibfnamefont{J.~S.} \bibnamefont{{Bullock}}},
  \bibinfo{author}{\bibfnamefont{J.~R.} \bibnamefont{{Primack}}},
  \bibinfo{author}{\bibfnamefont{A.~V.} \bibnamefont{{Kravtsov}}},
  \bibnamefont{and} \bibinfo{author}{\bibfnamefont{A.}~\bibnamefont{{Dekel}}},
  \bibinfo{journal}{\apj} \textbf{\bibinfo{volume}{568}}, \bibinfo{pages}{52}
  (\bibinfo{year}{2002}).

\bibitem[{\citenamefont{{Sheth} and {Tormen}}(2004)}]{sheth_tormen04}
\bibinfo{author}{\bibfnamefont{R.~K.} \bibnamefont{{Sheth}}} \bibnamefont{and}
  \bibinfo{author}{\bibfnamefont{G.}~\bibnamefont{{Tormen}}},
  \bibinfo{journal}{MNRAS} \textbf{\bibinfo{volume}{350}},
  \bibinfo{pages}{1385} (\bibinfo{year}{2004}).

\bibitem[{\citenamefont{{Benson} et~al.}(2005)\citenamefont{{Benson},
  {Kamionkowski}, and {Hassani}}}]{benson_etal05}
\bibinfo{author}{\bibfnamefont{A.~J.} \bibnamefont{{Benson}}},
  \bibinfo{author}{\bibfnamefont{M.}~\bibnamefont{{Kamionkowski}}},
  \bibnamefont{and} \bibinfo{author}{\bibfnamefont{S.~H.}
  \bibnamefont{{Hassani}}}, \bibinfo{journal}{\mnras}
  \textbf{\bibinfo{volume}{357}}, \bibinfo{pages}{847} (\bibinfo{year}{2005}),
  \eprint{astro-ph/0407136}.

\bibitem[{\citenamefont{{Li} et~al.}(2005)\citenamefont{{Li}, {Mo}, and {van
  den Bosch}}}]{li_etal05}
\bibinfo{author}{\bibfnamefont{Y.}~\bibnamefont{{Li}}},
  \bibinfo{author}{\bibfnamefont{H.~J.} \bibnamefont{{Mo}}}, \bibnamefont{and}
  \bibinfo{author}{\bibfnamefont{F.~C.} \bibnamefont{{van den Bosch}}},
  \bibinfo{journal}{\mnras Submitted, astro-ph/0510372}
  (\bibinfo{year}{2005}), \eprint{astro-ph/0510372}.

\bibitem[{\citenamefont{{Sheth}}(1998)}]{sheth98}
\bibinfo{author}{\bibfnamefont{R.~K.} \bibnamefont{{Sheth}}},
  \bibinfo{journal}{\mnras} \textbf{\bibinfo{volume}{300}},
  \bibinfo{pages}{1057} (\bibinfo{year}{1998}).

\bibitem[{\citenamefont{{Monaco}}(1995)}]{monaco95}
\bibinfo{author}{\bibfnamefont{P.}~\bibnamefont{{Monaco}}},
  \bibinfo{journal}{\apj} \textbf{\bibinfo{volume}{447}}, \bibinfo{pages}{23}
  (\bibinfo{year}{1995}), \eprint{astro-ph/9406029}.

\bibitem[{\citenamefont{{Bond} and {Myers}}(1996)}]{bond_myers96}
\bibinfo{author}{\bibfnamefont{J.~R.} \bibnamefont{{Bond}}} \bibnamefont{and}
  \bibinfo{author}{\bibfnamefont{S.~T.} \bibnamefont{{Myers}}},
  \bibinfo{journal}{\apjs} \textbf{\bibinfo{volume}{103}}, \bibinfo{pages}{1}
  (\bibinfo{year}{1996}).

\bibitem[{\citenamefont{{Monaco}}(1997{\natexlab{a}})}]{monaco97a}
\bibinfo{author}{\bibfnamefont{P.}~\bibnamefont{{Monaco}}},
  \bibinfo{journal}{\mnras} \textbf{\bibinfo{volume}{287}},
  \bibinfo{pages}{753} (\bibinfo{year}{1997}{\natexlab{a}}),
  \eprint{astro-ph/9606027}.

\bibitem[{\citenamefont{{Monaco}}(1997{\natexlab{b}})}]{monaco97b}
\bibinfo{author}{\bibfnamefont{P.}~\bibnamefont{{Monaco}}},
  \bibinfo{journal}{\mnras} \textbf{\bibinfo{volume}{290}},
  \bibinfo{pages}{439} (\bibinfo{year}{1997}{\natexlab{b}}),
  \eprint{astro-ph/9606029}.

\bibitem[{\citenamefont{{Audit} et~al.}(1997)\citenamefont{{Audit}, {Teyssier},
  and {Alimi}}}]{audit_etal97}
\bibinfo{author}{\bibfnamefont{E.}~\bibnamefont{{Audit}}},
  \bibinfo{author}{\bibfnamefont{R.}~\bibnamefont{{Teyssier}}},
  \bibnamefont{and} \bibinfo{author}{\bibfnamefont{J.-M.}
  \bibnamefont{{Alimi}}}, \bibinfo{journal}{\aa}
  \textbf{\bibinfo{volume}{325}}, \bibinfo{pages}{540} (\bibinfo{year}{1997}).

\bibitem[{\citenamefont{{Lee} and {Shandarin}}(1998)}]{lee_shandarin98}
\bibinfo{author}{\bibfnamefont{J.}~\bibnamefont{{Lee}}} \bibnamefont{and}
  \bibinfo{author}{\bibfnamefont{S.}~\bibnamefont{{Shandarin}}},
  \bibinfo{journal}{\apj} \textbf{\bibinfo{volume}{500}}, \bibinfo{pages}{14}
  (\bibinfo{year}{1998}).

\bibitem[{\citenamefont{{Furlanetto} et~al.}(2004)\citenamefont{{Furlanetto},
  {Zaldarriaga}, and {Hernquist}}}]{furlanetto_etal04}
\bibinfo{author}{\bibfnamefont{S.}~\bibnamefont{{Furlanetto}}},
  \bibinfo{author}{\bibfnamefont{M.}~\bibnamefont{{Zaldarriaga}}},
  \bibnamefont{and}
  \bibinfo{author}{\bibfnamefont{L.}~\bibnamefont{{Hernquist}}},
  \bibinfo{journal}{\apj} \textbf{\bibinfo{volume}{613}}, \bibinfo{pages}{1}
  (\bibinfo{year}{2004}).

\bibitem[{\citenamefont{{Zhang} and {Hui}}(2006)}]{zhang_hui06}
\bibinfo{author}{\bibfnamefont{J.}~\bibnamefont{{Zhang}}} \bibnamefont{and}
  \bibinfo{author}{\bibfnamefont{L.}~\bibnamefont{{Hui}}},
  \bibinfo{journal}{\apj} \textbf{\bibinfo{volume}{641}}, \bibinfo{pages}{641}
  (\bibinfo{year}{2006}), \eprint{astro-ph/0508384}.

\bibitem[{\citenamefont{{Arfken} and {Weber}}(1995)}]{arfken_weber95}
\bibinfo{author}{\bibfnamefont{G.~B.} \bibnamefont{{Arfken}}} \bibnamefont{and}
  \bibinfo{author}{\bibfnamefont{H.~J.} \bibnamefont{{Weber}}},
  \bibinfo{journal}{Materials and Manufacturing Processes}
  (\bibinfo{year}{1995}).

\bibitem[{\citenamefont{{Gao} et~al.}(2005)\citenamefont{{Gao}, {Springel}, and
  {White}}}]{gao_etal05}
\bibinfo{author}{\bibfnamefont{L.}~\bibnamefont{{Gao}}},
  \bibinfo{author}{\bibfnamefont{V.}~\bibnamefont{{Springel}}},
  \bibnamefont{and} \bibinfo{author}{\bibfnamefont{S.~D.~M.}
  \bibnamefont{{White}}}, \bibinfo{journal}{MNRAS}
  \textbf{\bibinfo{volume}{363}}, \bibinfo{pages}{L66} (\bibinfo{year}{2005}).

\bibitem[{\citenamefont{{Wechsler} et~al.}(2006)\citenamefont{{Wechsler},
  {Zentner}, {Bullock}, {Kravtsov}, and {Allgood}}}]{wechsler_etal06}
\bibinfo{author}{\bibfnamefont{R.~H.} \bibnamefont{{Wechsler}}},
  \bibinfo{author}{\bibfnamefont{A.~R.} \bibnamefont{{Zentner}}},
  \bibinfo{author}{\bibfnamefont{J.~S.} \bibnamefont{{Bullock}}},
  \bibinfo{author}{\bibfnamefont{A.~V.} \bibnamefont{{Kravtsov}}},
  \bibnamefont{and}
  \bibinfo{author}{\bibfnamefont{B.}~\bibnamefont{{Allgood}}},
  \bibinfo{journal}{ApJ, In Press (astro-ph/0512416)}  (\bibinfo{year}{2006}).

\bibitem[{\citenamefont{{Wetzel} et~al.}(2006)\citenamefont{{Wetzel}, {Cohn},
  {White}, {Holz}, and {Warren}}}]{wetzel_etal06}
\bibinfo{author}{\bibfnamefont{A.~R.} \bibnamefont{{Wetzel}}},
  \bibinfo{author}{\bibfnamefont{J.~D.} \bibnamefont{{Cohn}}},
  \bibinfo{author}{\bibfnamefont{M.}~\bibnamefont{{White}}},
  \bibinfo{author}{\bibfnamefont{D.~E.} \bibnamefont{{Holz}}},
  \bibnamefont{and} \bibinfo{author}{\bibfnamefont{M.~S.}
  \bibnamefont{{Warren}}}, \bibinfo{journal}{ArXiv Astrophysics e-prints}
  (\bibinfo{year}{2006}), \eprint{astro-ph/0606699}.

\bibitem[{\citenamefont{{Maccio'} et~al.}(2006)\citenamefont{{Maccio'},
  {Dutton}, {van den Bosch}, {Moore}, {Potter}, and {Stadel}}}]{maccio_etal06}
\bibinfo{author}{\bibfnamefont{A.~V.} \bibnamefont{{Maccio'}}},
  \bibinfo{author}{\bibfnamefont{A.~A.} \bibnamefont{{Dutton}}},
  \bibinfo{author}{\bibfnamefont{F.~C.} \bibnamefont{{van den Bosch}}},
  \bibinfo{author}{\bibfnamefont{B.}~\bibnamefont{{Moore}}},
  \bibinfo{author}{\bibfnamefont{D.}~\bibnamefont{{Potter}}}, \bibnamefont{and}
  \bibinfo{author}{\bibfnamefont{J.}~\bibnamefont{{Stadel}}},
  \bibinfo{journal}{\mnras Submitted, astro-ph/0608157}
  (\bibinfo{year}{2006}), \eprint{astro-ph/0608157}.

\bibitem[{\citenamefont{{Berlind} et~al.}(2006)\citenamefont{{Berlind},
  {Kazin}, {Blanton}, {Pueblas}, {Scoccimarro}, and {Hogg}}}]{berlind_etal06}
\bibinfo{author}{\bibfnamefont{A.~A.} \bibnamefont{{Berlind}}},
  \bibinfo{author}{\bibfnamefont{E.}~\bibnamefont{{Kazin}}},
  \bibinfo{author}{\bibfnamefont{M.~R.} \bibnamefont{{Blanton}}},
  \bibinfo{author}{\bibfnamefont{S.}~\bibnamefont{{Pueblas}}},
  \bibinfo{author}{\bibfnamefont{R.}~\bibnamefont{{Scoccimarro}}},
  \bibnamefont{and} \bibinfo{author}{\bibfnamefont{D.~W.}
  \bibnamefont{{Hogg}}}, \bibinfo{journal}{ArXiv Astrophysics e-prints}
  (\bibinfo{year}{2006}), \eprint{astro-ph/0610524}.

\bibitem[{\citenamefont{{Wang} et~al.}(2006)\citenamefont{{Wang}, {Mo}, and
  {Jing}}}]{wang_etal06}
\bibinfo{author}{\bibfnamefont{H.~Y.} \bibnamefont{{Wang}}},
  \bibinfo{author}{\bibfnamefont{H.~J.} \bibnamefont{{Mo}}}, \bibnamefont{and}
  \bibinfo{author}{\bibfnamefont{Y.~P.} \bibnamefont{{Jing}}},
  \bibinfo{journal}{\mnras, Submitted (astro-ph/0608690)}
  (\bibinfo{year}{2006}).

\bibitem[{\citenamefont{{Scoccimarro} and {Sheth}}(2002)}]{scoccimarro_sheth02}
\bibinfo{author}{\bibfnamefont{R.~.} \bibnamefont{{Scoccimarro}}}
  \bibnamefont{and} \bibinfo{author}{\bibfnamefont{R.~K.}
  \bibnamefont{{Sheth}}}, \bibinfo{journal}{\mnras}
  \textbf{\bibinfo{volume}{329}}, \bibinfo{pages}{629} (\bibinfo{year}{2002}).

\bibitem[{\citenamefont{{Scoccimarro}}(1998)}]{scoccimarro98}
\bibinfo{author}{\bibfnamefont{R.}~\bibnamefont{{Scoccimarro}}},
  \bibinfo{journal}{\mnras} \textbf{\bibinfo{volume}{299}},
  \bibinfo{pages}{1097} (\bibinfo{year}{1998}).

\end{thebibliography}

\end{document}